\documentclass{aa}  
\usepackage{graphicx}
\usepackage{txfonts}
\usepackage[colorlinks=true,allcolors=cyan]{hyperref}
\usepackage{color}
\usepackage{xcolor}
\usepackage{multirow}
\usepackage[normalem]{ulem}
\usepackage{array}
\newcolumntype{P}[1]{>{\centering\arraybackslash}p{#1}}

\begin{document} 

\title{Detecting clusters of galaxies and active galactic nuclei in an eROSITA all-sky survey digital twin \thanks{Tables \ref{tab:column_descr}, \ref{tab:column_descr_esass}, and \ref{tab:column_descr_events} are only available in electronic form
at the CDS via anonymous ftp to cdsarc.u-strasbg.fr (130.79.128.5)
or via \url{http://cdsweb.u-strasbg.fr/cgi-bin/qcat?J/A+A/}}}
\author{
R. Seppi\inst{1}\thanks{E-mail: rseppi@mpe.mpg.de} \and 
J. Comparat\inst{1} \and 
E. Bulbul\inst{1} \and
K. Nandra\inst{1} \and
A. Merloni\inst{1} \and
N. Clerc\inst{2} \and
T. Liu\inst{1} \and
V. Ghirardini\inst{1} \and
A. Liu\inst{1} \and
M. Salvato\inst{1} \and
J. S. Sanders\inst{1} \and
J. Wilms\inst{3} \and
T. Dwelly\inst{1} \and 
T. Dauser\inst{3} \and
O. K\"{o}nig\inst{3} \and 
M. E. Ramos-Ceja\inst{1} \and
C. Garrel\inst{1} \and
T. H. Reiprich\inst{4}
}
\institute{
Max-Planck-Institut f\"{u}r extraterrestrische Physik (MPE), Giessenbachstrasse 1, D-85748 Garching bei M\"unchen, Germany
\and
IRAP, Université de Toulouse, CNRS, UPS, CNES, Toulouse, France \and
Dr. Karl-Remeis-Sternwarte and ECAP, Sternwartstr. 7, 96049 Bamberg, Germany \and
Argelander-Institut f\"{u}r Astronomie (AIfA), Universit\"{a}t Bonn, Auf dem H\"{u}gel 71, 53121, Bonn, Germany
}

\date{Accepted XXX. Received YYY; in original form ZZZ}

\abstract{
The extended ROentgen Survey with an Imaging Telescope Array (eROSITA) onboard the Spectrum-Roentgen-Gamma (SRG) observatory is revolutionizing X-ray astronomy. The mission provides unprecedented samples of active galactic nuclei (AGN) and clusters of galaxies, with the potential of studying astrophysical properties of X-ray sources and measuring cosmological parameters using X-ray-selected samples with higher precision than ever before.  }
{
%
We aim to study the detection, and the selection of AGN and clusters of galaxies in the first eROSITA all-sky survey, and to characterize the properties of the source catalog.  }
{
%
We produced a half-sky simulation at the depth of the first eROSITA survey (eRASS1), by combining models that truthfully represent the population of clusters and AGN. In total, we simulated 1\,116\,758 clusters and 225\,583\,320 AGN. We ran the standard eROSITA detection algorithm, optimized for extragalactic sources. We matched the input and the source catalogs with a photon-based matching algorithm.
}
{%
We perfectly recovered the bright AGN and clusters. We detected half of the simulated AGN with flux larger than 2$\times$10$^{-14}$ erg/s/cm$^2$ as point sources and half of the simulated clusters with flux larger than 3$\times$10$^{-13}$ erg/s/cm$^2$ as extended sources in the 0.5 -- 2.0 keV band. 
We quantified the detection performance in terms of completeness, false detection rate, and contamination.
We studied the population in the source catalog according to multiple cuts of source detection and extension likelihood. We find that the latter is suitable for removing contamination, and the former is very efficient in minimizing the false detection rate. 
We find that the detection of clusters of galaxies is mainly driven by flux and exposure time. It additionally depends on secondary effects, such as the size of the clusters on the sky plane and their dynamical state. The cool core bias mostly affects faint clusters classified as point sources, while its impact on the extent-selected sample is small. We measured the fraction of the area covered by our simulation as a function of limiting flux. We measured the X-ray luminosity of the detected clusters and find that it is compatible with the simulated values.}
{
We discuss how to best build samples of galaxy clusters for cosmological purposes, accounting for the nonuniform depth of eROSITA.
This simulation provides a digital twin of the real eRASS1.
}

\maketitle

\section{Introduction}

Our knowledge of the large-scale structure (LSS) of the Universe has dramatically improved in the past decades thanks to a variety of surveys at different wavelengths. A wealth of information about the matter distribution on cosmological scales is obtained by optical data from galaxy clustering, measured by the Two-degree-Field Galaxy Redshift Survey \citep[2dFGRS, ][]{Colless2001_2dF}, the Galaxy and Mass Assembly (GAMA) Survey \citep[][]{Driver2009A&G....50e..12D}, the VIMOS Public Extragalactic Redshift Survey \citep[VIPERS, ][]{delaTorre2013A&A_VIMOS}, the Dark Energy Survey \citep[DES, ][]{Abbott2018DES1_clustering}, the Kilo-Degree Survey \citep[KiDS,][]{Joudaki2018KIDSclustering}, the Hyper Suprime-Cam Subaru Strategic Program \citep[HSC-SSP,][]{Hikage2019PASJHSC_shear}, and the Sloan Digital Sky Survey \citep[SDSS,][]{Alam2021PhRvD.103h3533A_SDSS}. Complementary data in the millimeter range trace the large-scale distribution of matter thanks to the lensing of the cosmic microwave background \citep[CMB, ][]{Sherwin2012PhRvDlensingcmb, Planck2014A&A_LSS_lensing}. In addition, large samples of extragalactic sources are provided by X-ray surveys, such as ROSAT \citep[][]{Boller2016A&AROSAT_2rxs} and the extended ROentgen Survey with an Imaging Telescope Array \citep[eROSITA, ][]{Merloni2012, Predehl2021A&A...647A...1P}. It is important to consider both galaxy clusters and active galactic nuclei (AGN) in this context: they both trace the LSS. They are fundamental to shedding light on the hot and energetic large-scale structure of the Universe.

Clusters of galaxies populate the most massive bound dark matter haloes in the Universe. They are the largest known virialized structures \citep{Kravtsov2012ARA&ABorgani, Pratt2019SSRv..215...25P}. 
In the context of hierarchical structure formation \citep[][]{White1991ApJ_hiercarchical}, they assemble at late times and reside in the nodes of the cosmic large-scale structure \citep[][]{Lacey_Cole1993MNRAS.262..627L, Springel2005b, Angulo2012, Klypin2016, Ishiyama2021MNRAS.506.4210I_UCHUU}. Their abundance as a function of mass and redshift (i.e., the measure of the halo mass function) is dependent on cosmological parameters \citep{Tinker2008, Allen2011, Lesci2022A&A...659A..88L, Clerc2022arXiv220311906C_review}. This makes them a great tool for cosmological studies. 
Galaxy clusters are 
observed in optical data as an over-density of red galaxies \citep[e.g.,][]{Rykoff2014redmapper, Abbott2020DESY1_clusters} or as peaks in weak-lensing convergence maps \citep[e.g.,][]{Miyazaki2018PASJ_WLclu}, by distortion of the CMB due to the Sunyaev–Zel’dovich (SZ) effect in the millimeter band \citep[e.g.,][]{Staniszewski2009ApJspt_clusters, Planck2016A&A_SZ_sources} and by extended emission in the X-ray band  \citep[e.g.,][]{Bohringer2004A&Areflex, Adami2018A&A...620A...5A, Finoguenov2020A&A...638A.114F, 2022A&A_LiuAng_eFEDS_clu}. 
The combination of multiwavelength data is key for a complete description of galaxy clusters.
On the one hand, optical surveys have the highest source density, which provides the largest samples of clusters using photometric data \citep[][]{Oguri2014MNRASCAMIRA, Bleem2015ApJS_Blanco}. 
On the other hand, pointed observations with interferometers in the radio and millimeter bands provide observations with extremely high angular resolution \citep[][]{Pasini2021arXiv210614524P}. In addition, SZ surveys with telescopes such as \emph{Planck} \citep[][]{Planck_2014}, the South Pole Telescope \citep[SPT, ][]{Bleem2015ApJS..216...27B}, or the Atacama Cosmology Telescope \citep[ACT, ][]{Hilton2021ApJACT} are effective in detecting high-redshift objects, thanks to the redshift-independent SZ signal. 
X-ray observations are particularly suitable to study clusters of galaxies. Clusters are the brightest extragalactic extended sources in the X-ray band \citep{Rosati2002xray_cluster_evo}, they emit mainly due to thermal bremsstrahlung from the hot intra-cluster medium \citep[][]{CavaliereFuscoFermiano1976A&A....49..137C} and their emissivity depends on the radial density profile. 

Active galactic nuclei (AGN) are very luminous objects, powered by the accretion of rich gas reservoirs onto super-massive black holes, and constitute the majority of the extragalactic sources detected in X-ray surveys \citep[see][for a review]{Padovani2017A&ARv..25....2P}. A large sample of AGN enables studies of the general evolution of supermassive black holes \citep[][]{Kauffmann2000MNRAS.311..576K}, the properties of the host galaxy \citep{Ferrarese2000ApJ...539L...9F}, the AGN clustering properties \citep[][]{Koutoulidis2013MNRAS.428.1382K, Viitanen2019A&A...629A..14V}, and their link to the underlying dark matter large-scale structure \citep{Fanidakis2011MNRAS.410...53F, Georgakakis2019agn}, as well as different channels through which these objects are formed \citep{Mayer2019RPPh...82a6901M}, and the mechanisms triggering bursts of X-ray radiation \citep{Arcodia2021Natur.592..704A}. 

With eROSITA onboard Spectrum-Roentgen-Gamma (SRG), a new era in X-ray astronomy is now unfolding \citep[][]{Merloni2012, Predehl2021A&A...647A...1P}. 
It has seven telescope modules with 54 nested mirror shells each. The Half Energy Width (HEW) of the point spread function (PSF) is about 15\arcsec\ for each module. 
eROSITA will scan the full X-ray sky eight times in four years, resulting in a set of eight all-sky surveys. The sensitivity of the final cumulative all-sky survey (eRASS:8) will be 25 times higher than its predecessor the ROSAT all-sky survey \citep[][]{Voges1999RASS,Boller2016A&AROSAT_2rxs}. During its performance verification phase, SRG-eROSITA successfully completed a mini-survey in the $\sim$140 square degrees eROSITA Final Equatorial Depth Survey \citep[eFEDS,][]{Brunner2022_efedscat}. 
Since December 2019, eROSITA is performing all-sky surveys. The sky is split in half between the German (eROSITA\_DE) and Russian consortium (eROSITA\_RU). The eROSITA\_DE area is split into 2\,447 tiles with a small overlap for data processing purposes. Of these, 2\,248 are uniquely owned by the German consortium, and the additional 199 are shared. Each tile covers a unique area of $\sim$8.7 square degrees.\\
eROSITA is predicted to ultimately detect a total of about 10$^{5}$ clusters of galaxies after the final cumulative all-sky survey (eRASS:8), the largest sample of X-ray-selected galaxy clusters to date. This will allow a variety of studies involving the cluster X-ray luminosity function \citep[][]{Mullis2004ApJ...607..175M, Koens2013MNRAS.435.3231K, Finoguenov2015A&A_CDFS, Adami2018A&A...620A...5A, Clerc2020MNRAS.497.3976C, 2022A&A_LiuAng_eFEDS_clu}, the clustering of galaxy clusters \citep[][]{Veropalumbo2014MNRAS.442.3275V, Marulli2018A&A...620A...1M, Marulli2021ApJ...920...13M, Lindholm2021A&A...646A...8L}, and provide powerful constraints on cosmological parameters such as the normalization of the power spectrum $\sigma_8$ and the matter content of the Universe $\Omega_M$ \citep{Borgani2008LNP...740..287B, Vikhlinin2009ApJ...692.1060V, Mantz2015cosmology, Pierre2016XLL, Schellenberger2017reiprich, Pacaud2018XLL_cosmology, IderChitham2020MNRAS.499.4768I, Garrel2021arXiv210913171G}. A prediction of the eROSITA cluster count cosmology capabilities is studied by \citet{Pillepich2012MNRAS.422...44P, Pillepich2018}.
A total number of about three million sources, most of which are AGN, are expected to be detected in eRASS:8, a factor of 20 better than ROSAT.  

An efficient and accurate detection of extragalactic sources is key to properly sampling the cosmic web and making the most out of the large samples provided by eROSITA. \\ 
The identification of galaxy clusters in X-ray surveys like eROSITA is affected by Poisson count noise in the low photon count regime and by the redshift-dimming effect on the cluster surface brightness.
Cluster samples selected from X-ray surveys are primarily flux-limited \citep[e.g., REFLEX, ][]{Bohringer2004A&Areflex}. The detection of clusters also depends on secondary effects, such as their extent on the sky, or the low surface brightness of very extended objects \citep[][]{Pacaud2006XMM_det, Burenin2007ApJS..172..561B, Finoguenov2020A&A...638A.114F}. In this context, the cool core bias and the dynamical state of galaxy clusters have also been studied in recent years \citep[][]{Hudson2010, Eckert2011, Rossetti2016dyn_state, AndradeSantos2017ApJ...843...76A, Kafer2019, Ghirardini2021morph_pars}. Relaxed clusters develop an efficient cooling toward their center, which enhances the X-ray emission in the inner region. Such peaked surface brightness profiles possibly bias the detection toward relaxed structures. This has an impact on cosmological studies using the halo mass function \citep[][]{Seppi2021A&A...652A.155S}. \\
The cross-correlation between clusters and AGN in the LSS creates an interplay between point and extended sources in the detection process. A detailed understanding of the point sources is fundamental to investigate not only the X-ray background and the completeness of the observed sample \citep{Georgakakis2008MNRAS.388.1205G}, but also the fraction of clusters that are misclassified as a point source \citep[][]{Pacaud2006XMM_det, Burenin2007ApJS..172..561B}. This happens because of the small size of high redshift clusters, the peaked emission from compact nearby groups, or the presence of a central AGN in the cluster, which can boost the detection of high redshift clusters \citep{McDonald2012Natur.488..349M_Phoenix, Trudeau2020A&A...642A.124T}. 
This misclassification is mitigated by multiwavelength follow-up observations. For instance, 
\citet{Salvato2021arXiv210614520S_efedsfollowup} found 346 cluster candidates in the eFEDS point-source catalog by the identification of the red sequence using optical data. An extensive study of these objects is provided by \citet{Bulbul2021arXiv211009544B_clusters_disguise}.

An effective way of investigating the detection and selection effects in surveys is to simulate the observational process in its greatest detail. This approach has been explored using mocks in different wavelengths, from the optical band \citep{Jimeno2017MNRAS.466.2658J, Oguri2018PASJ...70S..20O}, to the X-rays \citep{Liu2013A&A...549A.143L, Pierre2016XLL, Clerc2018A&A...617A..92C}, and the microwave sky \citep{Sehgal2010ApJ...709..920S_MWskysim}, or injecting simulated sources into real images \citep[][]{Suchyta2016MNRAS_Balrog_DES, Everett2022ApJ_Balrog_DES}. It allows accounting for instrumental effects and the observing strategy. 
Studying and quantifying effects that have an impact on the detection is then possible, comparing catalogs of simulated sources and the population that is detected in the simulation. 
Constant improvements in computational power and efficiency provide more detailed mocks. 
Recent progress in dark matter simulations allows to minimize the impact of cosmic variance thanks to the ability to simulate large volumes, but also resolve galaxy-like halos because of the small resolution \citep[e.g.,][]{Klypin2016, Chuang2019_UNIT, Ishiyama2021MNRAS.506.4210I_UCHUU}.\\ 
We study the eROSITA capabilities in the detection of extragalactic sources following this approach. Our goal is to understand the details of AGN and cluster detection and selection effects. These are two important subsequent steps. First, the detection should be optimized to maximize the ability to identify clusters and AGN, and make sure that the algorithm in question is detecting as many real sources as possible. After that, one can focus on selection criteria to clean the catalog of detected sources and obtain a certain sample according to the scientific goal.\\ 
In this paper, we use realistic end-to-end simulations to predict the population of objects observed by eROSITA, with a particular interest in extended sources, that are clusters of galaxies, and AGN. We focus on the eROSITA\_DE sky area. We start from the simulations described by \citet{Comparat2019agn, Comparat2020Xray_simulation}. 
We generate a half-sky simulation at the depth of the first eROSITA all-sky survey (eRASS1), the one reached after six months of operations. We follow the eROSITA scanning strategy. 
Photons are generated for 2438 eROSITA\_DE tiles. The background is directly resampled from the eRASS1 observations. We extend the cluster model from \citet{Comparat2020Xray_simulation} to galaxy groups down to 2$\times$10$^{\rm 13}$ M$_{\odot}$ using the relation between X-ray luminosity and stellar mass \citep{Anderson2015Lx_stellarmass}. \citet{Comparat2022arXiv220105169C} showed that such correction allows matching the relation between projected luminosity around eFEDS central galaxies and their stellar mass remarkably well. 
We run the eSASS (extended Science Analysis Software System) detection algorithm described by \citet{Brunner2022_efedscat}. We build a one-to-one association between simulated objects and the source catalog using the source ID of each simulated photon \citep[][]{Liu2021teng_simulation}, properly linked to a cluster, AGN, star, or the background. 
We assess the performance of the detection in terms of completeness (fraction of simulated objects that are recovered in the source catalog) and purity (fraction of entries in the source catalogs that are assigned to the correct simulated object). Our study follows up on the work of \citet{Liu2021teng_simulation} on the eFEDS simulations. We take one step further, accounting for the larger variations of exposure and background level in eRASS1.\\
This paper is organized as follows. We summarize the main features of the simulation and the X-ray model in Sect. \ref{sec:simulation}. We describe the detection process, the handling of the catalogs, and the classification of the sources in Sect. \ref{sec:data}. We provide our results in Sect. \ref{sec:results}. We study the population in the source catalog, the cumulative number density of AGN and clusters as a function of flux, the completeness of these samples, their relation with purity and contamination, and measure the X-ray luminosity of clusters. We further discuss our results in Sect. \ref{sec:discussion}, including the best strategy to build samples of clusters detected by eROSITA, accounting for the different exposure across the sky.
Finally, we summarize our findings in Sect. \ref{sec:conclusions}. 

\section{Simulated data}
\label{sec:simulation}
We follow the approach described by \citet{Comparat2020Xray_simulation} and create all-sky simulations. A dark matter light cone is built with snapshots at different redshifts. 
Cluster and AGN models are used to predict X-ray emission \citep{Comparat2019agn, Comparat2020Xray_simulation}. We upgrade the cluster model to the galaxy groups regime.
In this section, we review the main features of the simulations and models that are relevant for this analysis. 
The simulated data is released along with the article, see the description in appendix \ref{appendix:DR:sim}.

\subsection{Light cones from N-body dark matter simulations}
A light cone is created with the UNIT1i N-body simulations \citep{Chuang2019_UNIT}. These are computed in a Flat $\Lambda$CDM cosmology \citep{Planck2016cluster_cosmology}. The fiducial parameters are $\rm H_0 = 67.74\ km\ s^{-1}\ Mpc^{-1}$, $\rm \Omega_{m0} = 0.308900$, $\rm \Omega_{b0} = 0.048206$. The size of the simulation box is 1 Gpc/h\footnote{\textit{h} is the dimensionless Hubble constant, equal to the value of H$_0$/100.} and the mass resolution is 1.2$\times$10$^9$ M$_\odot$/h. It allows a detailed modeling of both clusters and AGN. It is suited for studying low mass structures down to 10$^{11}$ M$_\odot$, AGN up to z$\sim$6, and the eROSITA selection function \citep{Liu2021teng_simulation}.

\subsection{X-ray model components}
\label{subsec:model_components}
These simulations combine different source and X-ray background components. We describe each one of them in the following section.

\subsubsection{Galaxy clusters}
\citet{Comparat2020Xray_simulation} introduce a new method to simulate the X-ray emission from galaxy clusters. The principle is to build mock observations using real data as a starting point \citep[e.g.,][]{Kong2020MNRAS.499.3943K, Everett2020arXiv201212825E}. A total sample of 326 clusters is obtained by combining XMM-XXL \citep{Pierre2016XLL}, HIFLUGCS \citep{Reiprich2002HIFLUGCS}, X-COP \citep{Eckert2019xcop} and SPT-Chandra \citep{Sanders2018sptchandra}. Their combination constitutes a relatively fair benchmark for eROSITA observations. Their X-ray properties are well measured inside R$_{\rm 500c}$, the radius encompassing an average density that is 500 times the critical density of the Universe at the redshift of the cluster $\rho_c = 3H^2/8\pi G$, where H is the Hubble parameter and G is the universal gravitational constant. From these clusters, a covariance matrix between redshift, temperature, hydrostatic masses, and emissivity is constructed. 
Simulated emissivity profiles are drawn from the covariance matrix by a Gaussian random process. 
These profiles are assigned to dark matter haloes by a nearest neighbor process, considering mass and redshift. The brightness of the cluster core is linked to the dynamical state of the dark matter halo. 
The initial model 
is constructed using clusters with high counts and signal-to-noise ratio, making it reliable down to masses of $M_{500c} \sim$ 5$\times$10$^{13}$ M$_\odot$. 

In this article, we extend this model to galaxy groups for the eRASS1 simulation as follows. 
We use the relation between stellar mass and X-ray luminosity from \citet{Anderson2015Lx_stellarmass} as a reference. The stellar mass is assigned to halos by an abundance matching scheme \citep[see][and Section \ref{subsubsec:agnmodel}]{Comparat2019agn}.
We infer an average correction as a function of mass to align the scaling relation of the simulation to that of \citet{Anderson2015Lx_stellarmass}. 
The goal scaling relation between X-ray luminosity and the stellar mass of the central galaxy in each halo reads
\begin{equation}
    \log_{\rm 10}L_{\rm x,(0.5-2.0 \text{keV})} = 3\log_{\rm 10}M_{*} + 7.8.
    \label{eq:And15}
\end{equation}
This average correction bends the scaling relation predicted by \citet{Comparat2020Xray_simulation} at low mass to predict lower luminosities for lower mass haloes. 
Importantly, it preserves the scatter in the L$_{\rm X}$--mass scaling relation. 
These values substitute the ones obtained by integrating the emissivity profiles from the original covariance matrix. 
For haloes with a mass larger than M$_{\rm 500c}$ > 10$^{14}$ M$_\odot$ the correction is negligible, but it becomes very important in the mass range $10^{13}$ -- $5\times10^{13}$ M$_\odot$. 
In appendix \ref{app:model_extension}, both panels in Fig. \ref{fig:logNlogS_model} highlight the improvement of the model after applying the correction. The number density of sources as a function of X-ray flux (logN--logS) predicted for masses above $\log_{\rm 10} M/M_{\odot}$ > 13 is in excellent agreement with observations \citep{Finoguenov2007ApJS..172..182F_cosmos, Finoguenov2015A&A_CDFS, Finoguenov2020A&A...638A.114F, 2022A&A_LiuAng_eFEDS_clu, Chiu2021arXiv210705652C_efeds, Bahar2021arXiv211009534B_efeds_scalingrel}. With the eFEDS sample, the method is further validated. 
It offers a more complete picture of the cluster population. 
The relation between X-ray luminosity and $M_{\rm 500c}$ in the second panel of Fig. \ref{fig:logNlogS_model} shows the impact of the correction, especially for groups. 
The predicted values of $\log_{\rm 10}L_{\rm x}$ reach reasonable values of $\sim$ 41 (and below) at $\log_{\rm 10}M/M_{\odot} \sim$ 13.
The improved model is in line with different sets of observations, considering that these are flux-limited samples, whereas the orange curve is built with the complete simulated clusters population \citep{Lovisari2015A&A...573A.118L_scalingrel, Schellenberger2017MNRAS.469.3738S_scalingrel, Bulbul2019ApJ...871...50B_scalingrel, Lovisari2020ApJ...892..102L_scalingrel, Chiu2021arXiv210705652C_efeds, Bahar2021arXiv211009534B_efeds_scalingrel}. 
In general, our correction provides an excellent agreement between the new model and eFEDS clusters sample. 
We provide further details in Appendix \ref{app:model_extension}. In total, we simulate 1\,116\,758 clusters.

\subsubsection{Active galactic nuclei}
\label{subsubsec:agnmodel}
Active galactic nuclei are simulated by an empirical model that reliably reproduces their number density as a function of X-ray luminosity, clustering, and redshift \citep{Georgakakis2019agn, Comparat2019agn}. It is based on stellar mass to halo mass relations \citep{Moster2013MNRAS.428.3121M} and abundance matching to reproduce the hard X-ray AGN luminosity function \citep[][]{Aird2015MNRAS.451.1892A, Buchner2015ApJ} and their number density as a function of flux up to z = 6. It matches the observed AGN duty cycle (fraction of galaxies hosting an active nucleus) by construction \citep[][]{Georgakakis2019agn}. The model extends to very low X-ray fluxes $\sim$1$\times$10$^{-17}$ erg/s/cm$^2$, well under the eROSITA flux limit, which enable a prediction of the X-ray background due to faint AGN. 
For the construction of the AGN population in the eRASS1 simulation, the sky is first divided into 768 HEALPix\footnote{\url{https://healpix.sourceforge.io/}} fields, which ensures faster processing, but also a smaller volume, sampling the luminosity function down to about 10$^{-7}$ sources per Mpc$^3$. This prevents the simulation of extremely bright sources.
The model of the AGN spectra is an absorbed power-law with Compton reflection and a soft scattered component by cold matter (in Xspec {\tt tbabs*(plcabs+pexrav)+zpowerlw)*tbabs}). The spectral index of the power-law is equal to $\Gamma$ = 1.9.
Finally, a fine-grained K-correction is applied to the AGN population \citep[][]{Hogg2002astro.ph.10394H_Kcorr}. The simulation accounts for a cross-correlation between clusters and AGN since they are both generated from the same N-body simulation. We neglect secondary effects regarding the population of halos hosting AGN in cluster environments. Further observational studies involving the fraction of active galaxies in clusters as a function of redshift and a comparison to field galaxies are required to develop such a model, \citep[see][]{Martini2013ApJagnCLU, Koulouridis2014A&AagnCLU, Noordeh2020MNRAS.498.4095N}. In total, we simulate 225\,583\,320 AGN, about 200 times more than the clusters. Among them, 93\,311\,810 produce at least one count within 60\arcsec\ from the center.

\subsubsection{Stars}
Fluxes to be assigned to stars are drawn from the eFEDS logN--logS. We assign them to GAIA DR2 \citep{Gaia2018A&A...616A...1G} true positions randomly. 
The spectrum is a 0.8 keV APEC model at redshift 0. This model is simple, but nonetheless sufficient to mimic the increase of stellar density toward the Milky Way for this simulation at the eRASS1 depth \citep[][]{Schneider2021arXivefedsstars, Salvato2021arXiv210614520S_efedsfollowup}. In total, we simulate 373\,316 stars.

\subsubsection{Background}

Our approach is similar to the one detailed by \citet{Liu2021teng_simulation}, who decompose and re-simulate the eFEDS background, subtracting the contribution from the simulated faint AGN, that partially contribute to the cosmic X-ray background (CXB). 
However, this is not feasible in eRASS1, due to the nonuniform coverage of the sky and background emission.
We update such a method for the eRASS1 simulation. Background photons are obtained by resampling the observed eROSITA background maps, masking identified point and extended sources. This allows the introduction of spatially varying background, that closely follows real data. 
We start from the eROSITA\_DE eRASS1 event lists and source catalogs. Following the masking scheme devised by Comparat et al. (in prep.), the photons are split into two groups. First, we consider source photons: events located within 1.4 times the source radius of a detected source (see Sect. \ref{sec:data} for a definition of the source radius). Secondly, we select background photons: events located further than 1.4 times the source radius of any detected source.
These thresholds guarantee conservative masking of the sources in the event list to obtain a background event list. The complementary set of events constitutes the source event list. The whole dataset is mirrored in the eROSITA\_RU sky, to obtain an all-sky map. This is divided into 49\,125 HEALPix regions, each of them covering $\sim$ 0.84 $\deg^{2}$. The X-ray spectrum and the images of the background events are extracted from these regions. All the spectra are merged into a single mean background spectrum. These inputs are combined to generate a specific SIMPUT\footnote{\url{https://www.sternwarte.uni-erlangen.de/research/sixte/simput.php}, v-2.4.7} file for the mock background, that provides by construction a faithful reproduction of the observed eRASS1 background. 
\subsection{Mock observation}
Photons are simulated with the \textsc{SIXTE}\footnote{\url{https://www.sternwarte.uni-erlangen.de/research/sixte}, v-2.6.0} software \citep{Dauser2019A&A...630A..66D}, a dedicated end-to-end X-ray simulator. \textsc{SIXTE} is the official simulator for eROSITA. 
The result is a list of events with energy, position, and arrival time. 
This approach allows accounting for instrumental effects because the simulator relies on vignetting, energy-dependent PSF, ancillary response file (ARF), and redistribution matrix file (RMF) as input from calibration data. The setup follows the eROSITA all-sky scan strategy \citep{Merloni2012, Predehl2021A&A...647A...1P}.\\
We use the same attitude file from the real observations for the eRASS1 simulation. The attitude file specifies the details of the scanning by the spacecraft. It follows the planned observing strategy, scanning one full great circle every four hours. In addition, we use the same good time intervals (gti) of the real survey. This allows us to account for details such as orbit corrections, when the cameras are switched off, or camera failures, making the simulation an ideal digital twin of the real eRASS1. The total number of events in the simulation, covering about 20 618 square degrees, with energy of 0.2--10 keV is 187\,486\,754. There are 118\,905\,555 photons in the soft band (0.2--2.3 keV). These numbers are indeed very similar to the real data, respectively equal to 194\,350\,024 and 118 815 616 counts. The ratios between these numbers are 0.965 and 1.001 respectively.\\

\section{Data analysis method}
\label{sec:data}

In this section, we describe how the simulated event files are processed and analyzed. The final result is a catalog of sources identified by the detection algorithm. We refer to the latter as the source catalog in the rest of this work.
Only event files in the eROSITA\_DE sky are processed. 
We first generate the photons on the sky plane divided into 768 HEALPix regions and then create specific catalogs for each field. This way we do not simulate the same photons twice in the overlapping regions of different eROSITA tiles. Given our interest in cluster detection, we focus on a single band detection in the soft X-rays (0.2--2.3 keV), where the eROSITA effective area is the highest \citep{Predehl2021A&A...647A...1P}.

\begin{table}[]
    \centering
    \caption{Number of counts by sources detected with given values of detection and extension likelihood}    
    \begin{tabular}{|c|c|c|c|c|c|c|}
    \hline
    \hline
    \rule{0pt}{2.3ex} & \multicolumn{3}{c|}{\textbf{Clusters}} & \multicolumn{3}{c|}{\textbf{AGN}}\\
    \hline 
    \rule{0pt}{2.3ex} \multirow{2}{*}{DET\_LIKE} & \multicolumn{3}{c|}{N events <0.5$\times$R$_{\rm 500c}$} & \multicolumn{3}{c|}{N events <30$\arcsec$} \\
    \rule{0pt}{2.0ex} & ALL & CLU & BG & ALL & AGN & BG \\
    \hline
    \rule{0pt}{2.2ex} 5 & 21.3 & 8.7 & 10.1 & 5.1 & 3.7 & 1.3 \\
    8   & 25.8 & 11.4 & 11.4 & 6.6 & 5.0 & 1.5 \\
    10  & 30.4 & 13.8 & 13.0 & 7.7 & 6.0 & 1.6 \\
    15  & 42.6 & 21.3 & 16.9 & 10.1 & 8.4 & 1.7 \\
    20  & 48.8 & 25.4 & 17.9 & 12.5 & 10.6 & 1.7 \\ 
    25  & 74.4 & 36.6 & 23.6 &  14.6 & 12.7 & 1.8 \\
    50  & 100.8 & 62.3 & 29.8 &  25.2 & 22.6 & 2.4 \\
    75  & 152.4 & 96.3 & 42.5 &  35.6 & 32.3 & 3.0 \\
    100 & 209.0 & 127.2 & 61.5 & 46.5 & 42.1 & 3.7 \\
    \hline
    \end{tabular}
    \begin{tabular}{|c|c|c|c|}
    \hline
    \rule{0pt}{2.3ex} & \multicolumn{3}{c|}{\textbf{Clusters}} \\
    \hline
    \rule{0pt}{2.3ex} \multirow{2}{*}{EXT\_LIKE} & \multicolumn{3}{c|}{N events < 0.5$\times$R$_{\rm 500c}$} \\
    \rule{0pt}{2.0ex} & ALL & CLU & BG \\
    \hline
    \rule{0pt}{2.2ex} 6 & 69.4 & 31.4 & 31.2 \\
    8 & 78.5 & 42.7 & 27.2 \\
    10 & 89.7 & 46.9 & 33.2 \\
    15 & 103.9 & 59.1 & 34.0 \\
    20 & 139.9 & 67.5 & 32.5 \\
    25 & 144.7 & 90.9 & 42.3 \\
    50 & 275.9 & 168.6 & 77.8 \\
    75 & 405.9 & 284.2 & 95.6 \\
    100 & 530.1 & 376.5 & 119.1 \\
    \hline    
    \end{tabular}
    \vskip.1cm
    \footnotesize{\textbf{Notes.} The first column in the upper (lower) table reports the value of detection (extent) likelihood measured on a source. For clusters, the other columns show the total number of counts generated by all sources (ALL, includes photons from clusters, AGN, stars, and the background) inside half R$_{\rm 500c}$, the ones only generated by clusters (CLU) and the ones produced by the background (BG). For AGN, we report the total number of events within 30\arcsec, the ones generated by AGN and by the background.}
    \label{tab:counts_detlike_extlike}
\end{table}

\subsection{eSASS detection}

\begin{figure*}[h]
    \centering

    \includegraphics[width=1.95\columnwidth]{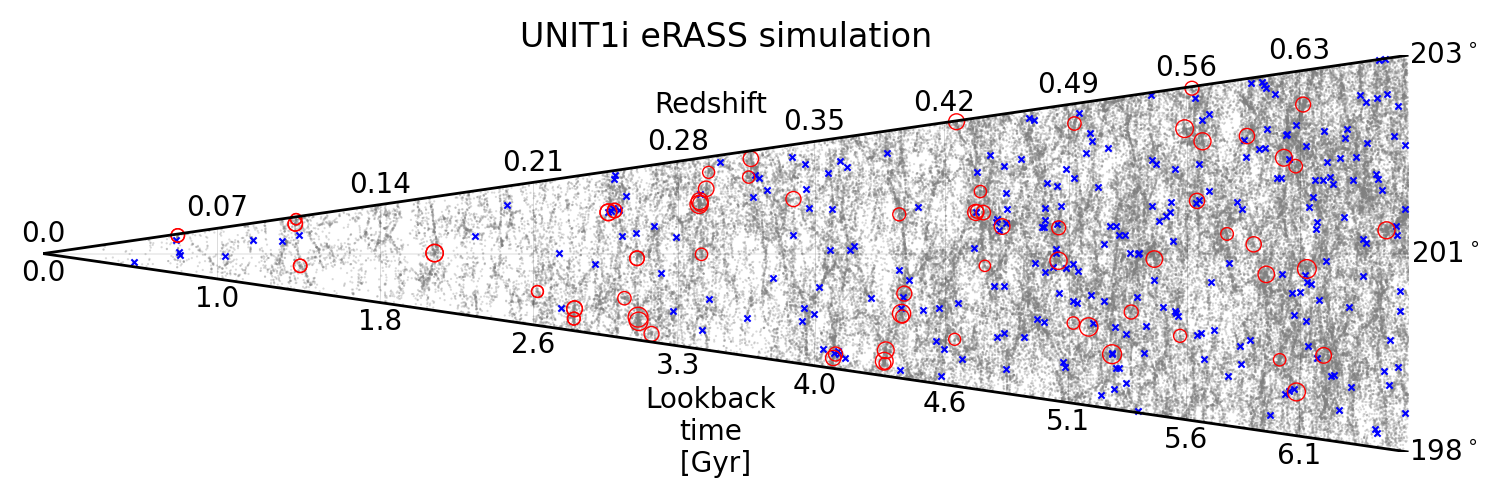}    
    \includegraphics[width=1.95\columnwidth]{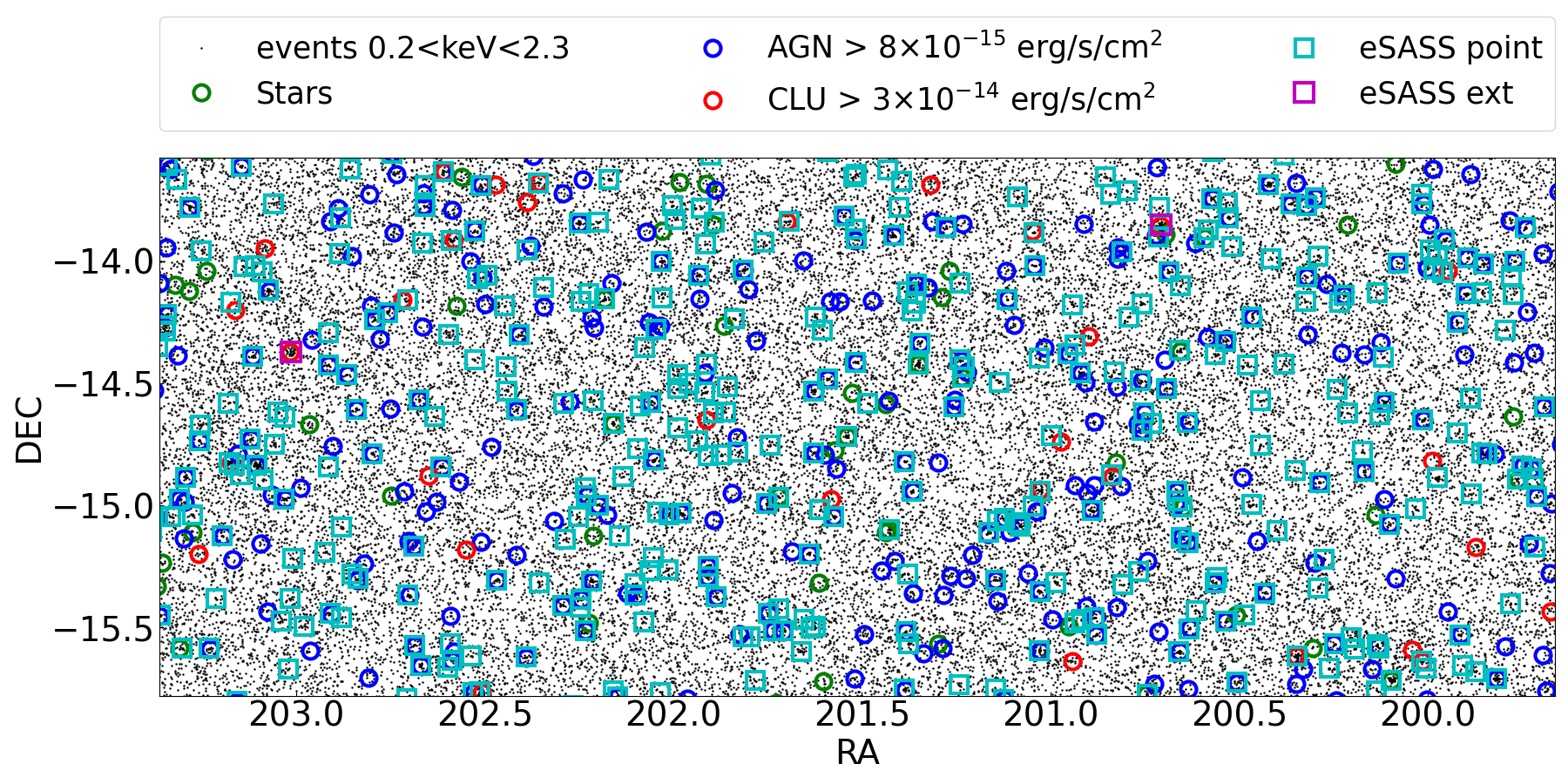}    
		\caption{Large scale distribution of extragalactic sources and their X-ray view in the simulation. \textbf{Top panel}: light cone of the UNIT1i-eRASS1 simulation. The wedge shows the fraction of the sky enclosed by the same RA and DEC of the bottom panel as a function of redshift and lookback time. The galaxies tracing the large-scale structure are shown in grey. The AGN are denoted in blue. The red circles show clusters and groups. The size of the circle is proportional to the mass of the object. 
		\textbf{Bottom panel}: central regions of tile 202105 of the eRASS1 simulation. This is the projection on the plane of the sky of light cone shown in the top panel. Photons with energies between 0.2 to 2.3 keV are shown by black dots, simulated stars by green circles, simulated AGN by blue circles, simulated clusters by red circles, eSASS extended detections by magenta squares, and eSASS point-like detections by cyan squares. 
	}
    \label{fig:evts_w_cats}
\end{figure*}

Each simulated tile is processed with the eROSITA Standard Analysis Software System (eSASS, version eSASSusers\_201009) \citep{Brunner2022_efedscat}. Starting from the calibrated event file, we produce 3.6$^\circ\times$3.6$^\circ$ images for the eRASS1 simulation and the corresponding exposure maps, using all 7 telescope modules, in the soft X-ray band 0.2--2.3 keV. The detection relies on a sliding box algorithm, that looks for overdensities of photons over the background map. It follows the subsequent steps.

1. {\tt erbox}: the image is scanned by a sliding cell, which marks potential sources if the signal-to-noise ratio is higher than a given threshold. This initial list of potential sources contains a large number of false detection, but maximizes the completeness. \\
2. {\tt erbackmap}: the potential sources are masked by constructing a detection mask and the image is interpolated to create an adaptively smoothed background image. This process is iterated three times, to converge toward a more robust background map \citep[][]{Brunner2022_efedscat, Liu2021teng_simulation}.\\
3. {\tt ermldet}: each box marked as a potential source is analyzed by a maximum likelihood PSF-fitting algorithm, based on the position, count rate, and extent of the source. It compares the distribution of counts to a $\beta$ model \citep{CavaliereFuscoFermiano1976A&A....49..137C} convolved with the eROSITA PSF. It allows a simultaneous fitting of multiple sources. Different choices of the minimum likelihood threshold control the purity of the sample, decreasing the false detection rate when increasing the threshold. This task produces a catalog of sources and a source map. \\
Sources are assigned a significance of the detection (detection likelihood), extension of the best fitting $\beta$ model (extent) and significance of the extended model over the point-like one (extension likelihood). These parameters are computed by minimizing the C-statistic \citep{Cash1979ApJ...228..939C} in Eq.\ref{eq:cstat}:
\begin{equation}
        C = 2 \sum_{i=1}^{N}(e_i - n_i\text{ln}e_i),
        \label{eq:cstat}
\end{equation}
where $n_i$ is the measured number of events in each pixel and $e_i$ is the expected value from the model. 
The significance of each source is computed by comparing the best fitting model to the zero count case $\Delta C = C_{\rm null} - C_{\rm fit}$ \citep[see][Sect. A.5]{Brunner2022_efedscat}. The probability that a source arises from a random background fluctuation is computed using the regularized incomplete
Gamma function P$_{\Gamma}$.
\begin{equation}
        P = 1 - P_{\Gamma}(\frac{\nu}{2}, \frac{\Delta C}{2}),
        \label{eq:prob_gamma}
\end{equation}
where $\nu$ is the number of degrees of freedom in the model. This is equal to three (four) for point (extended) sources, corresponding to positions on the pixels X and Y, count rate (and core radius of the $\beta$ model) for our study, which only uses one detection band. The likelihood for each source is finally related to the natural logarithm of such probability: 
\begin{equation}
        \mathcal{L}_{\rm det} =  -\text{ln}P.
        \label{eq:detlike}
\end{equation}
This gives a set of two fundamental parameters for each detection: DET\_LIKE ($\mathcal{L}_{\rm DET}$), and EXT\_LIKE ($\mathcal{L}_{\rm EXT}$). The first (second) one is related to the probability of identifying a spurious point (extended) source, exponentially proportional to $-$DET\_LIKE ($-$EXT\_LIKE).
The core radius of the best-fitting extended beta model is also provided. 
It is set to zero for point sources, its minimum and maximum values are 8\arcsec\ and 60\arcsec. A constant $\beta$ = 2/3 is assumed for the model so that the slope of the profile is equal to $-$3\footnote{S(r) = S$_0$[1+(r/r$_c$)$^2$]$^{-3\beta+1/2}$}. We show that on average our model generates profiles that are compatible with this assumption in Appendix \ref{app:model_extension}. The minimum thresholds of DET\_LIKE and EXT\_LIKE are extremely important in this step. They have a significant impact on the completeness and purity of the source catalog, see Sect. \ref{subsec:detection_efficieny}. We follow the same task processing as the eFEDS data, choosing values of detlikemin = 5 and extlikemin = 6 \citep{Brunner2022_efedscat}. The values of detection and extension likelihood are correlated to the number of events from a given source and from the local background by construction. AGN producing five counts on average are detected with DET\_LIKE = 10. Clusters of galaxies require a larger amount of events to be detected. A value of DET\_LIKE = 5 is measured for clusters with nine source counts and ten background counts inside half R$_{\rm 500c}$. Classifying the clusters as extended sources requires a larger number of events. A value of EXT\_LIKE = 6 is measured for clusters with about 30 counts inside half R$_{\rm 500c}$. When the ratio between source and background photons increases, the detection and extension likelihood rise as well. A value of EXT\_LIKE (DET\_LIKE) of 25 is measured on average for clusters with 91 (37) counts against 42 (24) events generated by the background. We provide a summary in Table \ref{tab:counts_detlike_extlike}. It shows the average number of counts generated by all sources, including clusters, agn, stars, and background, and the ones only generated by clusters (AGN) and background in the top left (right) panels at fixed values of detection likelihood. The bottom panel displays the counts at given extension likelihood value. \\
4. {\tt apetool}: we perform source aperture photometry and compute the sensitivity map for each simulated tile. This gives the minimum number of counts necessary to detect a point-like source as a function of position in the sky, and at a given Poisson false detection probability threshold.\\
5. {\tt srctool}: we measure the radius that maximizes the signal-to-noise ratio for each source. We refer to this parameter as source radius (srcRAD).\\
6. {\tt ersenmap}: we compute the sensitivity map for extended sources. This gives the minimum flux necessary for a source to be detected at a given DET\_LIKE threshold.\\
7. {\tt apetool}: we perform again source aperture photometry focusing on the extended sources and different apertures of 60, 90, 120, 150, 180, 240, 300, and 600\arcsec.

We perform the source detection in the soft (0.2--2.3 keV) X-ray band. 
In principle, one could choose specific detection and extension likelihood threshold according to different needs. We choose to characterize the extended sources without additional selections, using detlikemin = 5 and extlikemin = 6. This keeps our cluster catalog reasonably complete (down to some flux limit), without rejecting faint sources that are potentially interesting. Figure \ref{fig:evts_w_cats} shows an example of this whole process. It displays a wedge of the simulated light cone in the top panel, showing galaxies that trace the large-scale structure in grey and how this is populated by AGN in blue, and clusters and groups in red. The bottom panel shows the projection on the sky plane of the events emitted by the sources in the wedge. It displays simulated photons in the soft X-ray band (black dots), the simulated stars (green circles), AGN (blue circles), clusters (red circles), extended detections (magenta squares), and point-like detections (cyan squares). This tile gives a typical view of different possible cases. Red circles within a magenta square identify simulated clusters that are detected as extended, whereas red circles within a cyan square denote clusters detected as point sources. Similarly, input AGN and stars detected as point sources are shown by blue and green circles within cyan squares. Every circle (red, blue, or green) without a corresponding square denotes a simulated object that has not been detected. We show clusters and AGN respectively down to low flux limits of 3$\times$10$^{-14}$ erg/s/cm$^2$ and 8$\times$10$^{-15}$ erg/s/cm$^2$. This explains the undetected objects in Fig. \ref{fig:evts_w_cats}. Finally, background fluctuations that are detected as spurious sources are identified by squares without any circle.

The X-ray background drives the detection process, especially for faint sources. 
We compare the background maps computed on the simulation and on the real eRASS1 data. We find that the simulated background is overestimated by $\sim$ 10$\%$ compared to the observations. This is expected, because the cosmic X-ray background due to faint AGN is present both in the real eRASS1 background maps used to generate the background model, and as the simulated population of low-flux AGN.\\
We evaluate the impact of this 10$\%$ over-estimate of the background on the measured values of detection likelihood.
We consider a wide range of counts per pixel values generated by a source (between 0.04 and 0.4) and by the background (between 0.001 and 0.009). These intervals are compatible with the source maps and background maps produced by eSASS. We expand these counts on a grid of 5$\times$5 pixels, covering an area slightly larger than the eROSITA PSF. We compute the analytical value of detection likelihood by plugging these values into equations \ref{eq:cstat}, \ref{eq:prob_gamma}, and \ref{eq:detlike}. We repeat the process by increasing the background by 10$\%$, computing the new value of $\mathcal{L}_{\rm det}$, and comparing it to the initial result with the unbiased background.
We find that an overestimation of the background biases the detection likelihood to lower values. We measure a $\sim$ 4$\%$ negative impact on the calculation of detection likelihood for faint sources with DET\_LIKE $\sim$ 5 and a 2.5$\%$ negative impact on more clear sources with DET\_LIKE $\sim$ 20 due to a 10$\%$ overestimation of the background. We conclude that these effects have a minimal impact on the detection and characterization of faint sources around the detection limit, and do not significantly affect the study of more secure detections and the overall analysis of the population in the catalog.
We provide further details and figures in Appendix \ref{appendix:comp2data}.

\subsection{Catalog description}
\label{subsec:catalog_description}
\begin{figure*}
    \centering
    \includegraphics[width=2.0\columnwidth]{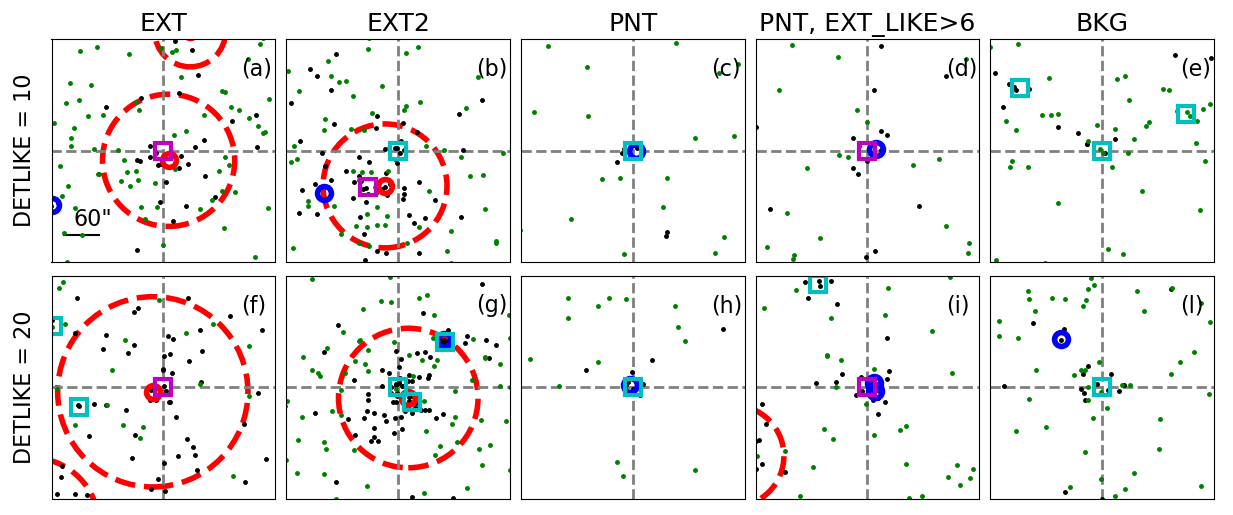}
	\caption{Examples of the eSASS catalog classification. Red (blue) solid circles show simulated clusters (AGN). Magenta (cyan) squares denote extended (point-like) eSASS entries, like in Fig. \ref{fig:evts_w_cats}. The dashed red circles enclose 0.5$\times$R$_{\rm 500c}$ of a simulated cluster. Soft X-ray photons from simulated sources are represented by black dots, the green ones come from the background. The first (second) row shows examples for sources with DET\_LIKE = 10 (20). Columns show respectively: an extended detection uniquely assigned to a simulated cluster, a secondary detection assigned to an input cluster, a point detection uniquely assigned to an AGN, an extended detection uniquely assigned to an AGN, and a detection without any simulated input. All panels have the same physical size. A ruler of 60 arcseconds is shown in the top-left one. 
	}
    \label{fig:sources}
\end{figure*}

We summarize the simulations and source catalog statistics in Table \ref{tab:cat_stat}. 
The catalogs described above have been further cleaned because of the following reasons. 
The generation of event files was not completed correctly because of numerical issues in 6 HEALPix fields in the simulation, covering about 320 square degrees. 
These have not been considered in the analysis presented in the rest of this work. 
In addition, an area of about 260 square degrees around the southern ecliptic pole (RA$\sim$93$^\circ$, DEC$\sim$-66$^\circ$, where the exposure is maximal due to the survey scan mode) has been masked in the eRASS1 simulation. The generation of cluster events was not successful.\\
We focus on the extragalactic sky, masking the areas with galactic latitude |g$_{\rm lat}$| < 10 deg. 
The final area taken into consideration corresponds to 17 703.4 
square degrees for the eRASS1 simulations. 

Following the example of \citet{Liu2021teng_simulation}, we merge simulated catalogs and source catalogs according to the integer identifier (ID) of each photon. Every simulated count has an ID that links it to the source that produced it. 
This method is more reliable than simply matching the catalogs (input and output) with coordinates, because it uses the origin of each simulated  photon: a cluster, AGN, star, or the background. We summarize the algorithm in the following paragraph.\\
First of all, we assess whether a detected source has a simulated counterpart or not. For point (extended) sources detected by eSASS, we study the photons within aperture radii of 20\arcsec\ (60\arcsec). Their origin is stored in each photon ID. The entry in the source catalog is associated with the simulated source that issued the largest number of photons in the aperture radius. This assigns the ID of the simulated counterpart to the entry in the source catalog. We call this ID\_Any. 
One caveat is that the simulation contains a large number of objects fainter than the eROSITA detection limit. Therefore, we only consider input sources that have at least two 
photons emitted during the mock observation. In addition, we set a lower counts threshold related to the local background counts, given by the counts corresponding to the 0.8 
percentile point of the Poisson distribution, whose mean is equal to the number of background photons inside the given aperture radius. \\
Secondly, if an additional simulated counterpart is found, the one emitting the highest number of photons is assigned to ID\_Any. The secondary counterpart is saved as ID\_Any2. \\
Finally, a simulated source can be split into multiple detected sources. This results in copies of the same ID\_Any. We select the detection where the simulated object provides the highest photons count and consider a unique matching between the two (ID\_Uniq). If ID\_Any does not refer to a unique counterpart, in cases where there are multiple entries in the source catalog pointing to the same ID\_Any, we use ID\_Any2 if it is available. A one-to-one matching between the simulated objects and the source catalog can be obtained with ID\_Uniq.
We divide the source catalog into five classes using the IDs just assigned, following the example of \citet{Liu2021teng_simulation}.

1. Primary counterpart of a simulated point source (PNT): detected source assigned to an ID\_Uniq of an AGN or star. This is a secure point source detection.

2. Primary counterpart of a simulated extended source (EXT): detected source assigned to an ID\_Uniq of a cluster. This is a secure cluster detection.

3. Secondary counterpart of a simulated point source (PNT2): detected source without an ID\_Uniq, but assigned to an ID\_Any of an AGN or star. This is a detection that corresponds to a fraction of a simulated point source but is not its primary counterpart. We refer to these as split sources corresponding to an AGN or star. 

4. Secondary counterpart of a simulated extended source (EXT2): detected source without an ID\_Uniq, but assigned to an ID\_Any of a cluster. This is a detection that corresponds to a fraction of a simulated extended source but is not its primary counterpart. We refer to these are split sources corresponding to a cluster;  

5. Background fluctuation (BKG): entry in the source catalog that is not associated with an ID\_Any. This is a false detection, due to a random fluctuation of the background, and is classified as a spurious source.

The first two classes are additionally divided into three subclasses to study whether the source emission is contaminated by a secondary source. To quantify this, we analyze the photons within 60\arcsec\ around every input source (denoted as ID\_1). If we find at least three photons emitted by a source different than the target, and this number of counts is larger than the square root of the target number counts, we consider the source emitting such photons as contaminating. In this case, we save the ID of the contaminating source as ID\_contam to the ID\_1 source. This allows separating isolated (not contaminated) sources from clusters and AGN contaminated by another cluster and or AGN. These cases potentially lead to source blending.\\
We summarize the simulations and source catalog statistics in Table \ref{tab:cat_stat}.
\begin{table*}[]
    \centering
    \caption{Summary statistics of eSASS catalog for the eRASS1 simulation}    
    \begin{tabular}{|c|c c|c|}
    \hline
    \hline
    \rule{0pt}{2.0ex} & \multicolumn{3}{c|}{eRASS1 simulation} \\
    \hline
    \rule{0pt}{2.0ex} \multirow{2}{*}{AREA} & \multirow{2}{*}{\textbf{Class}} &  \textbf{Full} & \textbf{Clean} \\
    & & 20\,617.8 deg$^2$ & 17\,703.4 deg$^2$ \\
    \hline
        \rule{0pt}{2.0ex} \multirow{4}{*}{CLUSTER} & EXT & 44\,440 & 38\,636 \\ 
        & EXT, $\mathcal{L_{\rm EXT}}$>6 & 5204 & 4220 \\ 
        & EXT2 & 8117 & 7300 \\
        & EXT2, $\mathcal{L_{\rm EXT}}$>6 & 177 & 148 \\
    \hline    
        \rule{0pt}{2.0ex} \multirow{4}{*}{AGN} & PNT & 708\,735 & 574\,733 \\ 
        & PNT, $\mathcal{L_{\rm EXT}}$>6 & 1653 & 1017 \\ 
        & PNT2 & 2296 & 1843 \\
        & PNT2, $\mathcal{L_{\rm EXT}}$>6 & 9 & 3 \\
    \hline    
        \rule{0pt}{2.0ex} \multirow{4}{*}{STAR} & PNT & 85\,004 & 49\,380 \\ 
        & PNT, $\mathcal{L_{\rm EXT}}$>6 & 313 & 178 \\ 
        & PNT2 & 561 & 361 \\ 
        & PNT2, $\mathcal{L_{\rm EXT}}$>6 & 1 & 1 \\
    \hline
        \rule{0pt}{2.0ex} BACKGROUND & BKG & 284 654 & 229 559 \\ 
        & BKG, $\mathcal{L_{\rm EXT}}$>6 & 374 & 48 \\
    \hline    
        \rule{0pt}{2.0ex} TOTAL & All & 1\,133\,807 & 901\,812 \\
        & $\mathcal{L_{\rm EXT}}$>6 & 7731 & 5615 \\
    \hline    
    \end{tabular}
    \vskip.15cm
    \footnotesize{\textbf{Notes.} The table reports the number of eSASS entries that are matched to a certain class of simulated objects (cluster, AGN, star, background). The catalog contains all sources with DET\_LIKE >= 5. Each line shows different sources identified in the eSASS catalogs: the number of all matches (point-like and extended), their subsample with EXT\_LIKE >= 6 (Extended), the ones relative to secondary matches (i.e., split sources, see EXT2 and PNT2), and the secondary matches that are classified as extended (EXT\_LIKE >= 6). The values in the second column include all the simulated tiles, the values in the third one account for the additionally cleaning (see Sect. \ref{subsec:catalog_description}).
    }
    \label{tab:cat_stat}
\end{table*}
We show different examples of classification of the sources in Fig. \ref{fig:sources}. The top left panel (a) shows an example of a simulated cluster that is detected as extended with DET\_LIKE = 10. The position of the detection is well aligned with the position of the simulated object. The dashed red circle encloses 0.5$\times$R$_{\rm 500c}$. The point detection in the center of the panel (b) is assigned to the bright simulated cluster just below, but it is not the primary detection, that is the extended one closer to the cluster center. This is the case of a split source. The third panel (c) highlights a simulated AGN (blue circle) properly detected as a point source (cyan square). The fourth panel (d) shows an example of contamination in the extent-selected catalog: an AGN detected as an extended source. Finally, the fifth panel (e) contains an extended detection without any simulated counterpart: a spurious source. In this case, most of the photons around the detection are coming from the background. This shows how background fluctuations end up decreasing the purity of the source catalog. The second row of the figure (panels f, g, h, i, and l) shows the same type of objects, but with a higher value of detection likelihood equal to 20. We notice how the distribution of photons around faint detected clusters or AGN and spurious sources is very similar. \\
We compare the eSASS source catalog from the eRASS1 simulation to real eRASS1 data in Appendix \ref{appendix:comp2data}.

\subsection{Imaging and spectral analysis}
\label{subsec:imaging_spectral_analysis}
We measure the temperature and luminosity of the simulated clusters detected as extended in the eRASS1 simulation, assuming the value of R$_{\rm 500c}$ from the simulation. We compare them to the simulated quantities. We focus on secure clusters detected with EXT\_LIKE > 20, spanning different ranges of exposure without additional selection on the sky area.
Our approach is the same as the one described by \citet{Ghirardini2021morph_pars, Ghirardini2021supercluster} and is summarized in this section.

1. \textbf{Source masking}: for each extended detection uniquely matched to a cluster, we mask every other source inside a circular region of $4\times R_{\rm 500c}$. For extended sources, the masking radius is equal to the extent measured by eSASS. For point-like ones, it corresponds to the point where the count rate convolved with the eROSITA PSF is consistent with the background within 1$\sigma$. 
This value is fixed to 10 arcseconds when it is lower than such threshold. \\
2. \textbf{Background extraction and modeling}: we use the {\tt srctool} command to extract the source spectrum in a circular region inside $R_{\rm 500c}$ and the background spectrum in a circular annulus between $3-4\times R_{\rm 500c}$. We model these two spectra simultaneously with the \textsc{xspec} software \citep[v 12.10.1f, ][]{Arnaud1996ASPC..101...17A}, using C-statistic \citep{Cash1979ApJ...228..939C}. The cluster emission is fitted by APEC model \citep{Smith2001apec} and the Galactic absorption is modeled by TBabs \citep{Wilms2000tbabs}.
The background model consists of a vignetted sky component and an unvignetted particle-induced one. The first describes photons focused by the telescope mirror and contains contributions from the Local Hot Bubble (apec), the Galactic Halo (tbabs$\times$apec), and faint unresolved AGNs (tbabs$\times$power-law). The second is due to instrumental effects and cosmic rays hitting the detector directly and is described by a combination of power-laws and Gaussian lines \citep{Liu2021teng_simulation}. We fix redshift and galactic column density to the simulated values and fit for temperature.\\
3. \textbf{Surface brightness fitting}: we proceed by measuring the cluster surface brightness inside $R_{\rm 500c}$ and fitting the density profile following \citet{Vikhlinin2006ApJ...640..691V} model, convolved with the PSF and projected onto the 2D image plane. The sky (particle) background model is folded with the vignetted (unvignetted) exposure map and added to the total model. The image
is fit using the Monte Carlo Markov chain (MCMC) code emcee \citep{Foreman-Mackey2013emcee}. We integrate the fitted 2D profile along the line of sight to obtain the surface brightness radial profile. \\
4. \textbf{Luminosity}: we finally convert the surface brightness radial profile to X-ray luminosity
using an absorbed apec model in \textsc{XSPEC}. Given the measured temperature of a cluster, this provides the conversion factor from count rate to luminosity.

\section{Results} 
\label{sec:results}
In this section, we present our main findings about the detection process. 
We start from the point of view of the catalog of sources detected by eSASS. We refer to it as the source catalog. We focus on the cleaned catalog, see Table \ref{tab:cat_stat} and Sect. \ref{subsec:catalog_description} for complete details. \\
We give an overview of how the source catalog is populated by clusters, AGN, stars, and spurious sources (Sect. \ref{sect:population_source_cat}). We then move to the standpoint of the simulated sources and study which of them are detected. We demonstrate how the method is able to recover clusters and AGN as a function of their simulated flux (Sect. \ref{subsect:logNlogS}, \ref{subsec:clu_compl}). 
We detail how the detection of galaxy clusters depends on size and dynamical state. \\
We then combine these two points of view, quantifying the performance of the method (completeness, contamination, and spurious fractions), also accounting for the uneven depth of the survey (Sect. \ref{subsec:detection_efficieny}).\\
We study the sensitive area in the eRASS1 simulation as a function of limiting flux (Sect. \ref{subsec:sensitivity}) and finally verify that our measurement of the X-ray luminosity of clusters are compatible with simulated values (Sect. \ref{subsec:LxTx}). 


\subsection{Population in the source catalog}
\label{sect:population_source_cat}

\begin{table*}[]
    \centering
    \caption{Population in the cleaned eSASS source catalog for different cuts of detection and extension likelihood.}    
    \begin{tabular}{|c|c|c|c|c|c|c|c|}
    \hline
    \hline
     & \multicolumn{7}{c|}{eRASS1}\\
    \hline 
    \rule{0pt}{2.3ex} CLASS, DET\_LIKE & > 5 & > 6 & > 7 & > 8 & > 10 & > 15 & > 25 \rule[-1.1ex]{0pt}{0pt} \\
    \hline
    \rule{0pt}{2.0ex} PNT & 69.214 & 80.087 & 86.830 & 90.717 & 93.850 & 94.843 & 94.056 \\
    EXT  & 4.285 & 4.615 & 4.718 & 4.715 & 4.635 & 4.779 & 5.698 \\
    PNT2  & 0.244 & 0.195 & 0.143 & 0.109 & 0.065 & 0.037 & 0.028 \\
    EXT2  & 0.810 & 0.749 & 0.661 & 0.568 & 0.437 & 0.298 & 0.217 \\
    BKG & 25.448 & 14.355 & 7.648 & 3.891 & 1.014 & 0.042 & 0.001 \\
    \hline
    \hline
    \rule{0pt}{2.5ex} & \multicolumn{7}{c|}{EXT\_LIKE > 6} \\
    \hline
    \rule{0pt}{2.0ex} CLASS, DET\_LIKE & > 5 & > 6 & > 7 & > 8 & > 10 & > 15 & > 25 \rule[-1.1ex]{0pt}{0pt} \\
    \hline
    \rule{0pt}{2.0ex}  PNT & 21.282 & 21.292 & 21.253 & 21.201 & 21.212 & 20.513 & 18.807 \\
    EXT & 75.156 & 75.299 & 75.452 & 75.634 & 75.951 & 77.175 & 79.763 \\
    PNT2 & 0.071 & 0.071 & 0.054 & 0.036 & 0.036 & 0.020 & 0.025 \\
    EXT2 & 2.636 & 2.624 & 2.614 & 2.589 & 2.456 & 2.096 & 1.405 \\
    BKG  & 0.855 & 0.714 & 0.627 & 0.539 & 0.346 & 0.196 & 0.000 \\

    \hline
    \end{tabular}\\
    \begin{tabular}{|c|c|c|c|c|c|c|}
    \hline
    \rule{0pt}{2.3ex} CLASS, EXT\_LIKE & > 6 & > 7 & > 8 & > 10 & > 15 & > 25 \rule[-1.1ex]{0pt}{0pt} \\
    \hline
    \rule{0pt}{2.0ex} PNT & 21.268 & 15.724 & 12.243 & 7.464 & 3.758 & 1.648 \\
    EXT & 75.169 & 80.803 & 84.383 & 89.137 & 92.863 & 95.655 \\
    PNT2 & 0.071 & 0.042 & 0.024 & 0.029 & 0.042 & 0.000 \\
    EXT2 & 2.636 & 2.787 & 2.831 & 3.108 & 3.252 & 2.697 \\
    BKG & 0.855 & 0.645 & 0.519 & 0.261 & 0.084 & 0.000 \\

    \hline    
    \end{tabular}
    \vskip.1cm
    \footnotesize{\textbf{Notes.} The five classes (PNT, EXT, PNT2, EXT2, BKG) are defined in Sect.\ref{sec:data}. The fractions are reported in percentage units. The table is divided into three main quadrants. The first one describes the full catalog for different cuts of detection likelihood. The second one focuses on the extent-selected catalog (EXT\_LIKE > 6) for different cuts of detection likelihood. Finally, the third one is about cuts of extension likelihood.
    }
    \label{tab:cat_stat_thresh}
\end{table*}

\begin{figure}[h]
    \centering
    \includegraphics[width=.923\columnwidth]{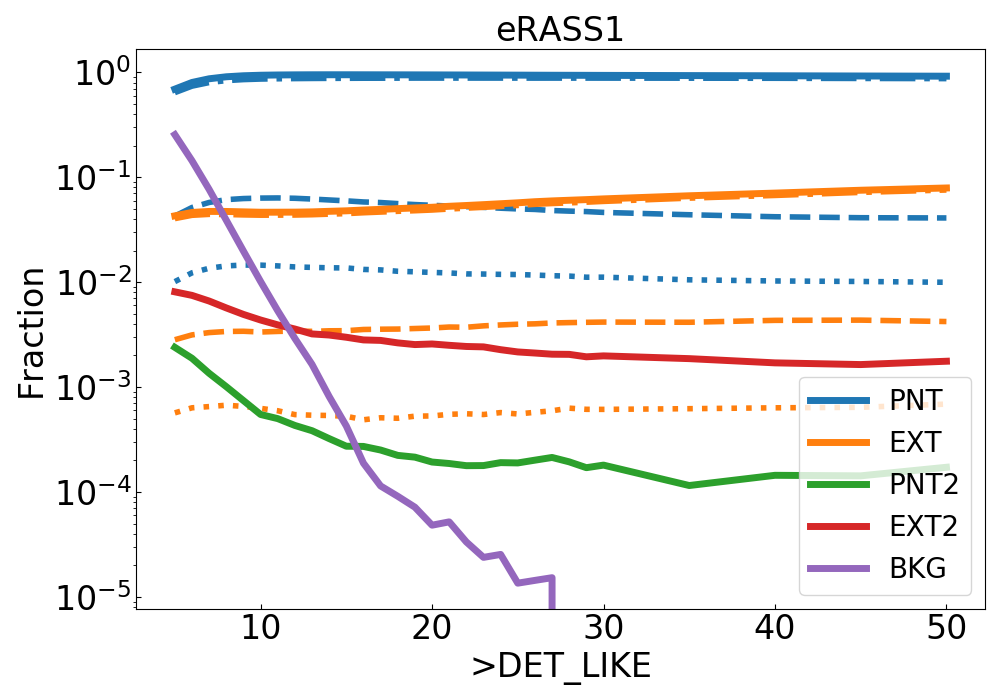}
    \includegraphics[width=.923\columnwidth]{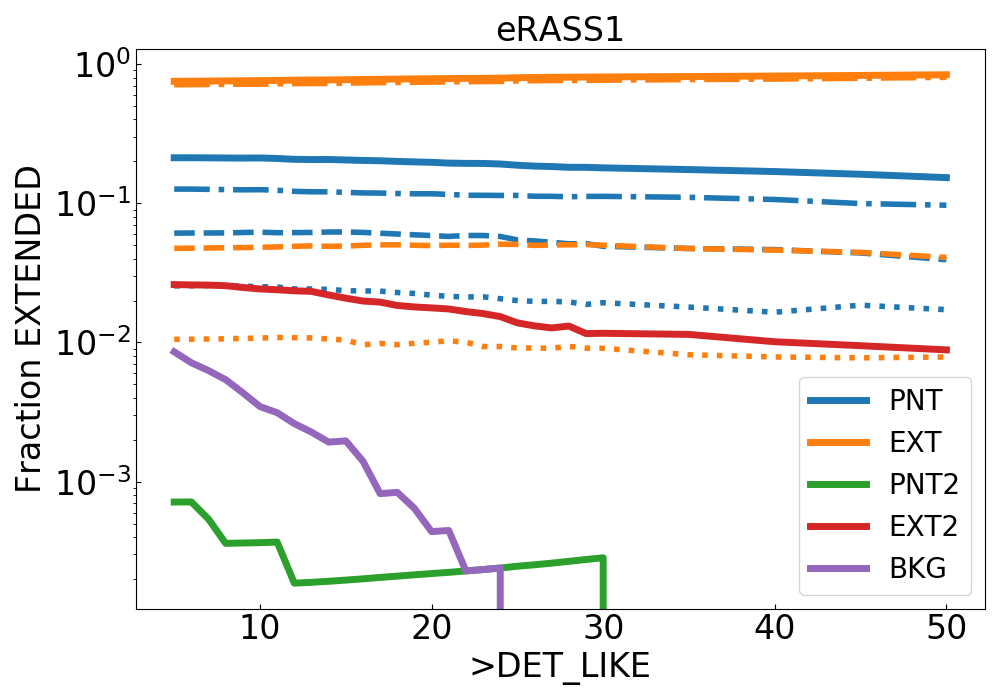}
    \includegraphics[width=.923\columnwidth]{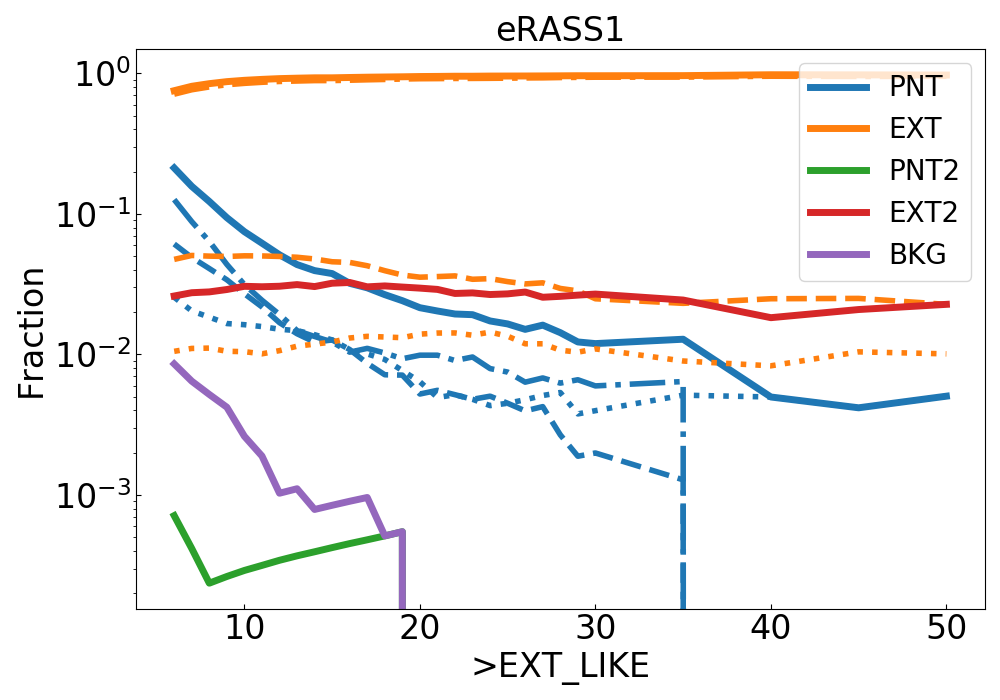}

	\caption{Population in the eSASS catalog. The total number of sources detected by eSASS in the eRASS1 simulation (cleaned, see \ref{subsec:catalog_description}) is 1\,133\,807 (901\,812). The number of extended sources is 7731 (5615).
	\textbf{Top panel}: Fraction of sources in the full catalog as a function of minimum detection likelihood.
	\textbf{Central panel}: Fraction of sources in the extent-selected sample (EXT\_LIKE >= 6) as a function of minimum detection likelihood. 
	\textbf{Bottom panel}: population in the source catalog as a function of minimum extension likelihood. Lines of different colors show the classes defined in Sect. \ref{sec:data}. The dash-dotted lines denote sources that are not contaminated by photons of a secondary source (no blending), the dashed ones identify sources contaminated by a point source, and the dotted ones show sources contaminated by a cluster.
	}
    \label{fig:population_detlike}
\end{figure}

We study the source population in the eSASS source catalog using fractions as a function of different cuts in detection and extension likelihood, using the classes defined in Sect. \ref{subsec:catalog_description}. We consider the full source catalog and the extent-selected sample (with positive values of EXT\_LIKE). The result is shown in Fig. \ref{fig:population_detlike}. 
We report the fraction corresponding to each class for different thresholds of detection and extension likelihood in Table \ref{tab:cat_stat_thresh}. The histograms of the total number of sources and the fractional histograms in linear scales are collected in Appendix \ref{appendix:histograms}.

\subsubsection{Full source catalog}

The cleaned source catalog of the eRASS1 simulation contains 901\,812 sources in total. Among them, 5615 are classified as extended.

\subsubsection*{Fraction of point sources}
The majority of the catalog consists of point sources, mostly AGN and a few stars. They make up 93.8$\%$ of the catalog for detection likelihood larger than 10 in the eRASS1 simulation. 
For detection likelihood greater than 25, this fraction increases to 94.1$\%$. This is driven by the predominant number density of the AGN population compared to other sources. In the whole cleaned catalog, 574\,733 entries are associated with an AGN. 

\subsubsection*{Fraction of clusters in source catalog}
In the eRASS1 simulation, 
clusters of galaxies only consist of about 4.3$\%$ of the whole catalog for DET\_LIKE > 5. 
Even when most of the false detections are removed, above DET\_LIKE = 25, this fraction remains low, at about 6$\%$. This difference between the fraction of AGN and clusters is driven by the intrinsic number density per square degree of these sources. For example, we simulate 18 clusters per square degree with flux larger than 10$^{\rm -14}$ erg/s/cm$^2$. At the same flux value, the input AGN are 100 per square degree (see Sect. \ref{subsect:logNlogS}). 
In addition, clusters need a larger amount of counts to be detected, especially as extended, compared to point sources: their extended emission requires a larger exposure to emerge over the background. 

\subsubsection*{Fraction of spurious detections}
A fraction of sources in the eSASS catalog is not matched to any input simulated object. These spurious detections are due to background fluctuations, that mimic the emission of a source.
The detection likelihood encodes by definition the probability for each entry in the source catalog of being a false detection, as explained in Sect. \ref{sec:data}. However, the analytical derivation does not account for additional effects during the measurement process. These include uncertainty in the background estimation, errors in the PSF-fitting, or issues related to hardware and calibration. Consequently, the false detection rate is larger than the one predicted by Equation \ref{eq:prob_gamma}.\\
The fraction of spurious sources drops significantly while increasing the detection likelihood threshold. 
We measure a spurious fraction  of 25.4$\%$ for DET\_LIKE > 5 and 14$\%$ for DET\_LIKE > 6. The false detection rate is further reduced to 4$\%$ at DET\_LIKE > 8 and 0.001$\%$ for DET\_LIKE > 25. Progressive cuts in detection likelihood are therefore efficient in removing background fluctuations from the source catalog.  

\subsubsection*{Fraction of split sources}
\begin{figure}
    \centering
    \includegraphics[width=0.49\columnwidth]{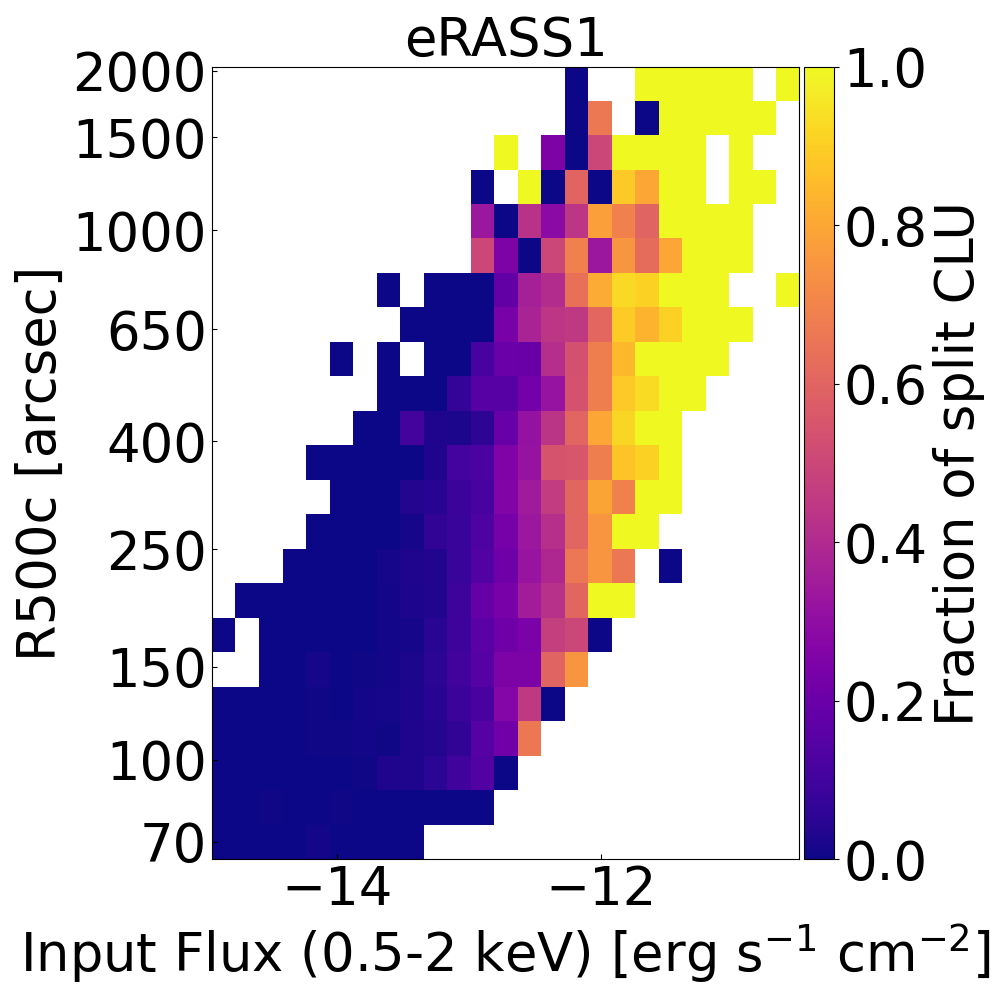}
    \includegraphics[width=0.49\columnwidth]{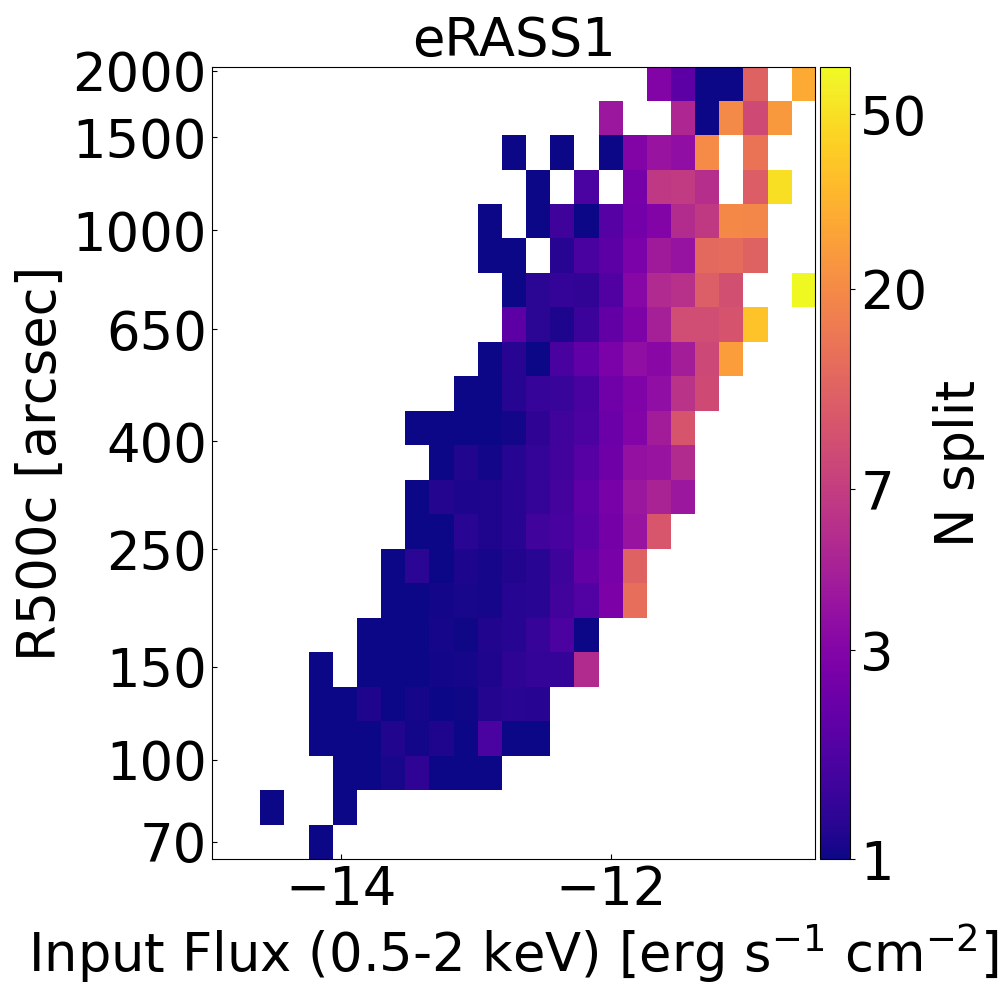}    
    \caption{Number of split sources as a function of flux and R$_{\rm 500c}$. The left-hand panel shows the fraction of detected clusters that are split into multiple sources, the right-hand one displays the average number of sources which a cluster with given flux and size is split into. The blank spaces contain no input clusters.
    }
    \label{fig:split_sources}
\end{figure}
Very bright and or extended input sources are possibly split into multiple detections. These are the one marked as secondary matches (PNT2, EXT2) in our classification scheme (see Section \ref{subsec:catalog_description}).
The fraction of entries in the source catalog marked as a secondary match to a point source (PNT2) is always under 0.5$\%$. Clusters are instead slightly more easily split into multiple sources, giving about 0.8$\%$ of entries cataloged as secondary matches to an extended source (EXT2). Together with decreasing the spurious fraction, increasing the DET\_LIKE threshold gets rid of these low significance secondary detections, as this fraction decreases to $\sim$0.2 $\%$ at DET\_LIKE > 25. A total of 4627 clusters are split into more than one (point or extended) source in the eRASS1 simulation. About 70$\%$ of these are split into only two sources. Among the clusters that are split, the average number of split sources is 2.76. We find that the number of sources into which a cluster is split mainly depends on its flux, and secondary the size on the sky plane of the cluster itself. For example, more than 90$\%$ of the clusters with R$_{\rm 500c}$ larger than 350 arcseconds are split into multiple sources. However, only the brightest objects are split into a large number of sources. A very bright and extended cluster with flux $\sim$ 10$^{-11}$ erg/s/cm$^2$ and R$_{\rm 500c}$ $\sim$500\arcsec\ is split into 24 sources by eSASS on average. 
There is one particular case of an extremely bright and extended cluster (F$_X$ = 3.10$\times$10$^{\rm -11}$ erg/s/cm$^2$, R$_{\rm 500c}$ = 13.5\arcmin) in the pole region that is split into 65 sources. These trends are highlighted in Fig. \ref{fig:split_sources}. The left-hand panel shows the fraction of clusters that are split into multiple sources as a function of flux and R$_{\rm 500c}$. The average number of sources that a cluster is split into is displayed on the right-hand panel. 

\subsubsection*{Blends}
We study the sources that are blended with a secondary one, according to the criteria defined in Sect. \ref{sec:data} to find objects whose emission is contaminated by another object. Most of the sources detected by eSASS are not contaminated by the emission of a secondary nearby object. 92$\%$ of the detected point sources are isolated. This number for clusters is 94$\%$.
In the full cleaned catalog, about 4$\%$ of the population consists of point sources contaminated by other point sources. This number increases to 6$\%$ for DET\_LIKE > 10, because of the drastic drop of spurious sources. For point sources contaminated by clusters (i.e., detections whose primary match is an AGN or a star, but that contain photons emitted by a cluster) this fraction reduces to 1$\%$. About 7$\%$ of the clusters in the full catalog are contaminated by other point sources. In such cases, the presence of the bright AGN enhances the emission from a physical source and helps the detection algorithm in the identification of a source at that position. The flux measured by eSASS for these sources will be biased \citep[see][]{Bulbul2021arXiv211009544B_clusters_disguise}. More detailed modeling of the cross-correlation between AGN and clusters is required to reach conclusive statements about blending.

\subsubsection{Extended source catalog}

We now focus on the extent-selected sample, selected by EXT\_LIKE >= 6. This is the minimum value of extension likelihood fixed by the choice of the parameter extlikemin = 6 (Sect. \ref{sec:data}). Different values of this parameter impact the properties of the extent-selected catalog. We detect a total of 5615 sources as extended in the full cleaned eSASS catalog of the eRASS1 simulation. 

\subsubsection*{Fraction of clusters}

The eRASS1 extent-selected catalog is dominated in numbers by clusters of galaxies: 75.2$\%$ of the eSASS sources are uniquely matched to a cluster, with a 21.2$\%$ point source contamination. 
When increasing the detection likelihood threshold to 25, clusters make up 79.8$\%$ of the catalog. These numbers increase more significantly when cutting in extension likelihood rather than detection likelihood. For EXT\_LIKE > 25, 95.6$\%$ of the eRASS1 sources are clusters. This is partially related to the significant decrement of background fluctuations, which is completely canceled at this value of extension likelihood. 
However, the main contribution is given by the drop of AGN that are mistakenly classified as extended sources, which reduces contamination significantly. This fraction changes from 21.2$\%$ for EXT\_LIKE > 6 to 1.6$\%$ for EXT\_LIKE > 25 in the eRASS1 simulation. 
\subsubsection*{Fraction of AGN}
The fraction of AGN in the extent-selected sample (EXT\_LIKE >= 6) is constant at around 20$\%$ as a function of detection likelihood cuts. Even for DET\_LIKE greater than 25, it still reaches 18.7$\%$. It means that progressive thresholds of detection likelihood are not efficient in reducing the fraction of AGN detected as extended.\\
The contribution of detections that contain a fraction of point source signal (PNT2, split point sources) is minimal in the extended select sample. It is well below 1$\%$ for any cut in detection or extension likelihood. The fraction of entries classified as cluster signal (EXT2, split clusters) is around 2.6$\%$ for eRASS1. Increasing the extension likelihood does not have a significant impact on this number. This is due to the fact that the scaling of these secondary matches with EXT\_LIKE is more similar to the one of primary matches, compared to the random background fluctuations. This is not true for cuts in detection likelihood, which keep a higher number of AGN in the extent-selected sample, reducing the relative contribution of both primary and secondary matches in the catalog. 
In fact, by increasing the DET\_LIKE threshold in the eRASS1 catalog from 5 to 25, the fraction of secondary matches also drops from 2.6 to 1.4$\%$ for  extended sources and from 0.071$\%$ to 0.025$\%$ for point-like ones respectively.

\subsubsection*{Fraction of spurious sources}
Random background fluctuations in the extent-selected catalog are efficiently removed using different thresholds of DET\_LIKE and EXT\_LIKE. For the former, the spurious fraction drops from 0.85$\%$ to 0.34$\%$ for detection likelihood larger than 5 and 10. The latter drops to 0.26$\%$ for EXT\_LIKE > 10. There are no spurious sources above detection likelihood larger than 25 and extension likelihood larger than 20 in the extent-selected sample of eRASS1. The decrement of the false detection rate is steeper as a function of EXT\_LIKE cuts. It means that, on top of reducing contamination, extension likelihood thresholds remove background fluctuations more efficiently than detection likelihood ones in the extent-selected sample.

   
\subsubsection*{Blends}
We study sources blended with another source in the extent-selected catalog (EXT\_LIKE >= 6). Focusing on the AGN that leak into the extent-selected sample, one can understand what caused the misclassification. For the whole extended eRASS1 sample, 6$\%$ of the catalog consists of AGN contaminated by another point source. This fraction is dominant over the ones contaminated by a cluster (2.5$\%$). However, when increasing the EXT\_LIKE threshold, the relation between these two classes changes significantly, to the point where for EXT\_LIKE > 40, all the detections assigned to an AGN by our matching algorithm are actually blended with a cluster. 
Follow-up observations in the optical band have the potential to confirm these clusters, which lowers our estimate of contamination in the extended select sample due to bright point sources by $\sim$1$\%$.
   
\subsection{Simulated and detected sources}
\label{subsect:logNlogS}

We now study which simulated sources are detected by eSASS. The detection process mainly depends on the net count rate of each source. Bright sources with large flux values provide a larger number of photons on the detector. Therefore, it is easier for the detection algorithm to identify them, compared to fainter objects dispersed in the local background. We investigate which simulated sources are detected by studying the number density as a function of the input flux threshold for AGN in Sect. \ref{subsubsec:logNlogS:agn}, and for 
clusters in Sect. \ref{subsubsec:logNlogS:clu}.

\begin{figure}
    \centering
    \includegraphics[width=\columnwidth]{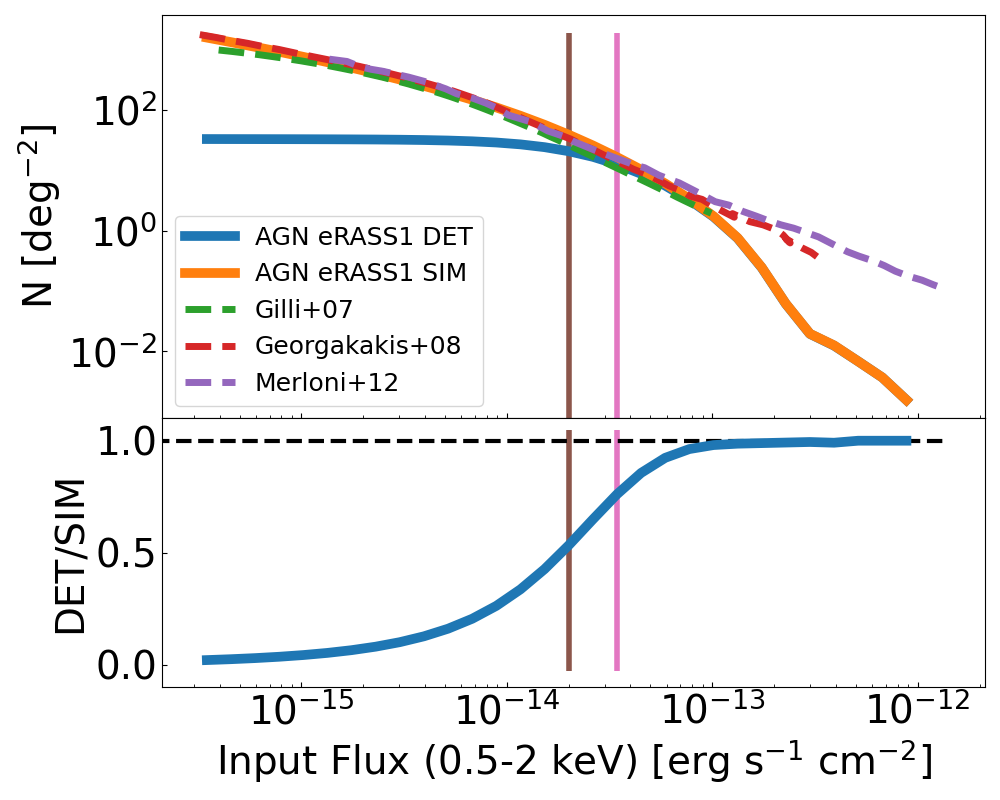}
	\caption{Cumulative number density of the AGN population. \textbf{Top panel}: The blue (orange) line shows the logN-logS built with the sample of detected (simulated) AGN. The green, red, violet dashed lines show the distributions from \citet{Gilli2007A&A...463...79G}, \citet{Georgakakis2008MNRAS.388.1205G} and \citet{Merloni2012}. The brown and pink vertical lines locate the eROSITA flux value where the ratio between the detected and simulated populations is equal to 0.5 and 0.8, respectively. \textbf{Bottom panel}: Ratio between the logN-logS of detected and simulated AGN. A black dashed line denotes a ratio equal to 1.0. 
	}
    \label{fig:AGN_logNlogS}
\end{figure}

\subsubsection{AGN logN--logS}
\label{subsubsec:logNlogS:agn}
We measure the cumulative number of detected AGN per square degree as a function of the input flux (0.5--2 keV band). 
We compare with the distribution of the simulated AGN \citep{Comparat2019agn}, with the observations from \citet{Gilli2007A&A...463...79G} and \citet{Georgakakis2008MNRAS.388.1205G}, and the collection from \citet{Merloni2012}. 
The result is shown in the upper panel of Fig. \ref{fig:AGN_logNlogS}. 
At the high flux end, the different shapes of the function denoting eRASS1 and other samples are expected due to the AGN simulation method in HEALPix fields as described in Sect. \ref{sec:simulation}. It reduces the volume probed by the model and the total number of the brightest AGN consequently decreases, but this method guarantees a significant gain in computation time. Given our goal of studying the simulated objects that are detected, this has no impact on our purpose. In the lower panel, we show the ratio between the logN--logS built with the detected and simulated populations of AGN.
Below the predicted eROSITA flux limits at $\sim$ 4$\times$10$^{-14}$ erg/s/cm$^{2}$ for eRASS1 \citep[see][Figure 4.3.1]{Merloni2012}, the number density of detected AGN deviates from the simulated one (solid curves depart from the dashed ones). 
Toward high fluxes, the number density of detected AGN converges to the simulated one. 
The ratio between these two curves reaches a value of 0.5 at $\sim$ 2$\times$10$^{\rm -14}$ erg/s/cm$^{\rm 2}$ for eRASS1. These numbers rise to $\sim$ 3.5$\times$10$^{\rm -14}$ erg/s/cm$^{\rm 2}$ and a ratio of 0.8 between the logN--logS of detected and simulated AGN. This is in excellent agreement with the prediction of the eRASS1 sensitivity for point sources in the same soft band 0.5--2.0 keV from \citep{Merloni2012}. 
The completeness of the source catalog behaves smoothly as a function of flux and is in line with the expectations. We study the completeness fraction of AGN in more detail and provide analytical fits in Appendix \ref{appendix:agn}.

\subsubsection{Cluster logN--logS}
\label{subsubsec:logNlogS:clu}

\begin{figure}
    \centering
    \includegraphics[width=\columnwidth]{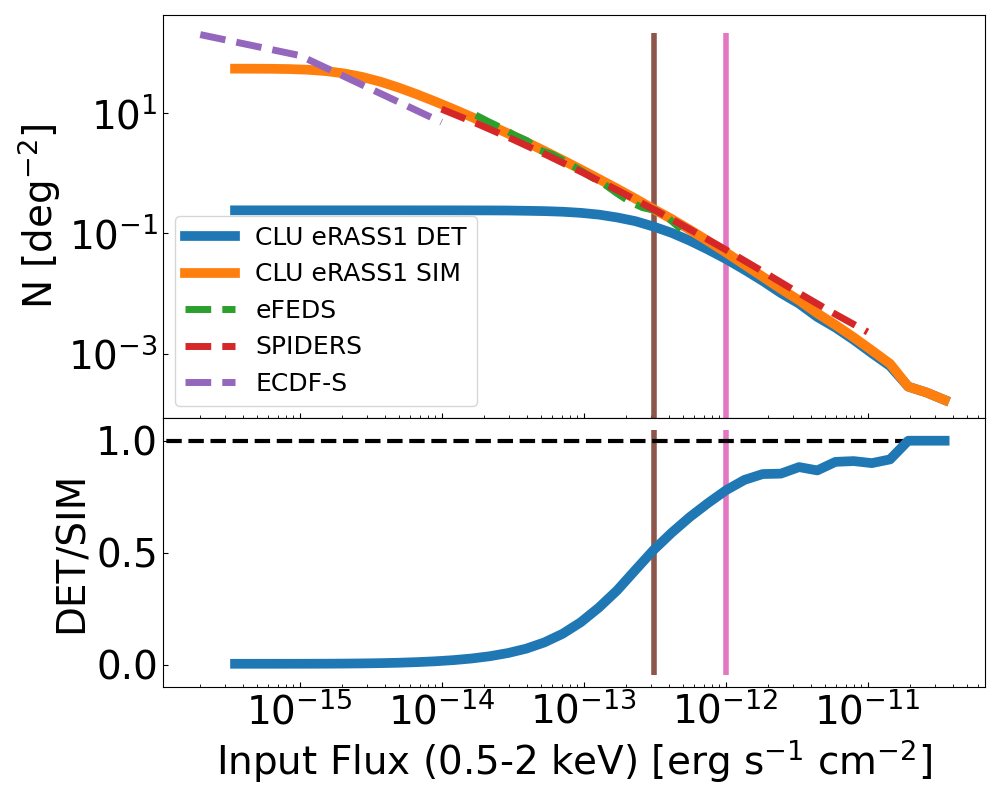}
	\caption{Cumulative number of clusters per square degree as a function of flux. \textbf{Top panel}: The solid blue (orange) line shows the logN-logS built with the sample of detected (simulated) clusters. The green dashed line shows the distributions of the eFEDS sample \citep{2022A&A_LiuAng_eFEDS_clu}, the red one denotes the SPIDERS sample \citep{Finoguenov2020A&A...638A.114F}, and the pink one the ECDF-S \citep{Finoguenov2015A&A_CDFS}. The brown and pink vertical lines locate the eROSITA flux value where the ratio between the detected and simulated populations is equal to 0.5 and 0.8, respectively. \textbf{Bottom panel}: Ratio between the logN-logS of detected and simulated clusters. A black dashed line denotes a ratio equal to 1.0. 
	}
    \label{fig:CLU_logNlogS}
\end{figure}

We study the cumulative number density of clusters as a function of the input flux. We detect 0.1 clusters per square degree with flux larger than 4$\times$10$^{\rm -13}$ erg/s/cm$^{\rm 2}$ in the eRASS1 simulation.
We detect all the clusters at the brightest flux end, as the ratio between the logN-logS built with detected and simulated clusters reaches a value of 1.0 for the eRASS1 simulation. 
It is equal to 0.5 for flux values of $\sim$ 3$\times$10$^{\rm -13}$ erg/s/cm$^{\rm 2}$. For the same flux limit, about 70$\%$ of the clusters with mass larger than $M_{\rm 500c}$>3$\times$10$^{14}$ M$_\odot$ are detected as extended sources. A ratio of 0.8 is reached for flux values of $\sim$ 1.5$\times$10$^{\rm -12}$ erg/s/cm$^{\rm 2}$ for eRASS1. 
These flux limit values are larger compared to the AGN ones. 
A different flux limit is thus expected between the two populations. 
The extension of the cluster model to galaxy groups allows a smooth transition between the faintest clusters that are not detected and the ones above the survey flux limit.
The detection method is able to fully recover the bright end of the cluster sample. 
Around the flux limit, additional selection effects, such as the cool core bias or the size of the object on the sky plane, influence the detection process. In addition, at fixed simulated flux, due to their spatial extent, clusters will be detected with a lower likelihood compared to a point source with the same flux.
We report the cumulative clusters number density as a function of flux in Fig. \ref{fig:CLU_logNlogS}. In the upper panel, we show the cluster logN--logS for eRASS1. The blue line denotes the detected cluster population, while the orange line the simulated one. The green one adds a comparison to the eFEDS logN--logS \citep{2022A&A_LiuAng_eFEDS_clu}. We additionally compare our result to The SPectroscopic IDentification of eROSITA Sources observational
program \citep[SPIDERS,][]{Clerc2016spiders, Finoguenov2020A&A...638A.114F} denoted by the red dashed line, and the Extended Chandra Deep Field South \citep[ECDF-S,][]{Finoguenov2015A&A_CDFS}, indicated by the purple dashed line. There is good agreement within these samples.  The bottom panel shows the ratio between the detected and the simulated populations. All clusters with high flux are detected as extended.
We present the challenges of the detection of extended sources in Sect. \ref{subsec:clu_compl}. 



\begin{table*}[h]
    \centering
    \caption{Different exposure and properties of the eRASS1 simulations.}
    \begin{tabular}{|c|c|c|c|c|c|c|c|}
    \hline
    \hline
    \rule{0pt}{2.2ex} & \multicolumn{6}{c}{eRASS1} \\
    \hline
    \rule{0pt}{2.3ex} Exposure & Area [deg$^2$] & N$_{\rm CLU}$/deg$^2$ & Flux CLU 50$\%$ & Flux CLU 80$\%$ & N$_{\rm AGN}$/deg$^2$ & Flux AGN 50$\%$ & Flux AGN 80$\%$  \\
    \hline
    \rule{0pt}{2.2ex} < 110 s & 6710 & 0.13 & 7.13$\times$10$^{\rm -13}$ & 3.39$\times$10$^{\rm -12}$ & 21.78 & 3.76$\times$10$^{\rm -14}$ & 7.02$\times$10$^{\rm -14}$\\
    \hline    
    \rule{0pt}{2.2ex} 110 s -- 150 s & 4543 & 0.22 & 4.67$\times$10$^{\rm -13}$ & 1.2$\times$10$^{\rm -12}$ & 29.41 & 3.01$\times$10$^{\rm -14}$ & 5.31$\times$10$^{\rm -14}$\\
    \hline    
    \rule{0pt}{2.2ex} 150 s -- 400 s & 6073 & 0.34 & 3.28$\times$10$^{\rm -13}$ & 9.72$\times$10$^{\rm -13}$ & 42.94 & 2.22$\times$10$^{\rm -14}$ & 3.98$\times$10$^{\rm -14}$\\
    \hline
    \rule{0pt}{2.2ex} > 400 s & 377 & 1.05 & 1.12$\times$10$^{\rm -13}$ & 4.75$\times$10$^{\rm -13}$ & 93.71 & 1.10$\times$10$^{\rm -14}$ & 1.93$\times$10$^{\rm -14}$\\
    \hline
    \end{tabular}

    \footnotesize{\textbf{Notes.} Each column denotes: exposure interval, Area covered with the given exposure, number density of clusters detected as extended, flux limit where the completeness is equal to 0.5 for clusters detected as extended, flux limit where the completeness is equal to 0.8 for clusters detected as extended, number density of AGN detected as point sources, flux limit where the completeness is equal to 0.5 for AGN detected as points, and flux limit where the completeness is equal to 0.8 for AGN detected as points. 
    }
    \label{tab:clu_density}
\end{table*}

\subsection{Cluster completeness}
\label{subsec:clu_compl}
\begin{figure}
    \centering
    \includegraphics[width=\columnwidth]{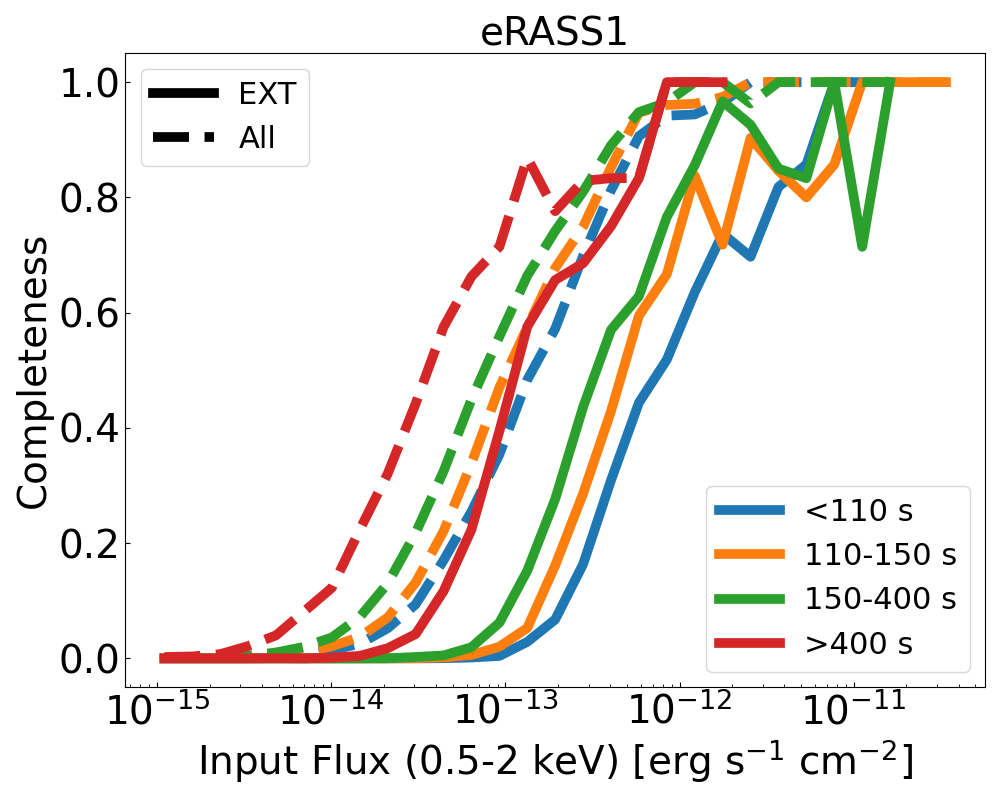}
	\caption{
	Fraction of simulated clusters with a counterpart 
	in the eSASS catalog as a function of simulated soft X-ray flux. We do not apply any additional likelihood selection. Each color identifies an exposure time range. Solid lines denote clusters only detected as extended, while dashed ones include the ones detected as point sources. 
	}
    \label{fig:completeness}
\end{figure}

The completeness is defined as the ratio between the number of detected and simulated objects, see Eq. \ref{eq:completeness}:
\begin{equation}
    C = \frac{N_{\rm DET}}{N_{\rm SIM}}.
    \label{eq:completeness}
\end{equation}
We measure the completeness of our source catalog as a function of the input flux in the 0.5--2 keV band. We study areas in the sky covered by different depths. We expect to measure higher completeness where the exposure is longer, which allows detecting a higher number of clusters. We consider four exposure time bins in this work, defining shallow, medium, deep, and pole regions. The respective intervals are < 110 s, 110 s -- 150 s, 150 s -- 400 s, > 400 s for the mock eRASS1. Such intervals are designed to identify three regions covering roughly a similar area on the sky, and a fourth, smaller one that encloses the pole with large exposure. Additional details are provided in Table \ref{tab:clu_density}. With this approach, we can quantify the gain of detected clusters thanks to deeper observations.
We show the result for eRASS1 in Fig. \ref{fig:completeness}. The lines are color-coded according to exposure time intervals. The solid lines show clusters of galaxies detected as extended, dashed ones additionally consider clusters detected as point sources with EXT\_LIKE = 0. Adding the latter population increases completeness at a fixed value of flux. 
Focusing on the objects detected as extended, we measure a completeness fraction of 0.5 at 3.3$\times$10$^{\rm -13}$ erg/s/cm$^2$ for regions around the average eRASS1 exposure of about 275 s, denoted by the green solid line. 
This result is comparable with previous predictions by \citet{Clerc2018A&A...617A..92C}, who measured a completeness value of 0.5 at $\sim$ 5$\times$10$^{\rm -14}$ erg/s/cm$^2$ in equatorial fields with eRASS:8 depth of about 2.0 ks. The decrement of completeness in the 150 s -- 400 s range is due to a merging system, where only one eSASS detection with EXT\_LIKE>6 is present. The latter is assigned to one of the two clusters, the one providing most of the counts around the detection. The second cluster is assigned to a nearby point-like detection instead. Adding the clusters detected as point sources increases completeness. For the depth interval 150 s -- 400 s in eRASS1, the 50$\%$ completeness is reached at flux equal to 8$\times$10$^{\rm -14}$ erg/s/cm$^2$. There is a flux difference of about 0.7 dex with the addition of this population.\\ 
The measure of completeness is positively correlated with exposure time.
In the eRASS1 simulation, the fraction of clusters with flux $\sim$ 5$\times$10$^{\rm -13}$ erg/s/cm$^2$ that are detected as extended goes from 0.39 (exposure < 110 s) to 0.8 (exposure > 400 s). 

\begin{figure}
    \centering
    \includegraphics[width=0.8\columnwidth]{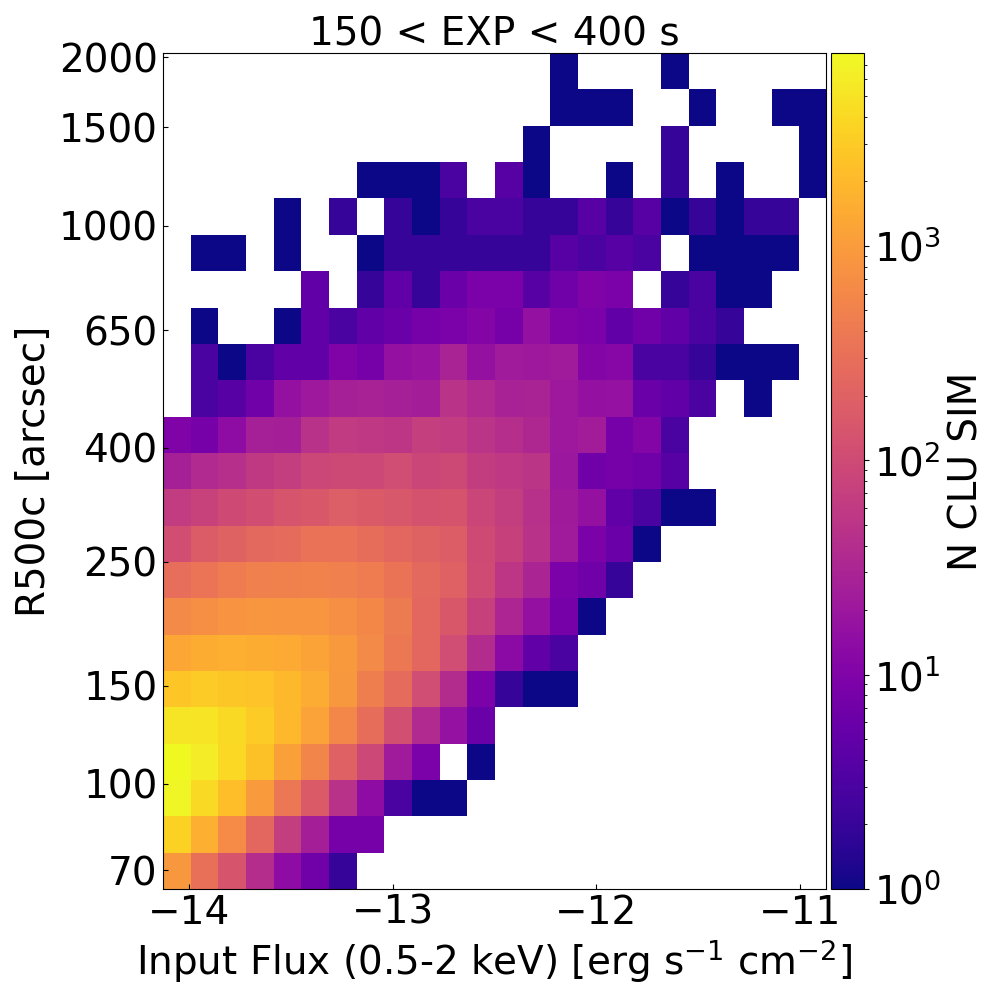}
    \includegraphics[width=0.8\columnwidth]{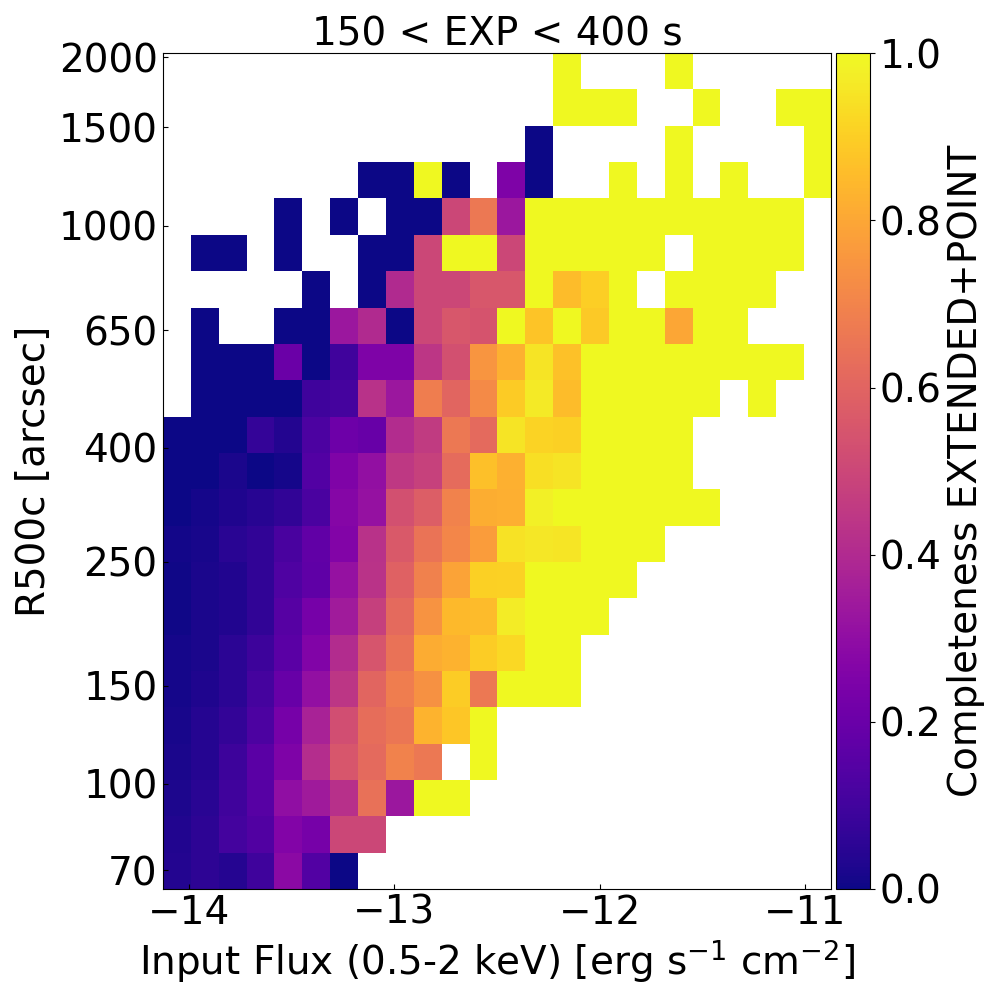}
    \includegraphics[width=0.8\columnwidth]{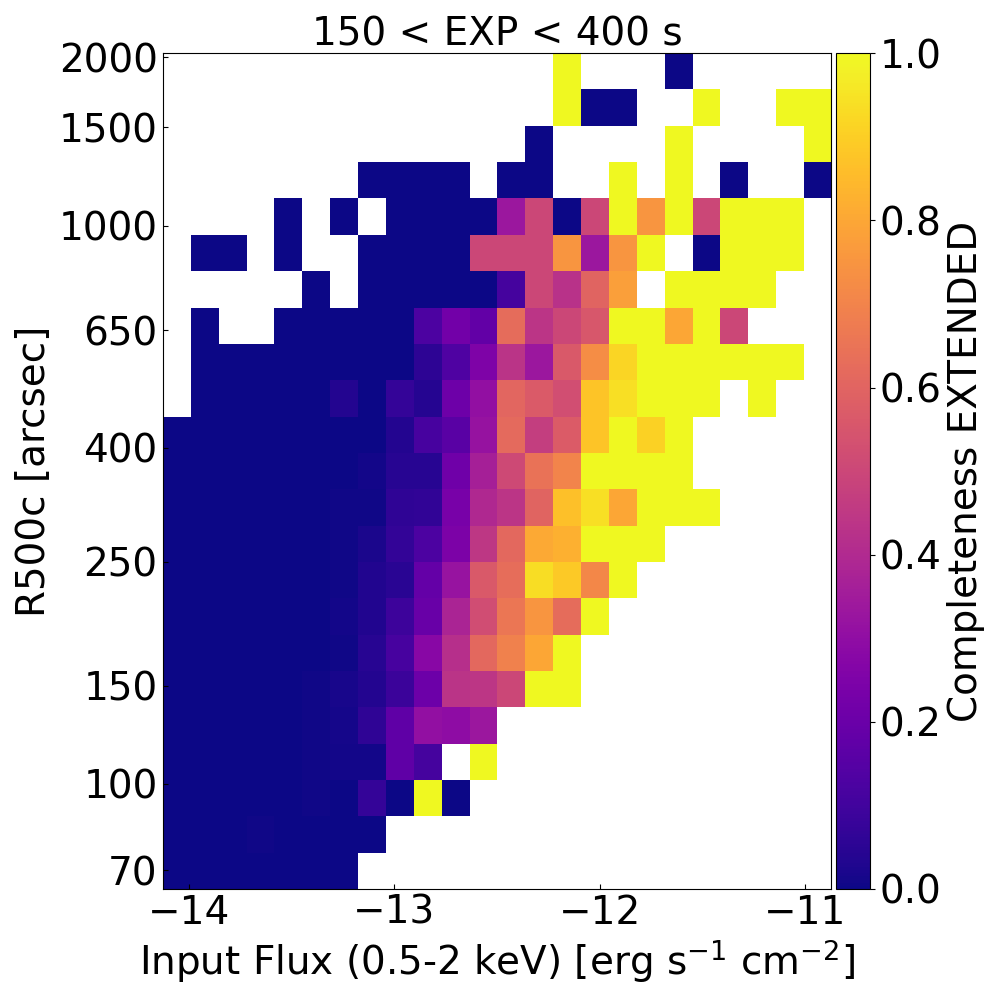}
    \caption{Simulated and detected clusters population as a function of the input flux and size on the sky. The figures refer to areas of the eRASS1 simulation covered by an exposure between 150 s and 400 s. The blank spaces contain no input clusters. \textbf{Top panel}: number of simulated clusters in the flux--R$_{\rm 500c}$ space. \textbf{Central panel}: fraction of simulated clusters that is detected by eSASS, either as extended or point source. \textbf{Bottom panel}: fraction of simulated clusters that is only detected as extended. 
    }
    \label{fig:completeness_r500c}
\end{figure}

The increase in the number of detected objects between the shallow and deep regions is expected, but nevertheless remarkable. It translates into an increment of the number density of clusters detected as extended with exposure time. In the former, we detect and properly classify as extended 0.13 clusters per square degree (exposure < 110 s). In the latter, such number increases to 1.05.
Our result is in agreement with previous works \citep[][]{Pacaud2006XMM_det, Clerc2012MNRAS.423.3545C, Clerc2018A&A...617A..92C}. This only means that we recover a larger number of simulated clusters in deep areas, not that the detection is necessarily more efficient. A different fraction of spurious sources is also detected in areas with large exposure because the background has lower fluctuations. Its overall level might be larger, but its lower variability may also reduce the false detection rate. Such deep areas additionally suffer from a higher degeneracy between blended point sources and proper extended ones, as well as between AGN in clusters and cluster substructures, which has an impact on the measure of contamination. A detailed discussion is presented in the next section \ref{subsec:detection_efficieny}. We do a similar study for AGN and provide details and analytical fits in Appendix \ref{appendix:agn}.

\subsubsection{Completeness and apparent cluster size}

We investigate the impact of the apparent physical size of the clusters on the sky 
on the detection. This information is encoded in the critical radius R$_{\rm 500c}$. 
We compare the number of detected objects to the simulated one on a 2D grid of flux and R$_{\rm 500c}$. 
Considering the angular size of the cluster on the sky (e.g., in arcseconds) instead of its physical size (in kpc) allows to additionally account for the impact of redshift, which makes distant massive large clusters appear smaller than nearby ones with similar mass.\\ 
We find that the detection of extended sources is not solely a simple function of flux and exposure time. At fixed flux and exposure time, the completeness varies as a function of the size of the clusters on the sky. In the eRASS1 simulation, bright clusters with flux$\sim$ 1$\times$10$^{\rm -12}$ erg/s/cm$^2$, located in an area covered by exposure 150 s -- 400 s, and R$_{\rm 500c}$ = 180\arcsec\ are detected as extended with a completeness of 0.75. The rest of these sources are actually detected but misclassified as point sources. In fact, the completeness reaches a value of 1.0 when adding the population of clusters detected as point-like objects. At the same value of flux, for larger objects with R$_{\rm 500c}$ = 300\arcsec, we measure a completeness of 0.84. The characterization of extremely large clusters is also challenging, because these can be split into multiple sources. In fact, the completeness decreases for large values of R$_{\rm 500c}$, above 400\arcsec\ and flux of 1$\times$10$^{\rm -12}$ erg/s/cm$^2$. As R$_{\rm 500c}$ increases, the surface brightness goes down rapidly. Therefore these cases represent the population of clusters which are very extended but with very low surface brightness, therefore they are harder to be detected.
This is shown in Fig. \ref{fig:completeness_r500c}.
It displays the number of the simulated clusters population in the upper panels, the fraction of these objects that are detected as extended or point sources in the central ones, and finally only the ones classified as extended in the lower panels.
It focuses on exposure intervals containing the average depth for our simulation, in the 150 s -- 400 s range for eRASS1.
This figure confirms the trends of increasing completeness with flux (see Fig \ref{fig:completeness}). 
In addition, it demonstrates how the selection of extended sources is not a simple function of flux and exposure, but also of the size the object on the sky, encoded in our measure of R$_{\rm 500c}$. 

\begin{figure}
    \centering
    \includegraphics[width=0.8\columnwidth]{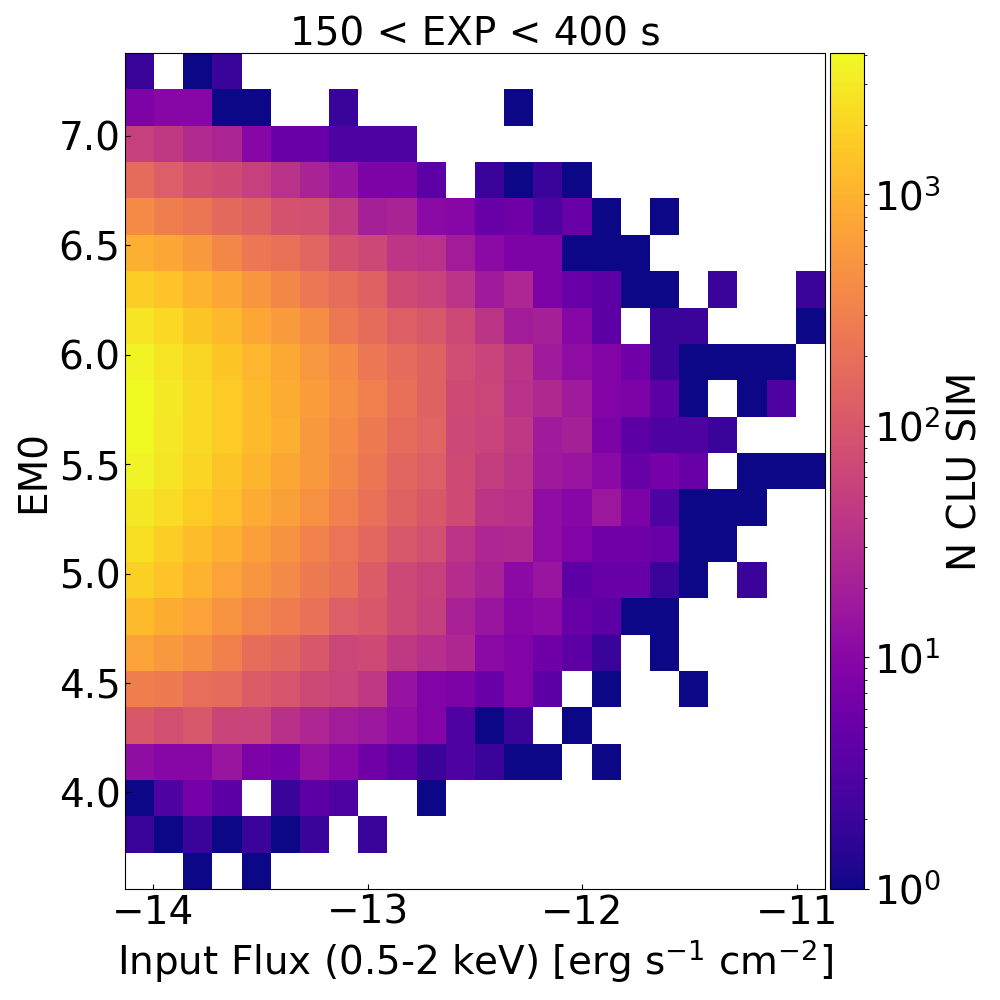}
    \includegraphics[width=0.8\columnwidth]{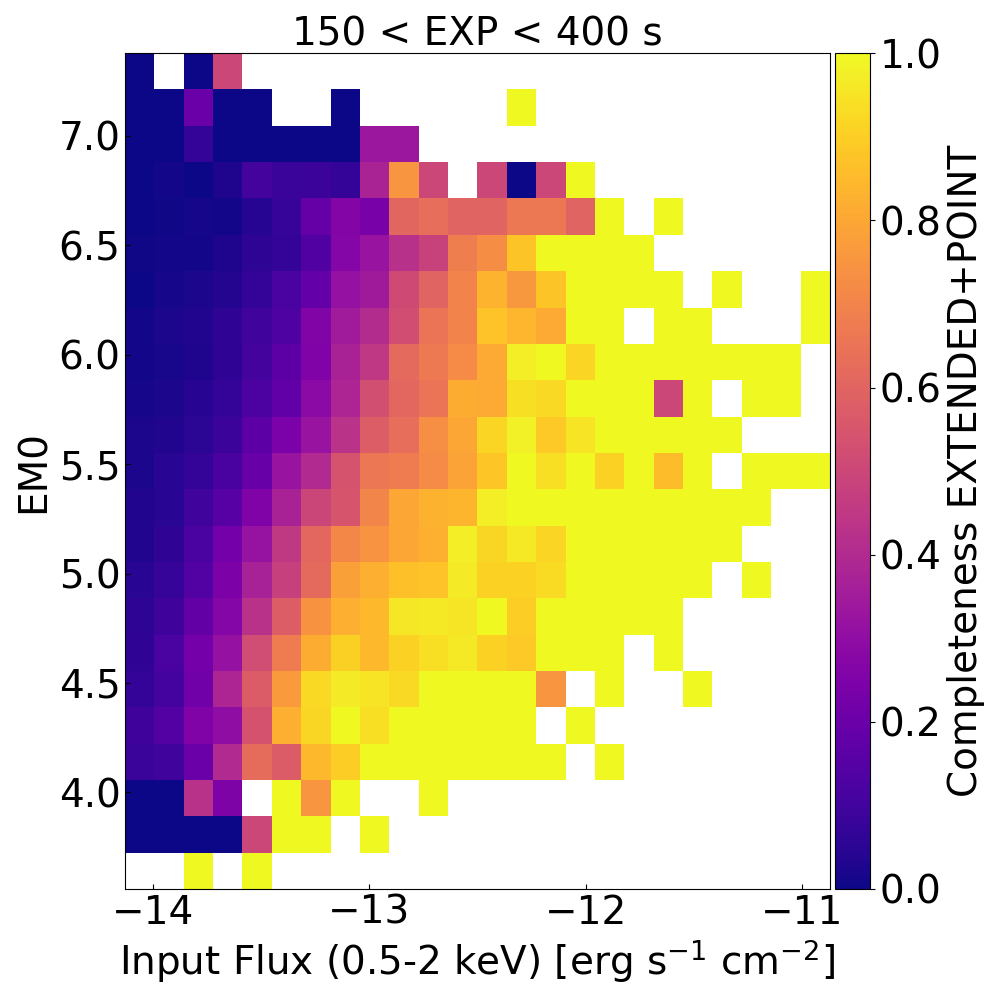}
    \includegraphics[width=0.8\columnwidth]{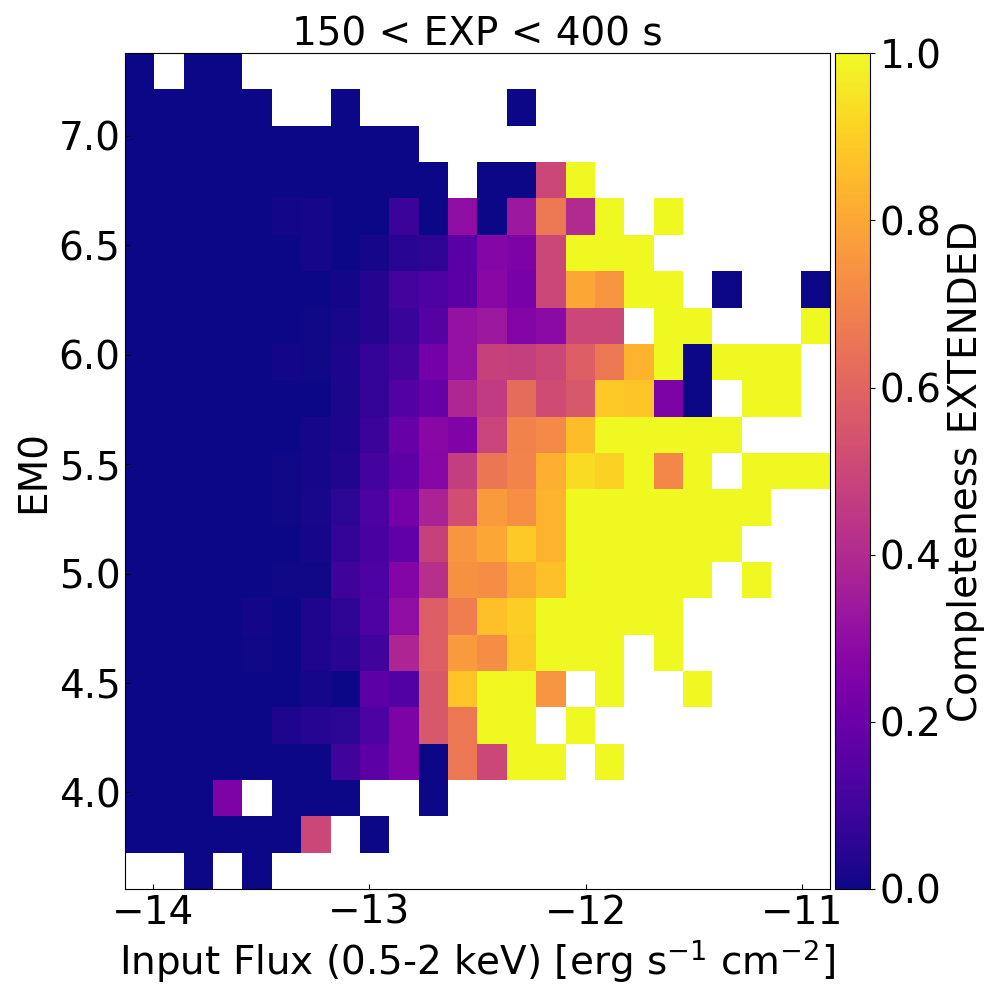}
    \caption{Population of simulated and detected clusters as a function of the input flux and dynamical state. The panels show areas of the eRASS1 simulation covered by an exposure between 150 s and 400 s. The blank spaces contain no input clusters. \textbf{Top panel}: number of simulated clusters in the flux--EM0 space. \textbf{Central panel}: fraction of simulated clusters that is detected by eSASS, either as extended or point source. \textbf{Bottom panel}: fraction of simulated clusters that is only detected as extended.}
    \label{fig:completeness_EM}
\end{figure}

\subsubsection{Completeness as a function of the central emissivity}
\label{subsec:coolcore}
We study the impact of the clusters dynamical state on the detection.
Such property is related to the central cluster emission. 
In the simulations, we relate the emissivity in the central region of the cluster to a parameter of the dark matter halo ($X_{\rm off}$) which encodes its dynamical state. 
The offset parameter, $X_{\rm off}$, is the displacement between the halo center of mass and its peak of the density profile \citep{Klypin2016, Seppi2021A&A...652A.155S}. The negative log$_{\rm 10}$ of the central emissivity (EM0) is proportionally related to X$_{\rm off}$ 
\citep[see][for more details]{Comparat2020Xray_simulation}. 
Dynamically relaxed dark matter halos (with low offset parameter) host clusters with peaked emissivity profiles (cool cores with high central emissivity, and low EM0 in this formulation). Conversely, disturbed halos (with large offset parameter) host noncool core clusters with flatter emissivity profiles. We measure the completeness fraction as a function of EM0 for clusters in different bins of flux (Fig. \ref{fig:completeness_EM}). This allows quantifying the impact of the cool core bias, which makes the detection more efficient toward clusters with a peaked emission in the core.
We describe the results for the eRASS1 simulation in the following paragraph.\\
We find that clusters with low flux are hardly detected as extended. In this regime, where few objects are detected, they are mostly characterized as point sources. About 25$\%$ of the simulated objects with a flux of $\sim$ 3$\times$10$^{\rm -14}$ erg/s/cm$^2$, EM0 $\sim$ 5, and covered by an exposure between 150 s and 400s are detected, but none of them is classified as extended. At these low fluxes, we see evidence of the cool core bias. In fact, we detect only 7$\%$ of the extremely unrelaxed simulated clusters with EM0 = 6 at this flux value, as the completeness drops by a factor of $\sim$ 3.5 from relaxed to disturbed structures. The detection is generally more efficient for brighter objects. 82$\%$ of disturbed structures (EM0 = 5.5) and flux $\sim$ 3$\times$10$^{\rm -13}$ erg/s/cm$^2$ are identified by eSASS and 39$\%$ of them are characterized as extended. In addition, in this regime the cool core clusters are still detected as extended sources. For instance, at the value of EM0 = 5, every cluster brighter than $\sim$ 1$\times$10$^{\rm -12}$ erg/s/cm$^2$ is properly classified as extended. There is a smooth transition between these two regimes: 85$\%$ of the extreme cool cores (EM0 = 4.5) with flux $\sim$1$\times$10$^{\rm -13}$ erg/s/cm$^2$ are detected, but only 14$\%$ is identified as extended. Moving to the bright end of the flux distribution, the sample becomes less affected by the cool core bias. Among the clusters with flux $\sim$ 1$\times$10$^{-12}$ erg/s/cm$^2$, relaxed (disturbed) ones with EM0 = 4.5 (EM0 = 5.5) are detected as extended in 100$\%$ (91$\%$) of the cases. This transition is clear by comparing the central and bottom panels of Fig. \ref{fig:completeness_EM}. They remark the different behavior of the completeness for simulated bright cool cores between the clusters only identified as extended and the sample with the addition of the point-like detections in the eRASS1 simulation. When including the clusters detected as point sources (central panel), the population is skewed toward lower values of EM0, especially at low flux. This effect is mitigated in the extent-selected sample (bottom panel). We further discuss an explanation in Sect. \ref{subsubsec:srcsize_dyn_state}. We conclude that the cool core bias strongly affects only the faint clusters detected as point sources. Its impact on brighter objects detected as extended is reduced. Our results suggest that a stricter selection focused on bright eROSITA clusters with larger values of extension likelihood provides a sample that is barely affected by the cool core bias. This is in agreement with \citet{Ghirardini2021morph_pars}, who do not find a clear preference for cool core clusters in the extent-selected eFEDS clusters, and \citet{Bulbul2021arXiv211009544B_clusters_disguise}, who find steeper emissivity profiles and more concentrated objects only within the sample of eFEDS clusters detected as point sources.  
\begin{figure*}
    \centering
    \includegraphics[width=1.\columnwidth]{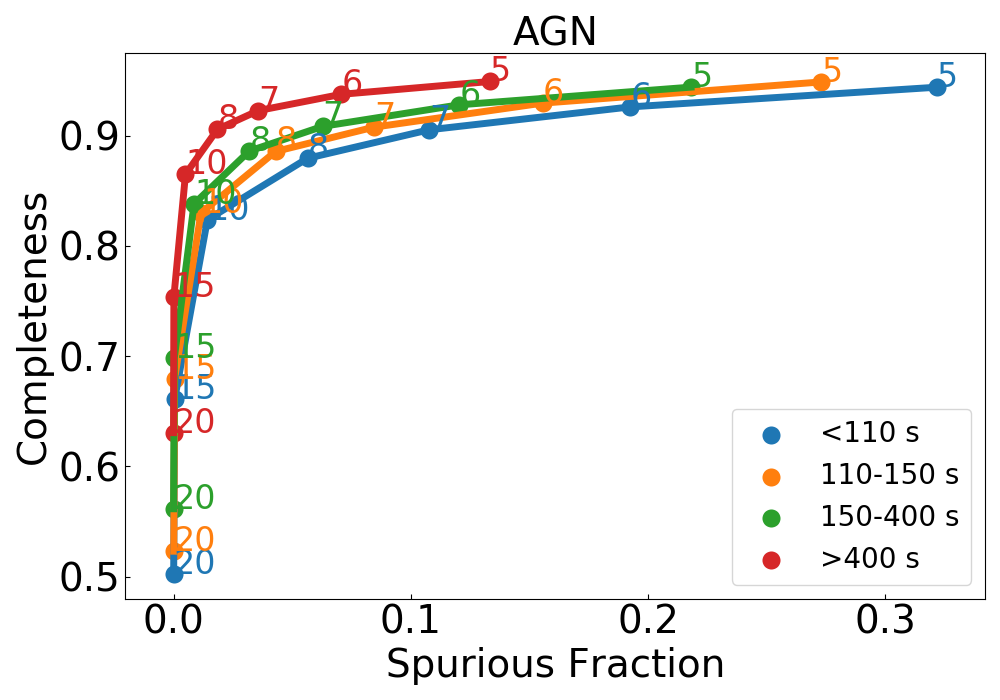}     
    \includegraphics[width=1.\columnwidth]{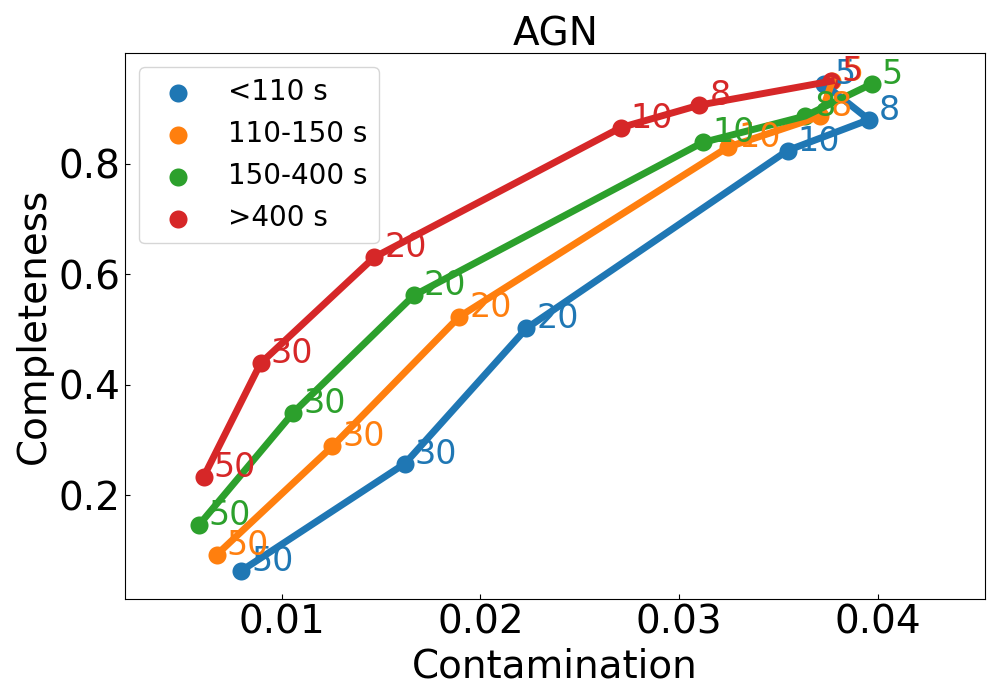}    
    \includegraphics[width=1.\columnwidth]{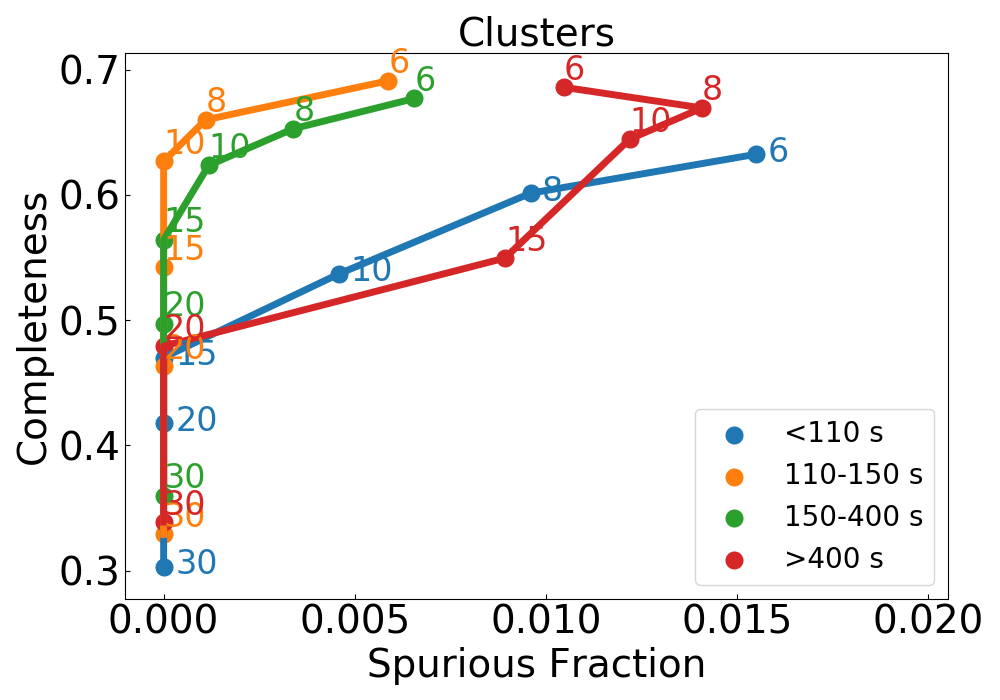}
    \includegraphics[width=1.\columnwidth]{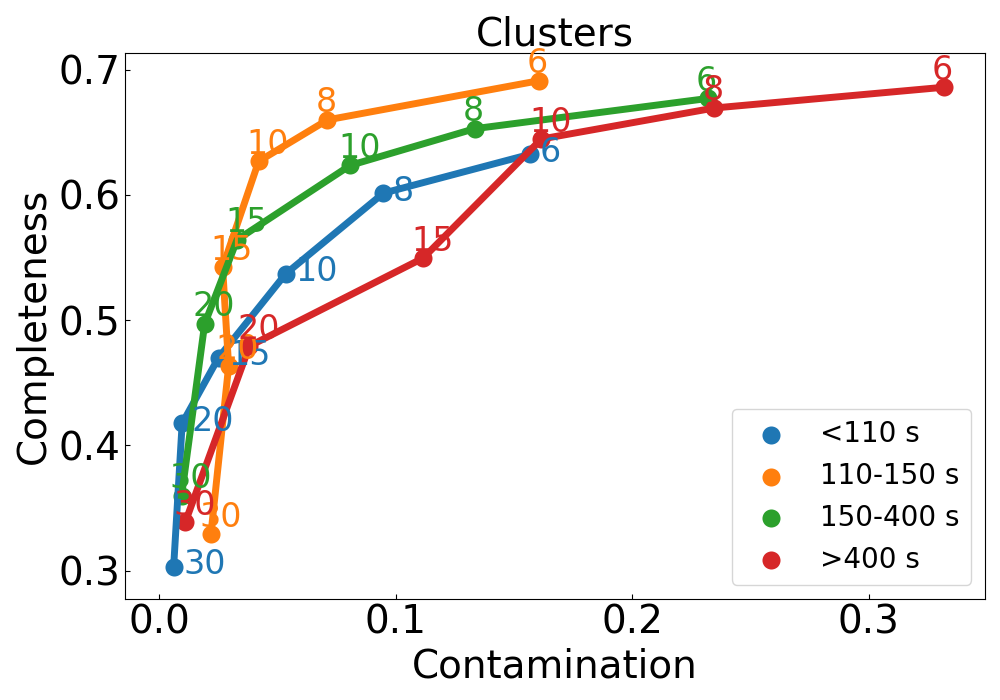}
	\caption{
	Efficiency of the eSASS detection for extragalactic sources in the eRASS1 simulation. The completeness is measured for simulated objects above the different flux limits for each exposure interval defined in Table \ref{tab:clu_density}. \textbf{Top panels}: Detection efficiency for AGN detected as point sources (EXT\_LIKE = 0). The numbers denote DET\_LIKE thresholds. \textbf{Bottom panels}: Detection efficiency for clusters. The numbers denote EXT\_LIKE thresholds. No additional cuts of DET\_LIKE are applied. \textbf{Left-hand panels}: Completeness as a function of spurious fraction. \textbf{Right-hand panels}: Completeness as a function of contamination. Different exposure intervals are shown in different colors. 
	}
    \label{fig:completness-contamination}
\end{figure*}

\subsection{Detection efficiency}
\label{subsec:detection_efficieny}

Different aspects come into play when evaluating the performance of a detection algorithm. We quantify the ability to recover simulated sources, to properly classify them as point-like or extended, and to minimize the false identification of background fluctuations. 

The first one is completeness, that is the fraction of simulated objects with a given flux that have been detected, see Eq. \ref{eq:completeness}. When measuring this number, we choose different flux thresholds according to exposure time in the two simulations. We consider the same depth bins described in previous sections (see Table \ref{tab:clu_density}). 
We work with exposure intervals and flux values identifying the 50$\%$ completeness for clusters detected as extended from Fig. \ref{fig:completeness}, and the 80$\%$ completeness for AGN. These are also reported in Table \ref{tab:clu_density}. We use these limits as thresholds and consider all the objects brighter than such values. \\ 
Secondly, one needs to account for contamination given by objects that should not be in the catalog of interest. For instance, contamination in a cluster catalog is given by bright AGN that are mistakenly classified as extended sources. This is measured by the fraction of entries in the extended source catalog that are assigned to a simulated AGN or star. For an AGN catalog instead, contamination is caused by faint and or cool core clusters that are erroneously detected as point sources.\\
Finally, it is important to consider the false detections, that are entries in the source catalog related to a random background fluctuation, not to a physical source. This causes a fraction of spurious sources in the eSASS catalog. 
Contamination and false detection rate are usually enclosed in the notion of purity. The purer a catalog, the fewer contaminants and spurious sources it contains.\\
We combine these aspects in a single concept: the detection efficiency, which encodes the completeness and purity of the source catalog. We measure completeness, contamination, and the fraction of spurious sources in the eRASS1 simulation for different intervals of exposure time, defined in Table \ref{tab:clu_density}. In addition, we account for different thresholds of the detection and extension likelihood to cut the catalogs and study how they impact the eSASS performance in terms of detection efficiency for AGN identified as point-like and clusters of galaxies characterized as extended. We report our results in the next paragraphs.

\subsubsection*{AGN}
Increasing exposure time allows detecting a fixed fraction of sources down to lower fluxes. In the full catalog with DET\_LIKE > 5, we measure similar values of the completeness fraction in distinct exposure bins, thanks to the choices of different flux limits. The values are larger than 90$\%$ for eRASS1. These numbers will depend on the given flux limit. Our choice of the value where the fraction of detected AGN is equal to 0.8 leads to measuring a higher completeness fraction when using such values as thresholds. In general, we measure a lower fraction of spurious sources in areas with larger exposure. This is because even though the total number of background photons is higher, their fluctuations are suppressed, which results in a lower false detection rate. In the shallow areas, about 32$\%$ of the full source catalog does not have a simulated counterpart and is classified as spurious. This number is reduced to 13$\%$ in regions around the southern ecliptic pole with the deepest exposure. We provide an analytical fit for the false detection rate as a function of DET\_LIKE cuts and exposure time in Appendix \ref{appendix:agn} (see Equation \ref{eq:spur_detlike_texp} and Table \ref{tab:spur_point_pars}). Progressive cuts in detection likelihood clean the source catalog from these false detections but reduce the fraction of simulated AGN that are detected. Given our choices of flux thresholds, the completeness drops from 94.4$\%$ (94.9$\%$) to 82.4$\%$ (86.5$\%$) from DET\_LIKE > 5 to DET\_LIKE > 10 in the shallow (polar) region of eRASS1. \\
The fraction of clusters that leak into the point source sample is around 4$\%$. A higher detection likelihood cut of 20 reduces this contamination to 2.2$\%$ in shallow areas and 1.5$\%$ in the pole, but this results in a significant loss in terms of completeness, which respectively drops to 50$\%$ and 63$\%$. At fixed completeness, we measure higher contamination in areas with lower depth. This means that a larger exposure time is key to properly distinguish AGN detected as point sources from clusters contaminating the point-like sample, that should be classified as extended ones instead. 
All these trends are clear in the top panels of Fig. \ref{fig:completness-contamination}. The left-hand one shows the fraction of detected AGN as a function of the false detection rate. The lines are color-coded by exposure time and the dots and numbers denote different cuts in detection likelihood. The right-hand panel displays the correlation between AGN completeness and the fraction of clusters wrongly detected as point-like objects.

\subsubsection*{Clusters}
We perform a similar analysis for the population of clusters in the source catalog at different cuts of extension likelihood. The choice of flux limits corresponding to the 50$\%$ completeness in each exposure interval translates into completeness values of around 65$\%$ when using them as thresholds. The qualitative efficiency trends are slightly different from the AGN ones. For instance, we progressively measure a lower false detection rate from regions with exposure lower than 110 seconds (1.5$\%$) to the ones covered by 150s -- 400s (0.6$\%$). However, the behavior of the spurious fraction in the pole region is different. In fact, it increases from extension likelihood larger than 6 to 8 and then drops as expected. This trend is related to the close interplay between the removal of false detections and bright AGN that leak into the extent-selected sample. Progressive cuts in extension likelihood are very effective for the latter case so that the number of spurious sources with respect to the total increases from EXT\_LIKE > 6 to EXT\_LIKE > 8, but it still decreases with respect to the number of real sources. Increasing to EXT\_LIKE > 10 brings the false detection rate to 0.5$\%$ in the shallow areas and 0.1$\%$ in deep regions.
Together with rejecting spurious sources, increasing EXT\_LIKE thresholds are very effective in reducing the fraction of contamination. The latter goes from about 30$\%$ for EXT\_LIKE > 6 in the pole region of eRASS1 to 4$\%$ at EXT\_LIKE > 20. In regions covered by the average depth of the survey, with exposure 150 s -- 400 s for eRASS1, contamination goes from 23$\%$ to 1$\%$ for the same extension likelihood cuts of 6 and 20. Deep regions suffer from contamination more than shallower ones. This value goes from about 15$\%$ in shallow areas to 32$\%$ in the pole region for eRASS1. This is due to the larger amount of bright AGN photons that can be mistaken for extended objects, but also due to the higher chance of merging nearby point sources into a single extended detection.\\
In eRASS1, we measure similar contamination of about 15$\%$ on shallow areas (< 110 s) cutting the catalog with EXT\_LIKE > 6 and the pole region (> 400 s) with EXT\_LIKE > 10. The completeness is also close to 60$\%$ with these cuts.
The average exposure of eRASS1 corresponding to roughly $\sim$ 275 s is included in the green curves. Cutting the catalogs at EXT\_LIKE > 20 provides about 50$\%$ of the simulated cluster above the chosen flux thresholds, with 2$\%$ contamination, and a null false detection rate in eRASS1.\\
The completeness-spurious fraction and contamination curves for clusters are shown in the bottom panels of Fig. \ref{fig:completness-contamination}. The figure is color-coded according to the exposure time. It highlights the results described above. Progressive EXT\_LIKE cuts of 6, 8, 10, 15, 20, and 30 are written as text. \\
These are key steps toward selecting a sample of clusters of galaxies to measure cosmological parameters with eROSITA, which has to be as pure and complete as possible. We discuss an alternative way of characterizing clusters of galaxies using the maximum signal-to-noise radius (srcRAD) in Appendix. \ref{appendix:clu_srcRAD}.

\subsection{Sensitivity}
\label{subsec:sensitivity}

We compute the sensitivity maps for point sources in each simulated tile, using the {\tt apetool} task, part of the eSASS chain described in Sect. \ref{sec:data}. The sensitivity map is related to the probability of identifying a detection in a given energy band and at a given position on the detector. Our sensitivity maps are given in units of counts. These values depend on the Poisson false detection probability, which is defined as the probability of detecting photons generated by a random background fluctuation inside a radius of a given value as a source. We set this threshold to a standard value of P = 4$\times$10$^{\rm -6}$, which corresponds to DET\_LIKE$\sim$12 (see Equation \ref{eq:detlike}). We consider apertures enclosing a local PSF encircled energy fraction equal to 60$\%$ \citep{Brunner2022_efedscat}. 
Given the definition of detection likelihood (Sect. \ref{sec:data}), these two quantities are related by DET\_LIKE = -$\ln$(P). The final sensitivity map depends on the estimated background map, the detection mask, and the exposure map. Additional details are provided by \citet{Georgakakis2008MNRAS.388.1205G}. 
For each simulated tile in the simulation, we obtain the lower count rate detection threshold by dividing sensitivity and exposure maps. We convert to flux by dividing 
the count rate by the energy conversion factor (ECF) in the soft X-ray band between 0.2 and 2.3 keV. The ECF is computed following \citet{Brunner2022_efedscat}, with an absorbed
power-law model of slope equal to 2.0 and varying galactic absorbing
column density (nH) equal to the average value in each tile. It is equal to 1.074$\times$10$^{\rm 12}$ cm$^2$/erg for an nH value of 3$\times$10$^{\rm 20}$ cm$^{-2}$. 
The result is the survey flux limit in areas of the sky covered with different exposure. We compute the cumulative distribution function of the flux limit for each tile and normalize it by the unique area covered by the sensitivity map. We sum up such quantity for all the simulated tiles. The result is the Area covered by the simulated first eROSITA all-sky survey as a function of limiting flux. 

\begin{figure}
    \centering
    \includegraphics[width=\columnwidth]{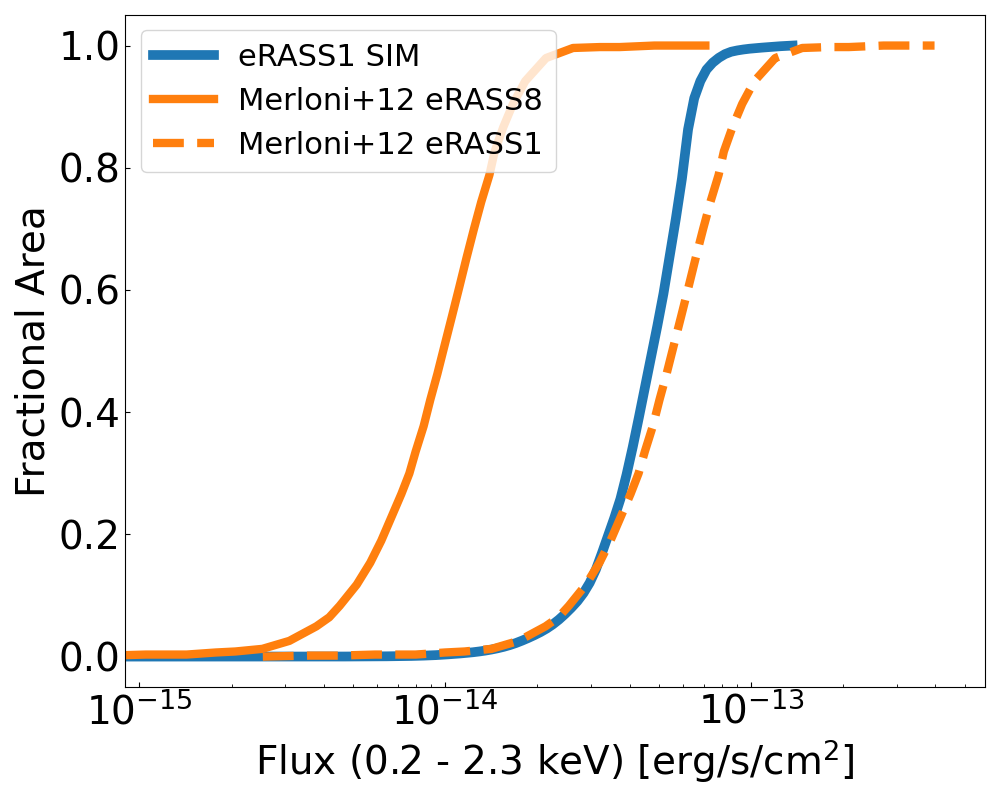}
	\caption{Simulated eROSITA fractional survey area as a function of flux limit. The eRASS1 simulation is denoted by the blue line, the prediction by \citet{Merloni2012} for eRASS:8 is shown in orange. The dashed line denotes an extrapolation of the eRASS:8 prediction to the depth of eRASS1. 
	}
    \label{fig:area_curve}
\end{figure}

We show the normalized survey area in figure \ref{fig:area_curve}. It displays the area curve for the eRASS1 simulation and a comparison to the eRASS:8 sensitivity prediction from \citet{Merloni2012}. The dashed orange line is an extrapolation of the eRASS:8 prediction to the depth of eRASS1, obtained by re-scaling the curve to the predicted eRASS1 limiting flux \citep[see Table 4.4.1 in][]{Merloni2012}, and multiplying by an additional factor of 1.403, converting the flux of an ideal absorbed power-law AGN model with N$_{\rm H}$ = 3$\times$10$^{20}$ cm$^{-2}$ and photon index $\Gamma$ = 1.8 from the 0.5--2.0 keV band to the 0.2--2.3 keV one. The agreement between the prediction and our measurement at the faint end is excellent. An offset at larger fluxes is expected because the former is based on an analytical derivation of the Poisson probability for false detections \citep[see Sect. 4.3.1 of ][]{Merloni2012}. A number of assumptions are taken into account to compute the final sensitivity, related to the background, the foreground absorption, and the exposure. Moreover, a further contribution is given by different Poisson probability thresholds, equal to 3$\times$10$^{\rm -7}$ against 4$\times$10$^{\rm -6}$ for this work. Our measure using the sensitivity maps computed by eSASS additionally accounts for a diverse and more realistic treatment of the X-ray background, as well as the true exposure derived from the real eRASS1 scanning process. This allows accounting for the nonuniform depth of the survey, compared to the prediction that is carried out at a fixed exposure equal to the average value for eRASS1 and does not include the higher sensitivity in deeper regions. This causes the difference between our measure and the prediction at the bright flux end. We find that 50$\%$ of the area is covered with a flux limit of 4.7$\times$10$^{\rm -14}$ 
erg/s/cm$^2$ in the 0.2--2.3 keV band in the eRASS1 simulation.

\subsection{Imaging and spectral analysis}
\label{subsec:LxTx}

\begin{figure}
    \centering
    \includegraphics[width=\columnwidth]{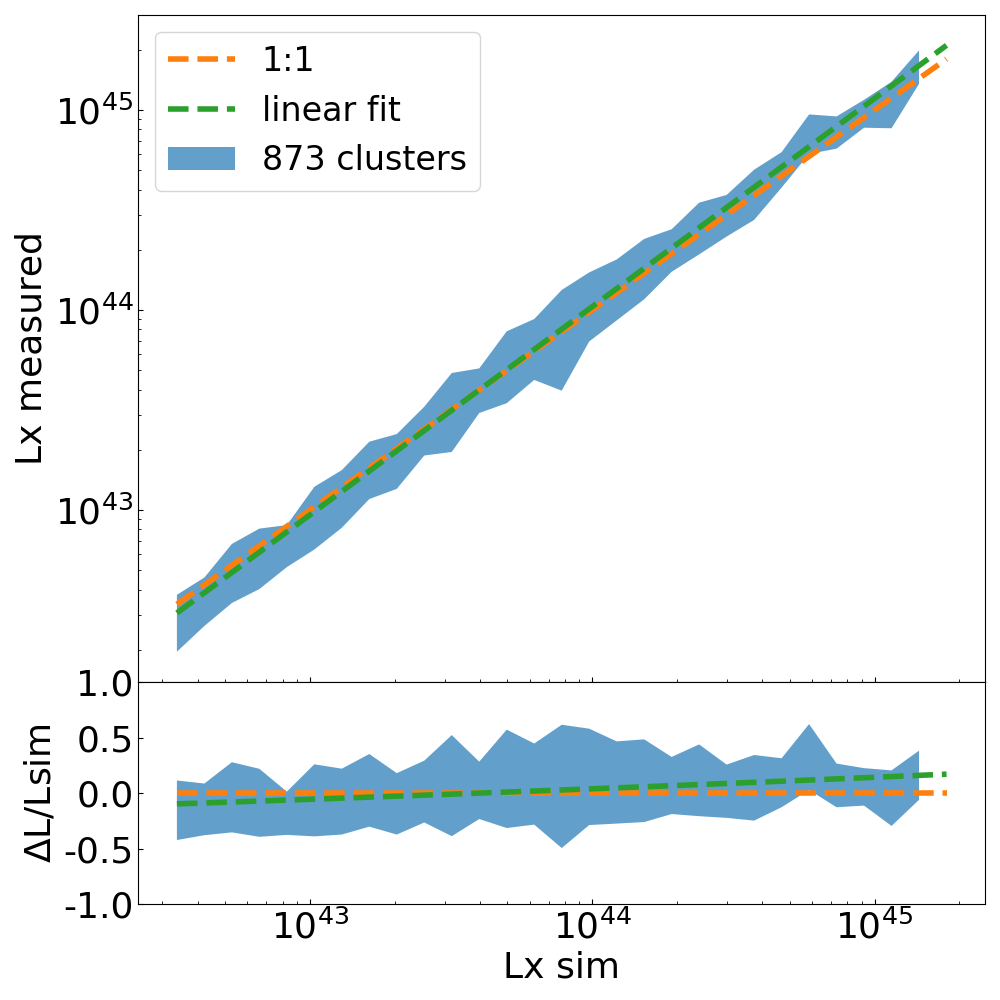}
	\caption{Measure of X-ray luminosity. \textbf{Top panel}: Comparison between average values of measured X-ray luminosity as a function of input ones. The blue shaded area encloses the average measured luminosity within 1$\sigma$ uncertainties. The dashed orange line shows a perfect one-to-one relation. \textbf{Lower panel}: Residual plot normalized by the input luminosity.}
    \label{fig:LxTx}
\end{figure}

We measure temperature and X-ray luminosity for a subsample of randomly selected clusters that have been detected as extended by eSASS. Our approach follows \citet{Ghirardini2021supercluster} and is described in Sect. \ref{subsec:imaging_spectral_analysis}.
This sample spans a wide range of exposure times, from equatorial shallow regions to deeper ones close to the southern ecliptic pole. It consists of 873 objects. In order to test our measurements, we compute a weighted mean of the measured luminosity in input luminosity bins with 0.1 dex width. We use weights that are equal to the inverse of the uncertainty on the value of measured X-ray luminosity.  
 The result is shown in Fig. \ref{fig:LxTx}.
The blue shaded area shows the average value of the recovered X-ray luminosity, enclosing the L$_{\rm X}$ standard deviation in each bin. This is always compatible with a perfect one-to-one relation, shown by the orange dashed line, and with a linear fit\footnote{\url{https://scipy.org/} \citep{Virtanen2020SciPy-NMeth}} in the form of log$_{\rm 10}$L$_{\rm X, M}$ = m log$_{\rm 10}$L$_{\rm X, SIM}$ + q, denoted by the green dashed line. The lower panel displays the residual plot. \\
We notice that the residuals of the linear fit slightly shifts from negative to positive values for increasing luminosity. The slope of the linear relation is m = 1.026$\pm$0.001. 
First of all, the fit of the density profile for faint clusters is more challenging, because they provide a lower amount of counts than bright objects. In addition, the temperature spectral fitting also requires a larger amount of photons to be precise. For fainter objects, this makes the conversion factor between count rate and luminosity more uncertain. This effect is partially mitigated by the fact that system with lower luminosity show also on average a lower gas temperature. These systems have more emission lines and a bremsstrahlung cutoff at a photon energy with high effective area, which partially reduces the number of counts needed to ultimately measure the temperature. The combination of these two factors biases the recovered X-ray luminosity toward lower values. The scatter slightly shifts toward positive values for luminous clusters. Bright structures are more probable to be extended on the sky, which increases the total net count of events that are not generated by the cluster itself but are potentially considered in the surface brightness fitting. This happens if point sources or other extended sources are not properly masked or if the background is not perfectly modeled. An additional component of the scatter in the relation between simulated and measured X-ray luminosity is given by bright nearby clusters. For these objects, the background extraction region is very large and can span areas with nH fluctuations. This may bias the temperature measure and ultimately the X-ray luminosity. These are all minor effects that do not affect our results on average. 
We find an excellent agreement between the input luminosity values and the measured ones. 












\section{Discussion}
\label{sec:discussion}
In the following section, we further discuss the importance of a proper characterization of source samples in the eROSITA surveys, and different strategies to build cluster samples for eRASS1.

\subsection{Biases of the survey sample properties}

Understanding the properties of large samples of sources from surveys such as eROSITA is crucial to exploit their scientific potential to the fullest. Accurate and precise detection and classification of sources in astronomical surveys is, therefore, an essential task. A multitude of factors make the process complex: the nature of the sources themselves, the characteristic of the telescope, and the detection pipeline. In general, it is important to understand and quantify the causes of errors and misclassification. \\
For instance, fluctuations of the X-ray background are potentially detected and classified as a source by eSASS. In this context, a biased measure of the X-ray background impacts not only the number of false detections, but also the detection likelihood of identified sources, because photons emitted from a source might be mistaken for background photons, or vice versa. An accurate estimate of the false detection rate is crucial to assess the fraction of spurious sources in a given sample. We showed that this can be achieved with realistic end-to-end simulations, identifying entries in the source catalog that are not matched to a simulated counterpart.

Another key factor is the contamination in the extent-selected sample (see Fig. \ref{fig:population_detlike}). It is important to figure out why it occurs and how it can be reduced.
Contamination is caused by different aspects.
The main contribution is given by bright point sources, that are classified as extended. In the cleaned eSASS catalog of our simulation, 1017 extended detections are assigned to a simulated AGN, about 18$\%$ of the total extent-selected catalog. Secondary effects include close pairs of bright AGN, that can mimic the emission of an extended object when the detection algorithm is not able to resolve and disentangle the point sources. 446 among the 1017 AGN detected as extended are contaminated by another point source in our simulation (see Sect. \ref{subsec:catalog_description}). In addition, bright nearby stars can appear extended on the sky, further contaminating the cluster sample. 178 extended detections are assigned to a star in the cleaned catalog of our simulation. Areas around bright known stars from the optical band can easily be masked in the real survey, which minimize the contamination due to stars. This effect is even magnified in areas with deep exposure, where random background fluctuations have a higher chance of being identified as an extended source. We find a total of 48 extended false detections.\\ 
Such cases end up in secure extended detections with large values of DET\_LIKE, which explains why choosing extremely high thresholds of detection likelihood do not lower contamination. Instead, a cut in EXT\_LIKE is needed to reduce the contamination due to point sources in the extent-selected catalog. This is clear in the bottom panel of Fig. \ref{fig:population_detlike}. It is possible to argue that a direct comparison of the population in the catalog after cutting at the same value of extension or detection likelihood is misleading, due to the intrinsic difference between them. In particular, EXT\_LIKE has typically smaller values than DET\_LIKE for extended detections. In the extent-selected sample, the 0.25, 0.5, 0.75, and 0.90 quantiles for EXT\_LIKE (DET\_LIKE) are 7.95 (22.69), 12.05 (37.19), 22.85 (65.22), 41.35 (114.99). For instance, if we focus on the 0.5 quantile, the AGN contamination is equal to 4.6$\%$ for EXT\_LIKE = 12.05 and to 18.2$\%$ at DET\_LIKE = 37.19. Therefore, we still conclude that applying extension likelihood cuts is a more efficient way of decreasing contamination. In observations, this can also be solved by a multiband approach, for example doing an optical follow-up of extended X-ray sources \citep[see][for an example]{Salvato2021arXiv210614520S_efedsfollowup}. This allows keeping all the cluster candidates in the catalog, that has the highest possible completeness. In a second step, one can look for overdensities of red galaxies around each X-ray detected cluster. If there is evidence of a red sequence, the cluster will be confirmed \citep[see][for an example]{Finoguenov2020A&A...638A.114F}. Otherwise, the catalog will be cleaned from a spurious or contaminating source, increasing the purity of the sample, while keeping the completeness level unchanged. This is a key ingredient toward precision cosmology with X-ray-selected clusters \citep{IderChitham2020MNRAS.499.4768I}. \\
With optical follow-up observations, one can not only find contaminating point sources classified as extended but also identify real clusters of galaxies that are misclassified as point sources \citep{Green10.1093/mnras/stw3059, Bulbul2021arXiv211009544B_clusters_disguise}. Understanding why extended sources are classified as point-like ones is key to correct this bias and properly characterize as many clusters as possible. A cluster ends up classified as a point source because of different reasons. The first one is brightness. These are usually faint objects, whose extended emission at the outskirts struggles to emerge over the local background. The second one is related to their cores. Clusters with a peaked emission in the center are possibly mistaken for point sources. In fact, we find that clusters with low flux and cool core are detected as point sources (see Fig. \ref{fig:completeness_EM}). Furthermore, high redshift clusters, even if intrinsically bright and extended, cover a tiny area on the sky, possibly smaller than the PSF of the telescope. Finally, clusters of galaxies hosting an AGN are potentially dominated by the emission of the latter. All these cases give rise to contamination and or misclassification for clusters of galaxies that leak into the point source sample. A purer cluster sample affected by less systematics may be obtained by a detection algorithm that excises the core region. This is because the cluster's outskirts have been shown to evolve in a self-similar way, with low scatter \citep{McDonald2017ApJ...843...28M, Kafer2019, Ghirardini2019A&A...621A..41G}. A more direct definition of the sample in terms of cluster mass is therefore achievable this way. This idea was implemented in clusters studies by \citet{Vikhlinin1998ApJ...502..558V}. A recent implementation is described by \citet{Kaefer2020wvdet}, where the X-ray images are filtered by a series of spatial wavelet filters with different scales, which allows isolating the extended emission from galaxy clusters. However, such a method requires a larger amount of counts to detect a cluster, which lowers the completeness of the sample. \\
The misclassification and contamination of clusters are additionally relevant for AGN. Simply selecting AGN from the point-like catalog means missing the bright objects that are mistakenly classified as extended, and accounting for faint clusters contaminating the point source catalog. However, we showed that this can be addressed by estimating completeness and contamination from realistic simulations, which provides the fraction of false detections, the contaminants, and the sources missed by the detection scheme according to desired selection criteria.

\subsubsection{Completeness purity trade-off}
Perfectly complete and pure samples of sources are ideally desired for astrophysical and cosmological studies. This means that, above the flux limit for a given experiment, a perfectly efficient detection and selection scheme should provide all the physical sources, making the source catalog as complete as technically feasible, identifying also very faint objects. 
Depending on the scientific application, it should also produce a catalog containing only the sources of interest, making the sample as pure and clean as possible. 
This means minimizing the rate of false detection: background fluctuations that are classified as physical sources. 
The concept of purity also includes contamination. For instance, in the extent-selected sample contamination is caused by bright AGN or stars, which should be classified as point sources instead. The number of such objects should also be minimized. 
The concepts of completeness and purity are closely related: maximizing the first means pushing the limits of the algorithm, and trying to identify the faintest physical objects. These are easily mistaken for random background fluctuations, which ultimately ends up costing a higher fraction of spurious sources in the final catalog. \\
In the context of the eROSITA surveys, the trade-off between completeness and purity is affected by various parameter choices made to select clusters. Different extension likelihood cuts are an example. 
Choosing a very low threshold will keep the catalog complete on the one hand, but on the other, the risk of introducing AGN in the catalog is higher, which increases contamination. Instead, higher likelihood thresholds will give a cleaner sample, at the cost of reducing the fraction of detected objects. This is evident in Fig. \ref{fig:completness-contamination}, where progressive EXT\_LIKE cuts degrade completeness, but improve purity, reducing the false detections and contamination. We stress that our choice of various flux limits for different exposure times guarantees a comparable benchmark between areas covered by varying depth.\\
Different choices regarding the parameters characterizing the source catalog should be taken according to the specific scientific goal. For example, if the goal is to work with a secure catalog from the start, higher thresholds should be chosen. This will minimize the spurious sources and the contamination, making such cluster sample pure. However, the completeness will also be reduced. Instead, if the goal is to select the highest possible number of clusters at first, a very low EXT\_LIKE limit is best. Such studies may involve the evolution of the luminosity function. A secondary step might then be required to clean the catalog, for example with multiwavelength observations such as an optical follow-up, allowing the confirmation of cluster candidates if there is evidence for a galaxy red sequence. This approach allows reducing contamination thanks to the multiwavelength information, while keeping the completeness level high, because no additional X-ray selection is applied. It also makes the cluster sample more secure, because it probes two distinct properties: the intra-cluster medium through X-rays, and the galaxy members in the optical and infrared bands. Samples defined in this way are particularly suitable for cosmological experiments. In this context, it is important to model the contamination and completeness levels together. For example, \citet{Aguena2018PhRvD..98l3529A} quantified the bias on the measure of cosmological parameters due to the imperfect modeling of completeness and purity in a cluster count experiment. They assumed a DES-like survey and found that a proper description of completeness and purity is key to measure unbiased cosmological parameters without degrading the constraining power especially when including low-mass clusters. A detailed description of the cluster selection (see Fig. \ref{fig:completness-contamination}) is therefore essential, since eROSITA will discover many new low-mass clusters and groups. Finally, other studies such as clustering may require a sample of objects contained in a well-defined volume. These can be constructed by rejecting faint and distant sources (see Sect. \ref{subsec:vollimsample}).

\subsubsection{Impact of source size and cool core bias}
\label{subsubsec:srcsize_dyn_state}
Given the morphological complexity of clusters of galaxies, their detection is not a simple function of flux and exposure time. \\
For example, it has been shown that the size of the cluster on the sky does have an impact on the identification \citep{Pacaud2006XMM_det, Burenin2007ApJS..172..561B, Clerc2018A&A...617A..92C}. On the one hand, the detection algorithm can easily detect bright nearby clusters and characterize them as extended. On the other hand, high redshift clusters, even if bright and large, may cover an area on the sky that is close to or smaller than the telescope PSF. The same holds for nearby groups with very low mass. Such objects are easily mistaken for point sources in the detection process. This makes the detection of clusters more complex. In Fig. \ref{fig:completeness_r500c}, we show that the fraction of clusters detected as extended is not only a function of flux and exposure, but it additionally depends on the size of R$_{\rm 500c}$ on the sky, even fixing the former two variables. This effect is more visible for clusters with a smaller radius, whose extended emission struggles to emerge over the background, compared to larger clusters with a similar flux. These objects are actually detected by eSASS, but classified as point sources, as expected. 
\\
Furthermore, the dynamical state of the clusters plays a role in the detection and classification.
Dynamically relaxed structures have had time to develop an efficient cooling toward the central regions, which enhances their central X-ray emission, resulting in a peaked surface brightness profile. This makes it easier for these types of objects to emerge over the background and biases the detection toward them. This is the notion of cool core bias \citep{Eckert2011}. 
However, clusters with a peaked profile can resemble the emission from a point source. In such cases, the peaked emission toward the central regions dominates over the tail at larger radii. This is not easily identified by eSASS, which ends up classifying the cluster as a point source. The net effect is that the detection is biased toward cool core clusters, but they might be easily misclassified as point sources. \\
The link between this effect, the exposure time, and the background has a significant impact on the detection process. 
On the one hand, a large exposure for a cool core cluster makes the large ratio between photons from the core and photons from the outskirts more clear over the background, making them look more similar to point sources than analogous objects covered by a shallow exposure. This will increase the probability to misclassify such clusters as point-like objects. On the other hand, with increasing depth the signal-to-noise ratio of the cluster outskirts will increase relative to the local background. In principle, a more accurate estimate of the background is also possible in this regime, thanks to the lower variability. These aspects should instead help the identification of clusters as extended. The characterization of the cool-core fraction in a cluster population also depends on the selection of the sample. \citet{Ghirardini2021morph_pars} measured the dynamical state of eFEDS clusters combining a set of quantities (such as concentration, central density, photons asymmetry, ellipticity) and did not find a prominent cool-core bias on the extent-selected sample. More detailed simulations at deeper exposures (e.g., eRASS:8) are needed to investigate this topic. Nonetheless, most of the brightest clusters are properly identified as extended (Fig. \ref{fig:completeness_EM}).
We conclude that the eSASS algorithm minimizes the impact of the cool core bias on the vast majority of the sample of clusters detected as extended sources. It mostly affects the low-flux clusters, where only the cool cores are detected, but classified as point sources. 

\subsection{Construction of volume-limited samples}
\label{subsec:vollimsample}
\begin{figure}
    \centering
    \includegraphics[width=1.0\columnwidth]{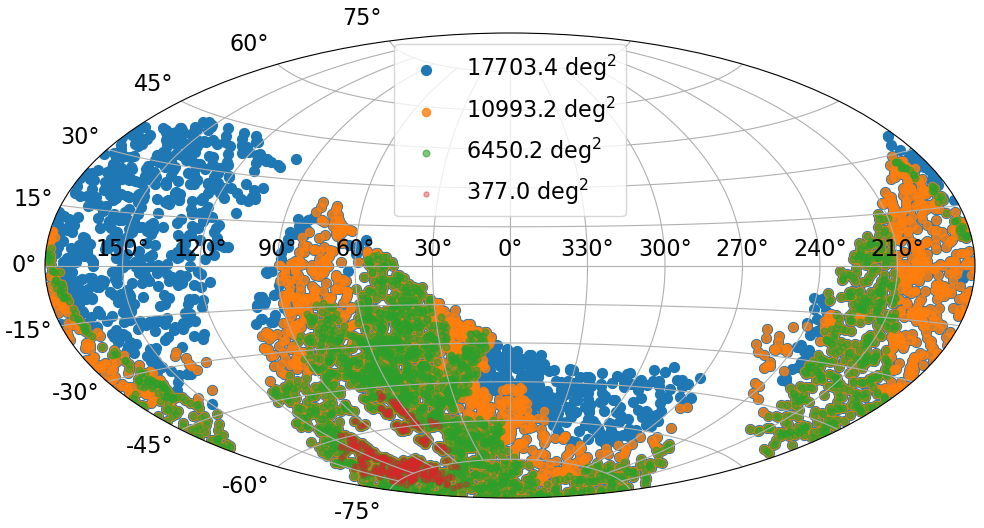}
    \includegraphics[width=.49\columnwidth]{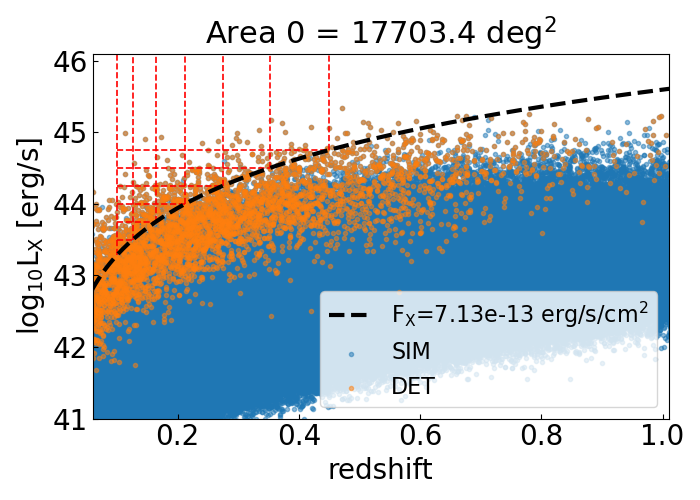}
    \includegraphics[width=.49\columnwidth]{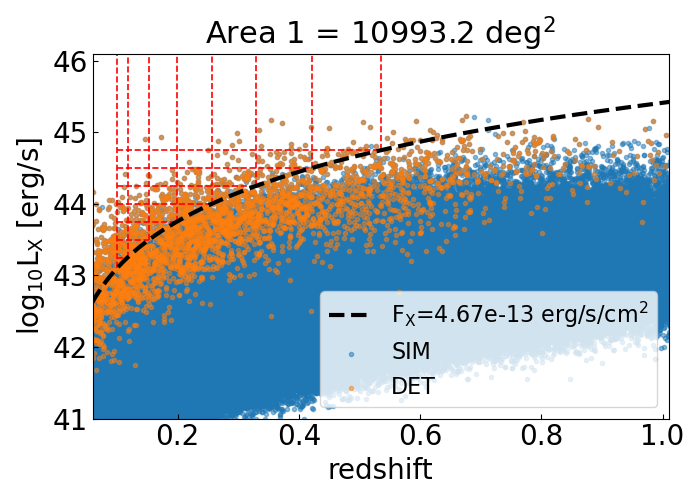}
    \includegraphics[width=.49\columnwidth]{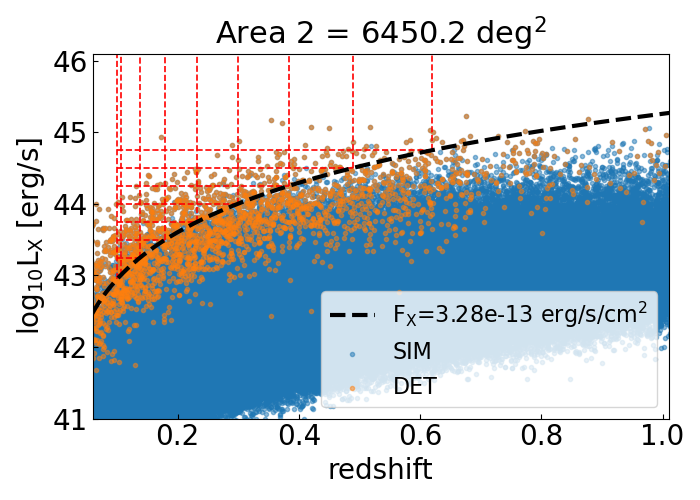}
    \includegraphics[width=.49\columnwidth]{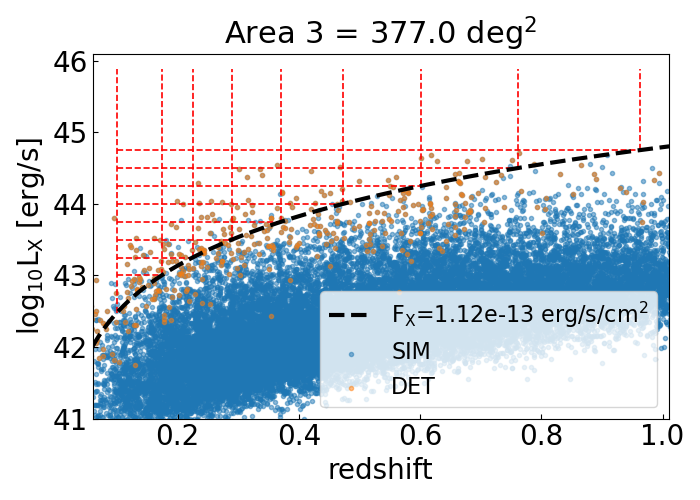}    
	\caption{Selection of a volume-limited cluster sample in the eRASS1 simulation. \textbf{Top panel}: sky map with the cluster population in areas covered by different depth. Areas 0, 1, 2, and 3 respectively cover regions with exposure larger than 0 s, 110 s, 150 s, and 400 s. They are cumulative areas with respect to the ones defined in Table \ref{tab:clu_density}.  \textbf{Bottom panels}: population of simulated and detected clusters in the luminosity--redshift plane. The black dashed lines denote the chosen flux threshold at each depth (see Table \ref{tab:clu_density}). The red dashed lines locate different areas above the given flux limits. The volume-limited sample is constructed with the objects within the regions delimited by these lines.
	}
    \label{fig:vol_lim_sample}
\end{figure}

\begin{table}[]
    \caption{Number of clusters in the volume-limited and flux-limited samples for areas covered with different depth.}
    \centering
    \begin{tabular}{|c|c|c|c|c|}
    \hline
        \rule{0pt}{2.3ex} & \multicolumn{4}{c|}{Number of clusters} \\
        \hline
        \rule{0pt}{2.3ex} Exposure [s] & \multicolumn{2}{c|}{Volume-limited} & \multicolumn{2}{c|}{Flux-limited} \\
        \hline
        \rule{0pt}{2.3ex} & DET & SIM & DET & SIM \\
         \hline
        \rule{0pt}{2.3ex} > 0 & 262 & 282 & 734 & 893 \\
         > 110 & 349 & 392 & 829 & 1044 \\
         > 150 & 361 & 414 & 752 & 992 \\
         > 400 & 80 & 100 & 146 & 200 \\
         \hline
    \end{tabular}
    \label{tab:NCLU_vollim}
\end{table} 

\begin{figure*}
    \centering
    \includegraphics[width=1.\columnwidth]{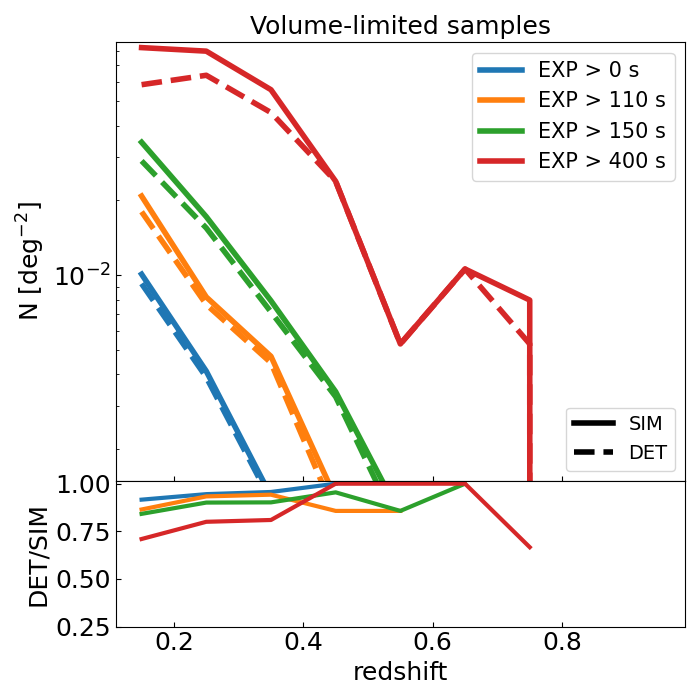}
    \includegraphics[width=1.\columnwidth]{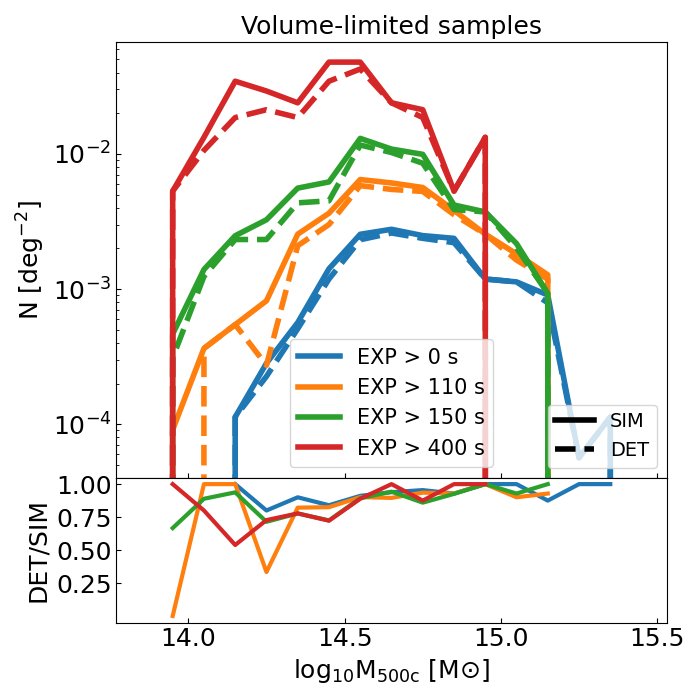}
    \includegraphics[width=1.\columnwidth]{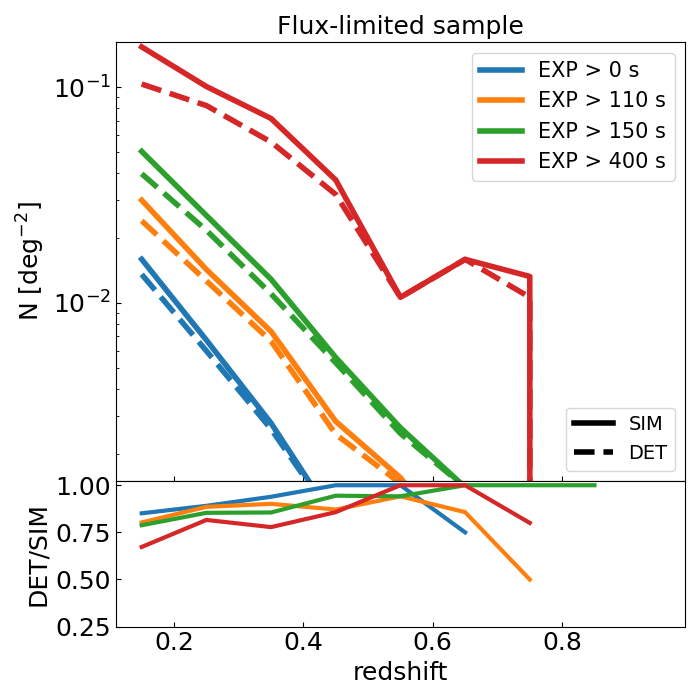}
    \includegraphics[width=1.\columnwidth]{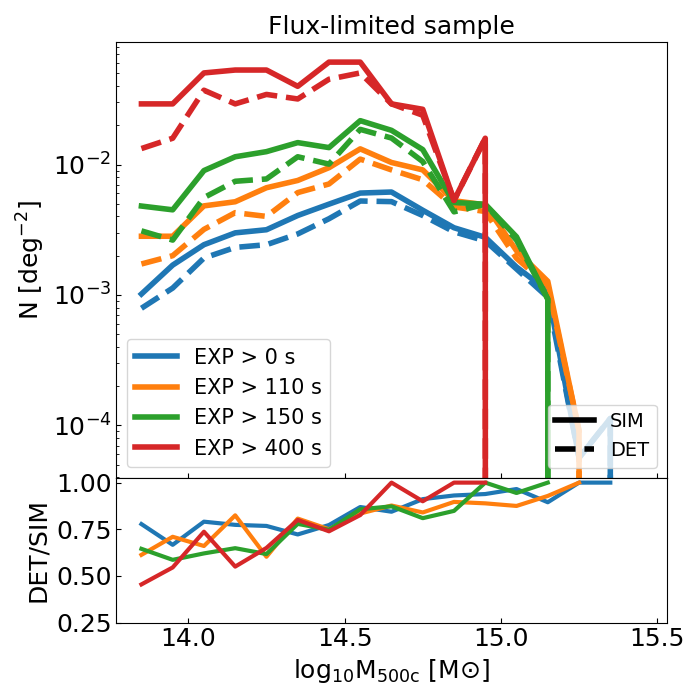}    
	\caption{Comparison between the volume-limited and flux-limited samples built with clusters detected as extended and simulated ones. The top panels display the volume-limited samples, the bottom panels show the flux-limited ones. \textbf{Left-hand panels}: relative contribution to the total cluster number density as a function of redshift for the four different populations. The lower plot shows the ratio between the N(z) built with the samples of detected and simulated clusters. \textbf{Right-hand panels}: relative contribution to the total cluster number density as a function of mass for the four different populations. The lower plot shows the ratio between the N(z) built with the samples of detected and simulated clusters.
	}
    \label{fig:vol_lim_sample2}
\end{figure*}

In the context of cosmological experiments, a well-defined sample of galaxy clusters is crucial. The eROSITA all-sky surveys naturally produce samples that are mainly flux-limited. Such samples are made up of clusters that reach the survey flux limit, which mainly depends on the telescope sensitivity and the scanning strategy. Therefore, a higher number density of objects is detected at low redshift and luminosity, compared to the high-z regime where only the brightest sources are detected. In addition, given that the sky coverage by eROSITA is not uniform in terms of depth, fainter objects can be detected in deep areas with a higher probability. Therefore, the source catalog will have different properties in regions with different exposure. These differences can be mitigated by building volume-limited samples, made up of clusters with a given set of properties inside a well-identified volume, that is encoded in the choice of the maximum distance (or redshift) up to which one is interested to build the sample. Then, in order to obtain an unbiased sample of objects, one can only consider sources whose luminosity is larger than the value corresponding to the flux limit at the chosen redshift. This provides a sample with a constant distribution of the number density as a function of redshift. Practically, a volume-limited sample is built from a flux-limited one by getting rid of the sources that are less luminous than a given threshold and further away than a certain distance (or redshift). The relation between these values of luminosity and redshift is set by the flux limit of the survey (see Eq.\ref{eq:Lum_z_fl}), where $d_{\rm L}$ denotes the luminosity distance, that depends on redshift:
\begin{equation}
    F = \frac{L}{4\pi d_{\rm L}^2(z)}.
    \label{eq:Lum_z_fl}
\end{equation}
Within such ranges of luminosity and redshift, a selection function built from simulations is less uncertain and allows unbiased results in cosmology experiments.
A volume-limited sample provides an even sampling of the large-scale structure, accounting for the observational limits of the survey. \\
Because eROSITA does not cover the sky with a uniform depth, we build different volume-limited samples applying higher flux limits in areas with lower exposure. We use z = 0.1 as the lower redshift boundary and consider the exposure intervals and flux limits corresponding to 50$\%$ completeness defined in Table \ref{tab:clu_density}. We account for the K-correction in the relation between flux and luminosity (Equation \ref{eq:Lum_z_fl}). It guarantees that the flux is always measured in the same energy band for clusters at different redshift. We start by considering all the clean extragalactic eROSITA\_DE sky (see Sect. \ref{subsec:catalog_description}) and applying the largest flux limit, identifying the 50$\%$ completeness in the shallowest areas of the survey. We proceed by reducing the area, gradually excluding shallower regions, and applying deeper flux cuts. The result is shown in Fig. \ref{fig:vol_lim_sample}. The first panel shows the cluster population in areas covered by a different exposure, that have been used to construct the volume-limited samples. The other four panels show how simulated (in blue) and detected clusters (in orange) populate the luminosity--redshift plane. The black dashed lines denote deeper flux limits as the area shrinks. The red dashed lines locate regions of this plane where the clusters are luminous enough to be above the flux limit at a given redshift. The volume-limited samples are finally built by considering the clusters inside the areas identified by these lines.
Table \ref{tab:NCLU_vollim} reports the number of clusters in the volume-limited samples and the corresponding flux-limited ones. \\
This method provides a collection of clusters detected as extended sources that are an even subsample of all the simulated clusters within the same ranges of luminosity and redshift.
This is shown in Fig. \ref{fig:vol_lim_sample2}.
The two top panels show how the volume-limited samples built with clusters detected as extended (dashed lines) compare to the one made up of simulated clusters (solid lines). The plot is color-coded according to the exposure time. The relation between the two samples in terms of redshift and M$_{\rm 500c}$ is shown in the right-hand and left-hand panels, respectively. The bottom panels show the corresponding flux-limited samples. The ratio between the number density distribution of detected and simulated clusters as a function of redshift is roughly constant for the volume-limited samples, and the majority of clusters with masses down to M$_{\rm 500c}$ $\sim$10$^{14}$M$_\odot$ within our selection are detected. This means that our method has the potential to identify a cluster sample that provides an even sampling of the large-scale structure at different redshift and exposure time. The same trends of the cluster number density as a function of mass and redshift for the flux-limited samples are qualitatively similar. For the second one, the completeness is lower compared to the volume-limited case, because the cuts in luminosity and redshift exclude clusters that are close to the flux limit and have a lower probability of being detected. Using the full flux-limited sample to measure cosmological parameters maximizes the number of clusters, but a robust selection function around the upper redshift limit of the survey is required. The advantage for the volume-limited samples is that they are contained in a well-defined cosmological volume. This potentially makes the definition of the survey volume less uncertain in a cluster counts cosmological experiment. 


\section{Summary and conclusions}
\label{sec:conclusions}

Thanks to the eROSITA X-ray telescope, we are detecting clusters of galaxies and active galactic nuclei in the X-ray band at an extraordinary rate. This has a multitude of science applications, from the evolution of accreting supermassive black holes \citep{Fanidakis2011MNRAS.410...53F}, to major steps forward in cosmological studies with X-ray-selected clusters samples \citep{Pillepich2018}. In this context, it is key to deeply understand the detection and selection of these objects, alongside the properties of the sources recovered by a given detection scheme. Using the models described by \citet{Comparat2019agn, Comparat2020Xray_simulation} we produce a half-sky simulation following the observational strategy of eROSITA, to the depth of the first all-sky survey. The simulated objects include clusters of galaxies, AGN, and stars. The population of simulated clusters is a truthful representation of real clusters of galaxies because the model is built from real observations. The background is simulated following an approach similar to the one detailed \citet{Liu2021teng_simulation}. For the eRASS1 simulation, we resample directly the real background maps. This provides an accurate digital twin of real eROSITA data. The result is shown in Fig. \ref{fig:evts_w_cats}. We run the eSASS detection algorithm on the simulation. We compare the background maps measured on the simulation and the real eRASS1 data. The simulated background is overestimated by $\sim$10$\%$, but this has a minor impact on the computation of the detection likelihood of each source (see Fig. \ref{fig:background}). We build a one-to-one correspondence between the source catalog and simulated objects thanks to a photon-based matching algorithm. We classify sources with five different labels: (i) uniquely identified with an AGN or star (PNT), (ii) uniquely identified with a cluster (EXT), (iii) fraction of AGN or star (PNT2), (iv) fraction of a cluster (EXT2), (v) background fluctuation (BKG). Various examples at values of detection likelihood equal to 10 and 20 are shown in Fig. \ref{fig:sources}. \\
We study the population in the source catalog as a function of different cuts in detection and extension likelihood. We find that the former is efficient in removing spurious sources from the catalog, which reduces the false detection rate. However, it does not reduce the contamination in the extended select sample due to bright AGN. Instead, progressive EXT\_LIKE thresholds are better suited for this task. In addition, at large values of EXT\_LIKE > 35 in eRASS1, all the contaminating AGN contain cluster emission. This reduces our estimate of contamination by $\sim$1$\%$. These results are shown in Fig. \ref{fig:population_detlike}. \\ 
Our detection algorithm perfectly recovers the bright end of the number density of objects as a function of flux for both clusters and AGN (see Fig. \ref{fig:AGN_logNlogS} and \ref{fig:CLU_logNlogS}). The eSASS detection scheme is suitable for detecting clusters of galaxies. We compare the number of simulated and detected clusters in four different intervals of exposure time. Three of them cover a similar sky area, the fourth one is smaller and centered around the ecliptic poles with extremely high depth. In areas covered by the average simulation depth, we detect half of the simulated structures at flux values of 3.3$\times$10$^{\rm -13}$ erg/s/cm$^2$ for eRASS1 (see Fig. \ref{fig:completeness}). We show how the selection of clusters is not a simple function of flux and exposure time, but the objects with different angular sizes on the sky plane are also detected differently, for instance clusters with smaller extent can be detected as point sources (see Fig. \ref{fig:completeness_r500c}). This is in agreement with previous work \citep{Pacaud2006XMM_det, Burenin2007ApJS..172..561B, Clerc2018A&A...617A..92C, Finoguenov2020A&A...638A.114F}. We study how the relaxation state impacts the detection, by exploiting the central emission parameter EM0. The detection is biased toward relaxed clusters with a low EM0. However, the effect is mostly relevant for clusters detected as point sources, as eSASS tends to classify some of these relaxed clusters with EXT\_LIKE = 0. This is particularly evident in the low flux regime around 10$^{\rm -13}$ erg/s/cm$^2$, where a high fraction of simulated objects has a counterpart in the source catalog, but such counterpart is extended (EXT\_LIKE >= 6) for only a few of them. The overall detection and characterization of clusters of galaxies are more efficient at the bright flux end, where they are still detected as extended sources (see Fig. \ref{fig:completeness_EM}). In the extent-selected sample, the impact of the cool core bias is minimal. These results are in agreement with the eFEDS sample \citep{Ghirardini2021morph_pars, Bulbul2021arXiv211009544B_clusters_disguise}. \\ 
We combine completeness and purity into the single concept of detection efficiency. We see how choosing specific flux thresholds for varying exposure times (see Table \ref{tab:clu_density}) allows detecting AGN and clusters with similar levels of completeness in areas covered with different depths. Figure \ref{fig:completness-contamination} shows that the false detection rate in shallower areas is larger. This is due to the lower signal-to-noise ratio, which causes higher relative fluctuations of the background. This is clear for the point-like sample. Progressive cuts in detection likelihood remove the majority of the spurious sources in the point-like sample. The false detection rate drops from 21.5$\%$ in areas covered by 150--400 s exposure at DET\_LIKE > 5 to < 1$\%$ for DET\_LIKE > 10. The fraction of clusters mistakenly assigned to the point sample is low, below 4$\%$ for every DET\_LIKE cut. Similar considerations can be done using thresholds of extension likelihood for clusters. In this case, the pole region shows different behavior of the false detection rate due to the cut in extension likelihood that is very efficient in removing spurious sources and also contaminating AGN in the extent-selected sample.  
Higher thresholds of extension likelihood are required to lower this value. Progressive EXT\_LIKE cuts are very effective in reducing contamination. In the region around the eRASS1 southern ecliptic pole, it drops by 26$\%$ from EXT\_LIKE > 6 to EXT\_LIKE > 20 (see Fig. \ref{fig:completness-contamination}). \\
We provide area curves as a function of limiting flux built from sensitivity maps in Fig. \ref{fig:area_curve}. Our measurement is in agreement with the prediction from \citet{Merloni2012}, especially at the faint end, but the ability of our method to account for the different exposures guarantees a better sensitivity at the bright flux end. We finally compute the X-ray luminosity of galaxy clusters in the eRASS1 simulation by fitting the surface brightness profile of each object, following the approach described by \citet{Ghirardini2021supercluster}. We show that on average we recover the simulated values of X-ray luminosity in Fig. \ref{fig:LxTx}.\\
We discuss how to best construct volume-limited samples applying different flux limits in areas covered with varying depth by eROSITA (see Fig. \ref{fig:vol_lim_sample}). This translates into different samples of clusters according to the values of luminosity and redshift. It guarantees an even sampling of the large-scale structure of the Universe also in a case of nonuniform coverage. The selection of these samples and their relative contribution to the cluster number density distribution as a function of redshift and mass is shown in Fig. \ref{fig:vol_lim_sample2}. 
\newline
We presented and analyzed a precise digital twin of the first eROSITA all-sky survey. Performing such a high-level simulation significantly increases our understanding of real data, allowing us to analyze how a realistically complex population of sources is observed by eROSITA. We studied the detection rate of galaxy clusters and AGN, accounting for the fraction of simulated objects that are detected by the eSASS pipeline, together with quantifying the false detection rate and contamination levels in the source catalog for point-like and extended sources. Using these results, one can control the fraction of false detections and or contaminants according to specific cuts of detection and extension likelihood in the real eRASS1 catalog. For example, this is useful for constructing different cluster samples, allowing for a precise contamination fraction. We addressed additional effects impacting the detection of clusters, such as their dynamical state and their physical size. This work is key toward characterizing the population of extragalactic sources in real eROSITA data.


\section*{Acknowledgements}

This work is based on data from eROSITA, the soft X-ray instrument aboard SRG, a joint Russian-German science mission supported by the Russian Space Agency (Roskosmos), in the interests of the Russian Academy of Sciences represented by its Space Research Institute (IKI), and the Deutsches Zentrum für Luft- und Raumfahrt (DLR). The SRG spacecraft was built by Lavochkin Association (NPOL) and its subcontractors, and is operated by NPOL with support from the Max Planck Institute for Extraterrestrial Physics (MPE).

The development and construction of the eROSITA X-ray instrument was led by MPE, with contributions from the Dr. Karl Remeis Observatory Bamberg \& ECAP (FAU Erlangen-Nuernberg), the University of Hamburg Observatory, the Leibniz Institute for Astrophysics Potsdam (AIP), and the Institute for Astronomy and Astrophysics of the University of Tübingen, with the support of DLR and the Max Planck Society. The Argelander Institute for Astronomy of the University of Bonn and the Ludwig Maximilians Universität Munich also participated in the science preparation for eROSITA.

The eROSITA data shown here were processed using the eSASS/NRTA
software system developed by the German eROSITA consortium.

The authors thank the referee for providing very useful comments that improved the paper.

\bibliographystyle{aa}
\bibliography{biblio} 

\begin{thebibliography}{139}
\expandafter\ifx\csname natexlab\endcsname\relax\def\natexlab#1{#1}\fi

\bibitem[{{Abbott} {et~al.}(2018){Abbott}, {Abdalla}, {Alarcon}, {Aleksi{\'c}},
  {Allam}, {Allen}, {Amara}, {Annis}, {Asorey}, {Avila}, \&
  et~al.}]{Abbott2018DES1_clustering}
{Abbott}, T.~M.~C., {Abdalla}, F.~B., {Alarcon}, A., {et~al.} 2018, \prd, 98,
  043526

\bibitem[{{Abbott} {et~al.}(2020){Abbott}, {Aguena}, {Alarcon}, {Allam},
  {Allen}, {Annis}, {Avila}, {Bacon}, {Bechtol}, {Bermeo}, {Bernstein},
  {Bertin}, {Bhargava}, {Bocquet}, {Brooks}, {Brout}, {Buckley-Geer}, {Burke},
  {Carnero Rosell}, {Carrasco Kind}, {Carretero}, {Castander}, {Cawthon},
  {Chang}, {Chen}, {Choi}, {Costanzi}, {Crocce}, {da Costa}, {Davis}, {De
  Vicente}, {DeRose}, {Desai}, {Diehl}, {Dietrich}, {Dodelson}, {Doel},
  {Drlica-Wagner}, {Eckert}, {Eifler}, {Elvin-Poole}, {Estrada}, {Everett},
  {Evrard}, {Farahi}, {Ferrero}, {Flaugher}, {Fosalba}, {Frieman},
  {Garc{\'\i}a-Bellido}, {Gatti}, {Gaztanaga}, {Gerdes}, {Giannantonio},
  {Giles}, {Grandis}, {Gruen}, {Gruendl}, {Gschwend}, {Gutierrez}, {Hartley},
  {Hinton}, {Hollowood}, {Honscheid}, {Hoyle}, {Huterer}, {James}, {Jarvis},
  {Jeltema}, {Johnson}, {Johnson}, {Kent}, {Krause}, {Kron}, {Kuehn},
  {Kuropatkin}, {Lahav}, {Li}, {Lidman}, {Lima}, {Lin}, {MacCrann}, {Maia},
  {Mantz}, {Marshall}, {Martini}, {Mayers}, {Melchior}, {Mena-Fern{\'a}ndez},
  {Menanteau}, {Miquel}, {Mohr}, {Nichol}, {Nord}, {Ogando}, {Palmese},
  {Paz-Chinch{\'o}n}, {Plazas}, {Prat}, {Rau}, {Romer}, {Roodman}, {Rooney},
  {Rozo}, {Rykoff}, {Sako}, {Samuroff}, {S{\'a}nchez}, {Sanchez}, {Saro},
  {Scarpine}, {Schubnell}, {Scolnic}, {Serrano}, {Sevilla-Noarbe}, {Sheldon},
  {Smith}, {Smith}, {Suchyta}, {Swanson}, {Tarle}, {Thomas}, {To}, {Troxel},
  {Tucker}, {Varga}, {von der Linden}, {Walker}, {Wechsler}, {Weller},
  {Wilkinson}, {Wu}, {Yanny}, {Zhang}, {Zhang}, {Zuntz}, \& {DES
  Collaboration}}]{Abbott2020DESY1_clusters}
{Abbott}, T.~M.~C., {Aguena}, M., {Alarcon}, A., {et~al.} 2020, \prd, 102,
  023509

\bibitem[{{Adami} {et~al.}(2018){Adami}, {Giles}, {Koulouridis}, {Pacaud},
  {Caretta}, {Pierre}, {Eckert}, {Ramos-Ceja}, {Gastaldello}, {Fotopoulou},
  {Guglielmo}, {Lidman}, {Sadibekova}, {Iovino}, {Maughan}, {Chiappetti},
  {Alis}, {Altieri}, {Baldry}, {Bottini}, {Birkinshaw}, {Bremer}, {Brown},
  {Cucciati}, {Driver}, {Elmer}, {Ettori}, {Evrard}, {Faccioli}, {Granett},
  {Grootes}, {Guzzo}, {Hopkins}, {Horellou}, {Lef{\`e}vre}, {Liske}, {Malek},
  {Marulli}, {Maurogordato}, {Owers}, {Paltani}, {Poggianti}, {Polletta},
  {Plionis}, {Pollo}, {Pompei}, {Ponman}, {Rapetti}, {Ricci}, {Robotham},
  {Tuffs}, {Tasca}, {Valtchanov}, {Vergani}, {Wagner}, {Willis}, \& {XXL
  Consortium}}]{Adami2018A&A...620A...5A}
{Adami}, C., {Giles}, P., {Koulouridis}, E., {et~al.} 2018, \aap, 620, A5

\bibitem[{{Aguena} \& {Lima}(2018)}]{Aguena2018PhRvD..98l3529A}
{Aguena}, M. \& {Lima}, M. 2018, \prd, 98, 123529

\bibitem[{{Aird} {et~al.}(2015){Aird}, {Coil}, {Georgakakis}, {Nandra},
  {Barro}, \& {P{\'e}rez-Gonz{\'a}lez}}]{Aird2015MNRAS.451.1892A}
{Aird}, J., {Coil}, A.~L., {Georgakakis}, A., {et~al.} 2015, \mnras, 451, 1892

\bibitem[{{Alam} {et~al.}(2021){Alam}, {Aubert}, {Avila}, {Balland},
  {Bautista}, {Bershady}, {Bizyaev}, {Blanton}, {Bolton}, {Bovy}, {Brinkmann},
  {Brownstein}, {Burtin}, {Chabanier}, {Chapman}, {Choi}, {Chuang}, {Comparat},
  {Cousinou}, {Cuceu}, {Dawson}, {de la Torre}, {de Mattia}, {Agathe}, {des
  Bourboux}, {Escoffier}, {Etourneau}, {Farr}, {Font-Ribera}, {Frinchaboy},
  {Fromenteau}, {Gil-Mar{\'\i}n}, {Le Goff}, {Gonzalez-Morales},
  {Gonzalez-Perez}, {Grabowski}, {Guy}, {Hawken}, {Hou}, {Kong}, {Parker},
  {Klaene}, {Kneib}, {Lin}, {Long}, {Lyke}, {de la Macorra}, {Martini},
  {Masters}, {Mohammad}, {Moon}, {Mueller}, {Mu{\~n}oz-Guti{\'e}rrez}, {Myers},
  {Nadathur}, {Neveux}, {Newman}, {Noterdaeme}, {Oravetz}, {Oravetz},
  {Palanque-Delabrouille}, {Pan}, {Paviot}, {Percival}, {P{\'e}rez-R{\`a}fols},
  {Petitjean}, {Pieri}, {Prakash}, {Raichoor}, {Ravoux}, {Rezaie}, {Rich},
  {Ross}, {Rossi}, {Ruggeri}, {Ruhlmann-Kleider}, {S{\'a}nchez}, {S{\'a}nchez},
  {S{\'a}nchez-Gallego}, {Sayres}, {Schneider}, {Seo}, {Shafieloo}, {Slosar},
  {Smith}, {Stermer}, {Tamone}, {Tinker}, {Tojeiro}, {Vargas-Maga{\~n}a},
  {Variu}, {Wang}, {Weaver}, {Weijmans}, {Y{\`e}che}, {Zarrouk}, {Zhao},
  {Zhao}, \& {Zheng}}]{Alam2021PhRvD.103h3533A_SDSS}
{Alam}, S., {Aubert}, M., {Avila}, S., {et~al.} 2021, \prd, 103, 083533

\bibitem[{{Allen} {et~al.}(2011){Allen}, {Evrard}, \& {Mantz}}]{Allen2011}
{Allen}, S.~W., {Evrard}, A.~E., \& {Mantz}, A.~B. 2011, \araa, 49, 409

\bibitem[{{Anderson} {et~al.}(2015){Anderson}, {Gaspari}, {White}, {Wang}, \&
  {Dai}}]{Anderson2015Lx_stellarmass}
{Anderson}, M.~E., {Gaspari}, M., {White}, S. D.~M., {Wang}, W., \& {Dai}, X.
  2015, \mnras, 449, 3806

\bibitem[{{Andrade-Santos} {et~al.}(2017){Andrade-Santos}, {Jones}, {Forman},
  {Lovisari}, {Vikhlinin}, {van Weeren}, {Murray}, {Arnaud}, {Pratt},
  {D{\'e}mocl{\`e}s}, {Kraft}, {Mazzotta}, {B{\"o}hringer}, {Chon},
  {Giacintucci}, {Clarke}, {Borgani}, {David}, {Douspis}, {Pointecouteau},
  {Dahle}, {Brown}, {Aghanim}, \& {Rasia}}]{AndradeSantos2017ApJ...843...76A}
{Andrade-Santos}, F., {Jones}, C., {Forman}, W.~R., {et~al.} 2017, \apj, 843,
  76

\bibitem[{{Angulo} {et~al.}(2012){Angulo}, {Springel}, {White}, {Jenkins},
  {Baugh}, \& {Frenk}}]{Angulo2012}
{Angulo}, R.~E., {Springel}, V., {White}, S.~D.~M., {et~al.} 2012, \mnras, 426,
  2046

\bibitem[{{Arcodia} {et~al.}(2021){Arcodia}, {Merloni}, {Nandra}, {Buchner},
  {Salvato}, {Pasham}, {Remillard}, {Comparat}, {Lamer}, {Ponti}, {Malyali},
  {Wolf}, {Arzoumanian}, {Bogensberger}, {Buckley}, {Gendreau}, {Gromadzki},
  {Kara}, {Krumpe}, {Markwardt}, {Ramos-Ceja}, {Rau}, {Schramm}, \&
  {Schwope}}]{Arcodia2021Natur.592..704A}
{Arcodia}, R., {Merloni}, A., {Nandra}, K., {et~al.} 2021, \nat, 592, 704

\bibitem[{{Arnaud}(1996)}]{Arnaud1996ASPC..101...17A}
{Arnaud}, K.~A. 1996, in Astronomical Society of the Pacific Conference Series,
  Vol. 101, Astronomical Data Analysis Software and Systems V, ed. G.~H.
  {Jacoby} \& J.~{Barnes}, 17

\bibitem[{{Bahar} {et~al.}(2021){Bahar}, {Bulbul}, {Clerc}, {Ghirardini},
  {Liu}, {Nandra}, {Pacaud}, {Chiu}, {Comparat}, {Ider-Chitham}, {Klein},
  {Liu}, {Merloni}, {Migkas}, {Okabe}, {Ramos-Ceja}, {Reiprich}, {Sanders}, \&
  {Schrabback}}]{Bahar2021arXiv211009534B_efeds_scalingrel}
{Bahar}, Y.~E., {Bulbul}, E., {Clerc}, N., {et~al.} 2021, arXiv e-prints,
  arXiv:2110.09534

\bibitem[{{Bleem} {et~al.}(2015{\natexlab{a}}){Bleem}, {Stalder}, {Brodwin},
  {Busha}, {Gladders}, {High}, {Rest}, \& {Wechsler}}]{Bleem2015ApJS_Blanco}
{Bleem}, L.~E., {Stalder}, B., {Brodwin}, M., {et~al.} 2015{\natexlab{a}},
  \apjs, 216, 20

\bibitem[{{Bleem} {et~al.}(2015{\natexlab{b}}){Bleem}, {Stalder}, {de Haan},
  {Aird}, {Allen}, {Applegate}, {Ashby}, {Bautz}, {Bayliss}, {Benson},
  {Bocquet}, {Brodwin}, {Carlstrom}, {Chang}, {Chiu}, {Cho}, {Clocchiatti},
  {Crawford}, {Crites}, {Desai}, {Dietrich}, {Dobbs}, {Foley}, {Forman},
  {George}, {Gladders}, {Gonzalez}, {Halverson}, {Hennig}, {Hoekstra},
  {Holder}, {Holzapfel}, {Hrubes}, {Jones}, {Keisler}, {Knox}, {Lee}, {Leitch},
  {Liu}, {Lueker}, {Luong-Van}, {Mantz}, {Marrone}, {McDonald}, {McMahon},
  {Meyer}, {Mocanu}, {Mohr}, {Murray}, {Padin}, {Pryke}, {Reichardt}, {Rest},
  {Ruel}, {Ruhl}, {Saliwanchik}, {Saro}, {Sayre}, {Schaffer}, {Schrabback},
  {Shirokoff}, {Song}, {Spieler}, {Stanford}, {Staniszewski}, {Stark}, {Story},
  {Stubbs}, {Vanderlinde}, {Vieira}, {Vikhlinin}, {Williamson}, {Zahn}, \&
  {Zenteno}}]{Bleem2015ApJS..216...27B}
{Bleem}, L.~E., {Stalder}, B., {de Haan}, T., {et~al.} 2015{\natexlab{b}},
  \apjs, 216, 27

\bibitem[{{B{\"o}hringer} {et~al.}(2004){B{\"o}hringer}, {Schuecker}, {Guzzo},
  {Collins}, {Voges}, {Cruddace}, {Ortiz-Gil}, {Chincarini}, {De Grandi},
  {Edge}, {MacGillivray}, {Neumann}, {Schindler}, \&
  {Shaver}}]{Bohringer2004A&Areflex}
{B{\"o}hringer}, H., {Schuecker}, P., {Guzzo}, L., {et~al.} 2004, \aap, 425,
  367

\bibitem[{{Boller} {et~al.}(2016){Boller}, {Freyberg}, {Tr{\"u}mper}, {Haberl},
  {Voges}, \& {Nandra}}]{Boller2016A&AROSAT_2rxs}
{Boller}, T., {Freyberg}, M.~J., {Tr{\"u}mper}, J., {et~al.} 2016, \aap, 588,
  A103

\bibitem[{{Borgani}(2008)}]{Borgani2008LNP...740..287B}
{Borgani}, S. 2008, {Cosmology with Clusters of Galaxies}, ed. M.~{Plionis},
  O.~{L{\'o}pez-Cruz}, \& D.~{Hughes}, Vol. 740, 24

\bibitem[{{Brunner} {et~al.}(2022){Brunner}, {Liu}, {Lamer}, {Georgakakis},
  {Merloni}, {Brusa}, {Bulbul}, {Dennerl}, {Friedrich}, {Liu}, {Maitra},
  {Nandra}, {Ramos-Ceja}, {Sanders}, {Stewart}, {Boller}, {Buchner}, {Clerc},
  \& {al, et}}]{Brunner2022_efedscat}
{Brunner}, H., {Liu}, T., {Lamer}, G., {et~al.} 2022, A\&A

\bibitem[{{Buchner} {et~al.}(2015){Buchner}, {Georgakakis}, {Nandra},
  {Brightman}, {Menzel}, {Liu}, {Hsu}, {Salvato}, {Rangel}, {Aird}, {Merloni},
  \& {Ross}}]{Buchner2015ApJ}
{Buchner}, J., {Georgakakis}, A., {Nandra}, K., {et~al.} 2015, \apj, 802, 89

\bibitem[{{Bulbul} {et~al.}(2019){Bulbul}, {Chiu}, {Mohr}, {McDonald},
  {Benson}, {Bautz}, {Bayliss}, {Bleem}, {Brodwin}, {Bocquet}, {Capasso},
  {Dietrich}, {Forman}, {Hlavacek-Larrondo}, {Holzapfel}, {Khullar}, {Klein},
  {Kraft}, {Miller}, {Reichardt}, {Saro}, {Sharon}, {Stalder}, {Schrabback}, \&
  {Stanford}}]{Bulbul2019ApJ...871...50B_scalingrel}
{Bulbul}, E., {Chiu}, I.~N., {Mohr}, J.~J., {et~al.} 2019, \apj, 871, 50

\bibitem[{{Bulbul} {et~al.}(2021){Bulbul}, {Liu}, {Pasini}, {Comparat},
  {Hoang}, {Klein}, {Ghirardini}, {Salvato}, {Merloni}, {Seppi}, {Wolf},
  {Anderson}, {Bahar}, {Brusa}, {Brueggen}, {Buchner}, {Dwelly},
  {Ibarra-Medel}, {Ider Chitham}, {Liu}, {Nandra}, {Ramos-Ceja}, {Sanders}, \&
  {Shen}}]{Bulbul2021arXiv211009544B_clusters_disguise}
{Bulbul}, E., {Liu}, A., {Pasini}, T., {et~al.} 2021, arXiv e-prints,
  arXiv:2110.09544

\bibitem[{{Burenin} {et~al.}(2007){Burenin}, {Vikhlinin}, {Hornstrup},
  {Ebeling}, {Quintana}, \& {Mescheryakov}}]{Burenin2007ApJS..172..561B}
{Burenin}, R.~A., {Vikhlinin}, A., {Hornstrup}, A., {et~al.} 2007, \apjs, 172,
  561

\bibitem[{{Cash}(1979)}]{Cash1979ApJ...228..939C}
{Cash}, W. 1979, \apj, 228, 939

\bibitem[{{Cavaliere} \&
  {Fusco-Femiano}(1976)}]{CavaliereFuscoFermiano1976A&A....49..137C}
{Cavaliere}, A. \& {Fusco-Femiano}, R. 1976, \aap, 500, 95

\bibitem[{{Chiu} {et~al.}(2021){Chiu}, {Ghirardini}, {Liu}, {Grandis},
  {Bulbul}, {Bahar}, {Comparat}, {Bocquet}, {Clerc}, {Klein}, {Liu}, {Li},
  {Miyatake}, {Mohr}, {Oguri}, {Okabe}, {Pacaud}, {Ramos-Ceja}, {Reiprich},
  {Schrabback}, \& {Umetsu}}]{Chiu2021arXiv210705652C_efeds}
{Chiu}, I.-N., {Ghirardini}, V., {Liu}, A., {et~al.} 2021, arXiv e-prints,
  arXiv:2107.05652

\bibitem[{{Chuang} {et~al.}(2019){Chuang}, {Yepes}, {Kitaura},
  {Pellejero-Ibanez}, {Rodr{\'\i}guez-Torres}, {Feng}, {Metcalf}, {Wechsler},
  {Zhao}, {To}, {Alam}, {Banerjee}, {DeRose}, {Giocoli}, {Knebe}, \&
  {Reyes}}]{Chuang2019_UNIT}
{Chuang}, C.-H., {Yepes}, G., {Kitaura}, F.-S., {et~al.} 2019, \mnras, 487, 48

\bibitem[{{Clerc} \& {Finoguenov}(2022)}]{Clerc2022arXiv220311906C_review}
{Clerc}, N. \& {Finoguenov}, A. 2022, arXiv e-prints, arXiv:2203.11906

\bibitem[{{Clerc} {et~al.}(2020){Clerc}, {Kirkpatrick}, {Finoguenov},
  {Capasso}, {Comparat}, {Damsted}, {Furnell}, {Kukkola}, {Ider Chitham},
  {Merloni}, {Salvato}, {Gueguen}, {Dwelly}, {Collins}, {Saro}, {Erfanianfar},
  {Schneider}, {Brownstein}, {Mamon}, {Padilla}, {Jullo}, \&
  {Bizyaev}}]{Clerc2020MNRAS.497.3976C}
{Clerc}, N., {Kirkpatrick}, C.~C., {Finoguenov}, A., {et~al.} 2020, \mnras,
  497, 3976

\bibitem[{{Clerc} {et~al.}(2016){Clerc}, {Merloni}, {Zhang}, {Finoguenov},
  {Dwelly}, {Nandra}, {Collins}, {Dawson}, {Kneib}, {Rozo}, {Rykoff},
  {Sadibekova}, {Brownstein}, {Lin}, {Ridl}, {Salvato}, {Schwope}, {Steinmetz},
  {Seo}, \& {Tinker}}]{Clerc2016spiders}
{Clerc}, N., {Merloni}, A., {Zhang}, Y.~Y., {et~al.} 2016, \mnras, 463, 4490

\bibitem[{{Clerc} {et~al.}(2012){Clerc}, {Pierre}, {Pacaud}, \&
  {Sadibekova}}]{Clerc2012MNRAS.423.3545C}
{Clerc}, N., {Pierre}, M., {Pacaud}, F., \& {Sadibekova}, T. 2012, \mnras, 423,
  3545

\bibitem[{{Clerc} {et~al.}(2018){Clerc}, {Ramos-Ceja}, {Ridl}, {Lamer},
  {Brunner}, {Hofmann}, {Comparat}, {Pacaud}, {K{\"a}fer}, {Reiprich},
  {Merloni}, {Schmid}, {Brand}, {Wilms}, {Friedrich}, {Finoguenov}, {Dauser},
  \& {Kreykenbohm}}]{Clerc2018A&A...617A..92C}
{Clerc}, N., {Ramos-Ceja}, M.~E., {Ridl}, J., {et~al.} 2018, \aap, 617, A92

\bibitem[{{Colless} {et~al.}(2001){Colless}, {Dalton}, {Maddox}, {Sutherland},
  {Norberg}, {Cole}, {Bland-Hawthorn}, {Bridges}, {Cannon}, {Collins}, {Couch},
  {Cross}, {Deeley}, {De Propris}, {Driver}, {Efstathiou}, {Ellis}, {Frenk},
  {Glazebrook}, {Jackson}, {Lahav}, {Lewis}, {Lumsden}, {Madgwick}, {Peacock},
  {Peterson}, {Price}, {Seaborne}, \& {Taylor}}]{Colless2001_2dF}
{Colless}, M., {Dalton}, G., {Maddox}, S., {et~al.} 2001, \mnras, 328, 1039

\bibitem[{{Comparat} {et~al.}(2020){Comparat}, {Eckert}, {Finoguenov},
  {Schmidt}, {Sanders}, {Nagai}, {Lau}, {K�fer}, {Pacaud}, {Clerc},
  {Reiprich}, {Bulbul}, {Chitham}, {Chiang}, {Ghirardini}, {Gonzalez-Perez},
  {Gozaliasl}, {Fitzpatrick}, {Klypin}, {Merloni}, {Nandra}, {Liu}, {Prada},
  {Ramos-Ceja}, {Salvato}, {Seppi}, {Tempel}, \&
  {Yepes}}]{Comparat2020Xray_simulation}
{Comparat}, J., {Eckert}, D., {Finoguenov}, A., {et~al.} 2020, The Open Journal
  of Astrophysics, 3, 13

\bibitem[{{Comparat} {et~al.}(2019){Comparat}, {Merloni}, {Salvato}, {Nandra},
  {Boller}, {Georgakakis}, {Finoguenov}, {Dwelly}, {Buchner}, {Del Moro},
  {Clerc}, {Wang}, {Zhao}, {Prada}, {Yepes}, {Brusa}, {Krumpe}, \&
  {Liu}}]{Comparat2019agn}
{Comparat}, J., {Merloni}, A., {Salvato}, M., {et~al.} 2019, \mnras, 487, 2005

\bibitem[{{Comparat} {et~al.}(2022){Comparat}, {Truong}, {Merloni},
  {Pillepich}, {Ponti}, {Driver}, {Bellstedt}, {Liske}, {Aird}, {Br{\"u}ggen},
  {Bulbul}, {Davies}, {Gonz{\'a}lez Villalba}, {Georgakakis}, {Haberl}, {Liu},
  {Maitra}, {Nandra}, {Popesso}, {Predehl}, {Robotham}, {Salvato}, {Thorne}, \&
  {Zhang}}]{Comparat2022arXiv220105169C}
{Comparat}, J., {Truong}, N., {Merloni}, A., {et~al.} 2022, arXiv e-prints,
  arXiv:2201.05169

\bibitem[{{Dauser} {et~al.}(2019){Dauser}, {Falkner}, {Lorenz}, {Kirsch},
  {Peille}, {Cucchetti}, {Schmid}, {Brand}, {Oertel}, {Smith}, \&
  {Wilms}}]{Dauser2019A&A...630A..66D}
{Dauser}, T., {Falkner}, S., {Lorenz}, M., {et~al.} 2019, \aap, 630, A66

\bibitem[{{de la Torre} {et~al.}(2013){de la Torre}, {Guzzo}, {Peacock},
  {Branchini}, {Iovino}, {Granett}, {Abbas}, {Adami}, {Arnouts}, {Bel},
  {Bolzonella}, {Bottini}, {Cappi}, {Coupon}, {Cucciati}, {Davidzon}, {De
  Lucia}, {Fritz}, {Franzetti}, {Fumana}, {Garilli}, {Ilbert}, {Krywult}, {Le
  Brun}, {Le F{\`e}vre}, {Maccagni}, {Ma{\l}ek}, {Marulli}, {McCracken},
  {Moscardini}, {Paioro}, {Percival}, {Polletta}, {Pollo}, {Schlagenhaufer},
  {Scodeggio}, {Tasca}, {Tojeiro}, {Vergani}, {Zanichelli}, {Burden}, {Di
  Porto}, {Marchetti}, {Marinoni}, {Mellier}, {Monaco}, {Nichol}, {Phleps},
  {Wolk}, \& {Zamorani}}]{delaTorre2013A&A_VIMOS}
{de la Torre}, S., {Guzzo}, L., {Peacock}, J.~A., {et~al.} 2013, \aap, 557, A54

\bibitem[{{Driver} {et~al.}(2009){Driver}, {Norberg}, {Baldry}, {Bamford},
  {Hopkins}, {Liske}, {Loveday}, {Peacock}, {Hill}, {Kelvin}, {Robotham},
  {Cross}, {Parkinson}, {Prescott}, {Conselice}, {Dunne}, {Brough}, {Jones},
  {Sharp}, {van Kampen}, {Oliver}, {Roseboom}, {Bland-Hawthorn}, {Croom},
  {Ellis}, {Cameron}, {Cole}, {Frenk}, {Couch}, {Graham}, {Proctor}, {De
  Propris}, {Doyle}, {Edmondson}, {Nichol}, {Thomas}, {Eales}, {Jarvis},
  {Kuijken}, {Lahav}, {Madore}, {Seibert}, {Meyer}, {Staveley-Smith},
  {Phillipps}, {Popescu}, {Sansom}, {Sutherland}, {Tuffs}, \&
  {Warren}}]{Driver2009A&G....50e..12D}
{Driver}, S.~P., {Norberg}, P., {Baldry}, I.~K., {et~al.} 2009, Astronomy and
  Geophysics, 50, 5.12

\bibitem[{{Eckert} {et~al.}(2019){Eckert}, {Ghirardini}, {Ettori}, {Rasia},
  {Biffi}, {Pointecouteau}, {Rossetti}, {Molendi}, {Vazza}, {Gastaldello},
  {Gaspari}, {De Grandi}, {Ghizzardi}, {Bourdin}, {Tchernin}, \&
  {Roncarelli}}]{Eckert2019xcop}
{Eckert}, D., {Ghirardini}, V., {Ettori}, S., {et~al.} 2019, \aap, 621, A40

\bibitem[{{Eckert} {et~al.}(2011){Eckert}, {Molendi}, \&
  {Paltani}}]{Eckert2011}
{Eckert}, D., {Molendi}, S., \& {Paltani}, S. 2011, \aap, 526, A79

\bibitem[{{Everett} {et~al.}(2020){Everett}, {Yanny}, {Kuropatkin}, {Huff},
  {Zhang}, {Myles}, {Masegian}, {Elvin-Poole}, {Allam}, {Bernstein},
  {Sevilla-Noarbe}, {Splettstoesser}, {Sheldon}, {Jarvis}, {Amon}, {Harrison},
  {Choi}, {Hartley}, {Alarcon}, {S{\'a}nchez}, {Gruen}, {Eckert}, {Prat},
  {Tabbutt}, {Busti}, {Becker}, {MacCrann}, {Diehl}, {Tucker}, {Bertin},
  {Jeltema}, {Drlica-Wagner}, {Gruendl}, {Bechtol}, {Carnero Rosell}, {Abbott},
  {Aguena}, {Annis}, {Bacon}, {Bhargava}, {Brooks}, {Burke}, {Carrasco Kind},
  {Carretero}, {Castander}, {Conselice}, {Costanzi}, {da Costa}, {Pereira}, {De
  Vicente}, {DeRose}, {Desai}, {Eifler}, {Evrard}, {Ferrero}, {Fosalba},
  {Frieman}, {Garc{\'\i}a-Bellido}, {Gaztanaga}, {Gerdes}, {Gutierrez},
  {Hinton}, {Hollowood}, {Honscheid}, {Huterer}, {James}, {Kent}, {Krause},
  {Kuehn}, {Lahav}, {Lima}, {Lin}, {Maia}, {Marshall}, {Melchior}, {Menanteau},
  {Miquel}, {Mohr}, {Morgan}, {Muir}, {Ogando}, {Palmese}, {Paz-Chinch{\'o}n},
  {Plazas}, {Rodriguez-Monroy}, {Romer}, {Roodman}, {Sanchez}, {Scarpine},
  {Serrano}, {Smith}, {Soares-Santos}, {Suchyta}, {Swanson}, {Tarle}, {To},
  {Troxel}, {Varga}, {Weller}, \& {Wilkinson}}]{Everett2020arXiv201212825E}
{Everett}, S., {Yanny}, B., {Kuropatkin}, N., {et~al.} 2020, arXiv e-prints,
  arXiv:2012.12825

\bibitem[{{Everett} {et~al.}(2022){Everett}, {Yanny}, {Kuropatkin}, {Huff},
  {Zhang}, {Myles}, {Masegian}, {Elvin-Poole}, {Allam}, {Bernstein},
  {Sevilla-Noarbe}, {Splettstoesser}, {Sheldon}, {Jarvis}, {Amon}, {Harrison},
  {Choi}, {Hartley}, {Alarcon}, {S{\'a}nchez}, {Gruen}, {Eckert}, {Prat},
  {Tabbutt}, {Busti}, {Becker}, {MacCrann}, {Diehl}, {Tucker}, {Bertin},
  {Jeltema}, {Drlica-Wagner}, {Gruendl}, {Bechtol}, {Rosell}, {Abbott},
  {Aguena}, {Annis}, {Bacon}, {Bhargava}, {Brooks}, {Burke}, {Kind},
  {Carretero}, {Castander}, {Conselice}, {Costanzi}, {da Costa}, {Pereira}, {De
  Vicente}, {DeRose}, {Desai}, {Eifler}, {Evrard}, {Ferrero}, {Fosalba},
  {Frieman}, {Garc{\'\i}a-Bellido}, {Gaztanaga}, {Gerdes}, {Gutierrez},
  {Hinton}, {Hollowood}, {Honscheid}, {Huterer}, {James}, {Kent}, {Krause},
  {Kuehn}, {Lahav}, {Lima}, {Lin}, {Maia}, {Marshall}, {Melchior}, {Menanteau},
  {Miquel}, {Mohr}, {Morgan}, {Muir}, {Ogando}, {Palmese}, {Paz-Chinch{\'o}n},
  {Plazas}, {Rodriguez-Monroy}, {Romer}, {Roodman}, {Sanchez}, {Scarpine},
  {Serrano}, {Smith}, {Soares-Santos}, {Suchyta}, {Swanson}, {Tarle}, {To},
  {Troxel}, {Varga}, {Weller}, {Wilkinson}, \&
  {Wilkinson}}]{Everett2022ApJ_Balrog_DES}
{Everett}, S., {Yanny}, B., {Kuropatkin}, N., {et~al.} 2022, \apjs, 258, 15

\bibitem[{{Fanidakis} {et~al.}(2011){Fanidakis}, {Baugh}, {Benson}, {Bower},
  {Cole}, {Done}, \& {Frenk}}]{Fanidakis2011MNRAS.410...53F}
{Fanidakis}, N., {Baugh}, C.~M., {Benson}, A.~J., {et~al.} 2011, \mnras, 410,
  53

\bibitem[{{Ferrarese} \& {Merritt}(2000)}]{Ferrarese2000ApJ...539L...9F}
{Ferrarese}, L. \& {Merritt}, D. 2000, \apjl, 539, L9

\bibitem[{{Finoguenov} {et~al.}(2007){Finoguenov}, {Guzzo}, {Hasinger},
  {Scoville}, {Aussel}, {B{\"o}hringer}, {Brusa}, {Capak}, {Cappelluti},
  {Comastri}, {Giodini}, {Griffiths}, {Impey}, {Koekemoer}, {Kneib},
  {Leauthaud}, {Le F{\`e}vre}, {Lilly}, {Mainieri}, {Massey}, {McCracken},
  {Mobasher}, {Murayama}, {Peacock}, {Sakelliou}, {Schinnerer}, {Silverman},
  {Smol{\v{c}}i{\'c}}, {Taniguchi}, {Tasca}, {Taylor}, {Trump}, \&
  {Zamorani}}]{Finoguenov2007ApJS..172..182F_cosmos}
{Finoguenov}, A., {Guzzo}, L., {Hasinger}, G., {et~al.} 2007, \apjs, 172, 182

\bibitem[{{Finoguenov} {et~al.}(2020){Finoguenov}, {Rykoff}, {Clerc},
  {Costanzi}, {Hagstotz}, {Ider Chitham}, {Kiiveri}, {Kirkpatrick}, {Capasso},
  {Comparat}, {Damsted}, {Dupke}, {Erfanianfar}, {Patrick Henry}, {Kaefer},
  {Kneib}, {Lindholm}, {Rozo}, {van Waerbeke}, \&
  {Weller}}]{Finoguenov2020A&A...638A.114F}
{Finoguenov}, A., {Rykoff}, E., {Clerc}, N., {et~al.} 2020, \aap, 638, A114

\bibitem[{{Finoguenov} {et~al.}(2015){Finoguenov}, {Tanaka}, {Cooper},
  {Allevato}, {Cappelluti}, {Choi}, {Heymans}, {Bauer}, {Ziparo}, {Ranalli},
  {Silverman}, {Brandt}, {Xue}, {Mulchaey}, {Howes}, {Schmid}, {Wilman},
  {Comastri}, {Hasinger}, {Mainieri}, {Luo}, {Tozzi}, {Rosati}, {Capak}, \&
  {Popesso}}]{Finoguenov2015A&A_CDFS}
{Finoguenov}, A., {Tanaka}, M., {Cooper}, M., {et~al.} 2015, \aap, 576, A130

\bibitem[{{Foreman-Mackey} {et~al.}(2013){Foreman-Mackey}, {Hogg}, {Lang}, \&
  {Goodman}}]{Foreman-Mackey2013emcee}
{Foreman-Mackey}, D., {Hogg}, D.~W., {Lang}, D., \& {Goodman}, J. 2013, \pasp,
  125, 306

\bibitem[{{Gaia Collaboration} {et~al.}(2018){Gaia Collaboration}, {Brown},
  {Vallenari}, {Prusti}, {de Bruijne}, {Babusiaux}, {Bailer-Jones}, {Biermann},
  {Evans}, {Eyer}, \& et~al.}]{Gaia2018A&A...616A...1G}
{Gaia Collaboration}, {Brown}, A.~G.~A., {Vallenari}, A., {et~al.} 2018, \aap,
  616, A1

\bibitem[{{Garrel} {et~al.}(2021){Garrel}, {Pierre}, {Valageas}, {Eckert},
  {Marulli}, {Veropalumbo}, {Pacaud}, {Clerc}, {Sereno}, {Umetsu},
  {Moscardini}, {Bhargava}, {Adami}, {Chiappetti}, {Gastaldello},
  {Koulouridis}, {Le Fevre}, \& {Plionis}}]{Garrel2021arXiv210913171G}
{Garrel}, C., {Pierre}, M., {Valageas}, P., {et~al.} 2021, arXiv e-prints,
  arXiv:2109.13171

\bibitem[{{Georgakakis} {et~al.}(2019){Georgakakis}, {Comparat}, {Merloni},
  {Ciesla}, {Aird}, \& {Finoguenov}}]{Georgakakis2019agn}
{Georgakakis}, A., {Comparat}, J., {Merloni}, A., {et~al.} 2019, \mnras, 487,
  275

\bibitem[{{Georgakakis} {et~al.}(2008){Georgakakis}, {Nandra}, {Laird}, {Aird},
  \& {Trichas}}]{Georgakakis2008MNRAS.388.1205G}
{Georgakakis}, A., {Nandra}, K., {Laird}, E.~S., {Aird}, J., \& {Trichas}, M.
  2008, \mnras, 388, 1205

\bibitem[{{Ghirardini} {et~al.}(2021{\natexlab{a}}){Ghirardini}, {Bahar},
  {Bulbul}, {Liu}, {Clerc}, {Pacaud}, {Comparat}, {Liu}, {Ramos-Ceja}, {Hoang},
  {Ider-Chitham}, {Klein}, {Merloni}, {Nandra}, {Ota}, {Predehl}, {Reiprich},
  {Sanders}, \& {Schrabback}}]{Ghirardini2021morph_pars}
{Ghirardini}, V., {Bahar}, E., {Bulbul}, E., {et~al.} 2021{\natexlab{a}}, arXiv
  e-prints, arXiv:2106.15086

\bibitem[{{Ghirardini} {et~al.}(2021{\natexlab{b}}){Ghirardini}, {Bulbul},
  {Hoang}, {Klein}, {Okabe}, {Biffi}, {Br{\"u}ggen}, {Ramos-Ceja}, {Comparat},
  {Oguri}, {Shimwell}, {Basu}, {Bonafede}, {Botteon}, {Brunetti}, {Cassano},
  {de Gasperin}, {Dennerl}, {Gatuzz}, {Gastaldello}, {Intema}, {Merloni},
  {Nandra}, {Pacaud}, {Predehl}, {Reiprich}, {Robrade}, {R{\"o}ttgering},
  {Sanders}, {van Weeren}, \& {Williams}}]{Ghirardini2021supercluster}
{Ghirardini}, V., {Bulbul}, E., {Hoang}, D.~N., {et~al.} 2021{\natexlab{b}},
  \aap, 647, A4

\bibitem[{{Ghirardini} {et~al.}(2019){Ghirardini}, {Eckert}, {Ettori},
  {Pointecouteau}, {Molendi}, {Gaspari}, {Rossetti}, {De Grandi}, {Roncarelli},
  {Bourdin}, {Mazzotta}, {Rasia}, \& {Vazza}}]{Ghirardini2019A&A...621A..41G}
{Ghirardini}, V., {Eckert}, D., {Ettori}, S., {et~al.} 2019, \aap, 621, A41

\bibitem[{{Gilli} {et~al.}(2007){Gilli}, {Comastri}, \&
  {Hasinger}}]{Gilli2007A&A...463...79G}
{Gilli}, R., {Comastri}, A., \& {Hasinger}, G. 2007, \aap, 463, 79

\bibitem[{Green {et~al.}(2016)Green, Edge, Ebeling, Burgett, Draper, Kaiser,
  Kudritzki, Magnier, Metcalfe, Wainscoat, \&
  Waters}]{Green10.1093/mnras/stw3059}
Green, T.~S., Edge, A.~C., Ebeling, H., {et~al.} 2016, Monthly Notices of the
  Royal Astronomical Society, 465, 4872

\bibitem[{{Hikage} {et~al.}(2019){Hikage}, {Oguri}, {Hamana}, {More},
  {Mandelbaum}, {Takada}, {K{\"o}hlinger}, {Miyatake}, {Nishizawa}, {Aihara},
  {Armstrong}, {Bosch}, {Coupon}, {Ducout}, {Ho}, {Hsieh}, {Komiyama},
  {Lanusse}, {Leauthaud}, {Lupton}, {Medezinski}, {Mineo}, {Miyama},
  {Miyazaki}, {Murata}, {Murayama}, {Shirasaki}, {Sif{\'o}n}, {Simet},
  {Speagle}, {Spergel}, {Strauss}, {Sugiyama}, {Tanaka}, {Utsumi}, {Wang}, \&
  {Yamada}}]{Hikage2019PASJHSC_shear}
{Hikage}, C., {Oguri}, M., {Hamana}, T., {et~al.} 2019, \pasj, 71, 43

\bibitem[{{Hilton} {et~al.}(2021){Hilton}, {Sif{\'o}n}, {Naess},
  {Madhavacheril}, {Oguri}, {Rozo}, {Rykoff}, {Abbott}, {Adhikari}, {Aguena},
  {Aiola}, {Allam}, {Amodeo}, {Amon}, {Annis}, {Ansarinejad}, {Aros-Bunster},
  {Austermann}, {Avila}, {Bacon}, {Battaglia}, {Beall}, {Becker}, {Bernstein},
  {Bertin}, {Bhandarkar}, {Bhargava}, {Bond}, {Brooks}, {Burke}, {Calabrese},
  {Carrasco Kind}, {Carretero}, {Choi}, {Choi}, {Conselice}, {da Costa},
  {Costanzi}, {Crichton}, {Crowley}, {D{\"u}nner}, {Denison}, {Devlin},
  {Dicker}, {Diehl}, {Dietrich}, {Doel}, {Duff}, {Duivenvoorden}, {Dunkley},
  {Everett}, {Ferraro}, {Ferrero}, {Fert{\'e}}, {Flaugher}, {Frieman},
  {Gallardo}, {Garc{\'\i}a-Bellido}, {Gaztanaga}, {Gerdes}, {Giles}, {Golec},
  {Gralla}, {Grandis}, {Gruen}, {Gruendl}, {Gschwend}, {Gutierrez}, {Han},
  {Hartley}, {Hasselfield}, {Hill}, {Hilton}, {Hincks}, {Hinton}, {Ho},
  {Honscheid}, {Hoyle}, {Hubmayr}, {Huffenberger}, {Hughes}, {Jaelani}, {Jain},
  {James}, {Jeltema}, {Kent}, {Knowles}, {Koopman}, {Kuehn}, {Lahav}, {Lima},
  {Lin}, {Lokken}, {Loubser}, {MacCrann}, {Maia}, {Marriage}, {Martin},
  {McMahon}, {Melchior}, {Menanteau}, {Miquel}, {Miyatake}, {Moodley},
  {Morgan}, {Mroczkowski}, {Nati}, {Newburgh}, {Niemack}, {Nishizawa},
  {Ogando}, {Orlowski-Scherer}, {Page}, {Palmese}, {Partridge},
  {Paz-Chinch{\'o}n}, {Phakathi}, {Plazas}, {Robertson}, {Romer}, {Carnero
  Rosell}, {Salatino}, {Sanchez}, {Schaan}, {Schillaci}, {Sehgal}, {Serrano},
  {Shin}, {Simon}, {Smith}, {Soares-Santos}, {Spergel}, {Staggs}, {Storer},
  {Suchyta}, {Swanson}, {Tarle}, {Thomas}, {To}, {Trac}, {Ullom}, {Vale}, {Van
  Lanen}, {Vavagiakis}, {De Vicente}, {Wilkinson}, {Wollack}, {Xu}, \&
  {Zhang}}]{Hilton2021ApJACT}
{Hilton}, M., {Sif{\'o}n}, C., {Naess}, S., {et~al.} 2021, \apjs, 253, 3

\bibitem[{{Hogg} {et~al.}(2002){Hogg}, {Baldry}, {Blanton}, \&
  {Eisenstein}}]{Hogg2002astro.ph.10394H_Kcorr}
{Hogg}, D.~W., {Baldry}, I.~K., {Blanton}, M.~R., \& {Eisenstein}, D.~J. 2002,
  arXiv e-prints, astro

\bibitem[{{Hudson} {et~al.}(2010){Hudson}, {Mittal}, {Reiprich}, {Nulsen},
  {Andernach}, \& {Sarazin}}]{Hudson2010}
{Hudson}, D.~S., {Mittal}, R., {Reiprich}, T.~H., {et~al.} 2010, \aap, 513, A37

\bibitem[{{Ider Chitham} {et~al.}(2020){Ider Chitham}, {Comparat},
  {Finoguenov}, {Clerc}, {Kirkpatrick}, {Damsted}, {Kukkola}, {Capasso},
  {Nandra}, {Merloni}, {Bulbul}, {Rykoff}, {Schneider}, \&
  {Brownstein}}]{IderChitham2020MNRAS.499.4768I}
{Ider Chitham}, J., {Comparat}, J., {Finoguenov}, A., {et~al.} 2020, \mnras,
  499, 4768

\bibitem[{{Ishiyama} {et~al.}(2021){Ishiyama}, {Prada}, {Klypin}, {Sinha},
  {Metcalf}, {Jullo}, {Altieri}, {Cora}, {Croton}, {de la Torre},
  {Mill{\'a}n-Calero}, {Oogi}, {Ruedas}, \&
  {Vega-Mart{\'\i}nez}}]{Ishiyama2021MNRAS.506.4210I_UCHUU}
{Ishiyama}, T., {Prada}, F., {Klypin}, A.~A., {et~al.} 2021, \mnras, 506, 4210

\bibitem[{{Jimeno} {et~al.}(2017){Jimeno}, {Broadhurst}, {Lazkoz}, {Angulo},
  {Diego}, {Umetsu}, \& {Chu}}]{Jimeno2017MNRAS.466.2658J}
{Jimeno}, P., {Broadhurst}, T., {Lazkoz}, R., {et~al.} 2017, \mnras, 466, 2658

\bibitem[{{Joudaki} {et~al.}(2018){Joudaki}, {Blake}, {Johnson}, {Amon},
  {Asgari}, {Choi}, {Erben}, {Glazebrook}, {Harnois-D{\'e}raps}, {Heymans},
  {Hildebrandt}, {Hoekstra}, {Klaes}, {Kuijken}, {Lidman}, {Mead}, {Miller},
  {Parkinson}, {Poole}, {Schneider}, {Viola}, \&
  {Wolf}}]{Joudaki2018KIDSclustering}
{Joudaki}, S., {Blake}, C., {Johnson}, A., {et~al.} 2018, \mnras, 474, 4894

\bibitem[{{K{\"a}fer} {et~al.}(2020){K{\"a}fer}, {Finoguenov}, {Eckert},
  {Clerc}, {Ramos-Ceja}, {Sanders}, \& {Ghirardini}}]{Kaefer2020wvdet}
{K{\"a}fer}, F., {Finoguenov}, A., {Eckert}, D., {et~al.} 2020, \aap, 634, A8

\bibitem[{{K{\"a}fer} {et~al.}(2019){K{\"a}fer}, {Finoguenov}, {Eckert},
  {Sanders}, {Reiprich}, \& {Nandra}}]{Kafer2019}
{K{\"a}fer}, F., {Finoguenov}, A., {Eckert}, D., {et~al.} 2019, \aap, 628, A43

\bibitem[{{Kauffmann} \& {Haehnelt}(2000)}]{Kauffmann2000MNRAS.311..576K}
{Kauffmann}, G. \& {Haehnelt}, M. 2000, \mnras, 311, 576

\bibitem[{{Klypin} {et~al.}(2016){Klypin}, {Yepes}, {Gottl{\"o}ber}, {Prada},
  \& {He{\ss}}}]{Klypin2016}
{Klypin}, A., {Yepes}, G., {Gottl{\"o}ber}, S., {Prada}, F., \& {He{\ss}}, S.
  2016, \mnras, 457, 4340

\bibitem[{{Koens} {et~al.}(2013){Koens}, {Maughan}, {Jones}, {Ebeling},
  {Horner}, {Perlman}, {Phillipps}, \& {Scharf}}]{Koens2013MNRAS.435.3231K}
{Koens}, L.~A., {Maughan}, B.~J., {Jones}, L.~R., {et~al.} 2013, \mnras, 435,
  3231

\bibitem[{{Kong} {et~al.}(2020){Kong}, {Burleigh}, {Ross}, {Moustakas},
  {Chuang}, {Comparat}, {de Mattia}, {du Mas des Bourboux}, {Honscheid}, {Lin},
  {Raichoor}, {Rossi}, \& {Zhao}}]{Kong2020MNRAS.499.3943K}
{Kong}, H., {Burleigh}, K.~J., {Ross}, A., {et~al.} 2020, \mnras, 499, 3943

\bibitem[{{Koulouridis} {et~al.}(2014){Koulouridis}, {Plionis}, {Melnyk},
  {Elyiv}, {Georgantopoulos}, {Clerc}, {Surdej}, {Chiappetti}, \&
  {Pierre}}]{Koulouridis2014A&AagnCLU}
{Koulouridis}, E., {Plionis}, M., {Melnyk}, O., {et~al.} 2014, \aap, 567, A83

\bibitem[{{Koutoulidis} {et~al.}(2013){Koutoulidis}, {Plionis},
  {Georgantopoulos}, \& {Fanidakis}}]{Koutoulidis2013MNRAS.428.1382K}
{Koutoulidis}, L., {Plionis}, M., {Georgantopoulos}, I., \& {Fanidakis}, N.
  2013, \mnras, 428, 1382

\bibitem[{{Kravtsov} \& {Borgani}(2012)}]{Kravtsov2012ARA&ABorgani}
{Kravtsov}, A.~V. \& {Borgani}, S. 2012, \araa, 50, 353

\bibitem[{{Lacey} \& {Cole}(1993)}]{Lacey_Cole1993MNRAS.262..627L}
{Lacey}, C. \& {Cole}, S. 1993, \mnras, 262, 627

\bibitem[{{Le Brun} {et~al.}(2014){Le Brun}, {McCarthy}, {Schaye}, \&
  {Ponman}}]{Lebrun_2014MNRAS.441.1270L_agnfeedback}
{Le Brun}, A. M.~C., {McCarthy}, I.~G., {Schaye}, J., \& {Ponman}, T.~J. 2014,
  \mnras, 441, 1270

\bibitem[{{Lesci} {et~al.}(2022){Lesci}, {Marulli}, {Moscardini}, {Sereno},
  {Veropalumbo}, {Maturi}, {Giocoli}, {Radovich}, {Bellagamba}, {Roncarelli},
  {Bardelli}, {Contarini}, {Covone}, {Ingoglia}, {Nanni}, \&
  {Puddu}}]{Lesci2022A&A...659A..88L}
{Lesci}, G.~F., {Marulli}, F., {Moscardini}, L., {et~al.} 2022, \aap, 659, A88

\bibitem[{{Lindholm} {et~al.}(2021){Lindholm}, {Finoguenov}, {Comparat},
  {Kirkpatrick}, {Rykoff}, {Clerc}, {Collins}, {Damsted}, {Ider Chitham}, \&
  {Padilla}}]{Lindholm2021A&A...646A...8L}
{Lindholm}, V., {Finoguenov}, A., {Comparat}, J., {et~al.} 2021, \aap, 646, A8

\bibitem[{{Liu} {et~al.}(2022){Liu}, {Bulbul}, {Ghirardini}, {Liu}, {Klein},
  {Clerc}, {{\"O}zsoy}, {Ramos-Ceja}, {Pacaud}, {Comparat}, {Okabe}, {Bahar},
  {Biffi}, {Brunner}, {Br{\"u}ggen}, {Buchner}, {Ider Chitham}, {Chiu},
  {Dolag}, {Gatuzz}, {Gonzalez}, {Hoang}, {Lamer}, {Merloni}, {Nandra},
  {Oguri}, {Ota}, {Predehl}, {Reiprich}, {Salvato}, {Schrabback}, {Sanders},
  {Seppi}, \& {Thibaud}}]{2022A&A_LiuAng_eFEDS_clu}
{Liu}, A., {Bulbul}, E., {Ghirardini}, V., {et~al.} 2022, \aap, 661, A2

\bibitem[{{Liu} {et~al.}(2021){Liu}, {Merloni}, {Comparat}, {Nandra},
  {Sanders}, {Lamer}, {Buchner}, {Dwelly}, {Freyberg}, {Malyali},
  {Georgakakis}, {Salvato}, {Brunner}, {Brusa}, {Klein}, {Ghirardini}, {Clerc},
  {Pacaud}, {Bulbul}, {Liu}, {Schwope}, {Robrade}, {Wilms}, {Dauser},
  {Ramos-Ceja}, {Reiprich}, {Boller}, \& {Wolf}}]{Liu2021teng_simulation}
{Liu}, T., {Merloni}, A., {Comparat}, J., {et~al.} 2021, arXiv e-prints,
  arXiv:2106.14528

\bibitem[{{Liu} {et~al.}(2013){Liu}, {Tozzi}, {Tundo}, {Moretti}, {Wang},
  {Rosati}, \& {Guglielmetti}}]{Liu2013A&A...549A.143L}
{Liu}, T., {Tozzi}, P., {Tundo}, E., {et~al.} 2013, \aap, 549, A143

\bibitem[{{Lovisari} {et~al.}(2015){Lovisari}, {Reiprich}, \&
  {Schellenberger}}]{Lovisari2015A&A...573A.118L_scalingrel}
{Lovisari}, L., {Reiprich}, T.~H., \& {Schellenberger}, G. 2015, \aap, 573,
  A118

\bibitem[{{Lovisari} {et~al.}(2020){Lovisari}, {Schellenberger}, {Sereno},
  {Ettori}, {Pratt}, {Forman}, {Jones}, {Andrade-Santos}, {Randall}, \&
  {Kraft}}]{Lovisari2020ApJ...892..102L_scalingrel}
{Lovisari}, L., {Schellenberger}, G., {Sereno}, M., {et~al.} 2020, \apj, 892,
  102

\bibitem[{{Mantz} {et~al.}(2016){Mantz}, {Allen}, {Morris}, {von der Linden},
  {Applegate}, {Kelly}, {Burke}, {Donovan}, \&
  {Ebeling}}]{Mantz2016_scaling_relation}
{Mantz}, A.~B., {Allen}, S.~W., {Morris}, R.~G., {et~al.} 2016, \mnras, 463,
  3582

\bibitem[{{Mantz} {et~al.}(2015){Mantz}, {von der Linden}, {Allen},
  {Applegate}, {Kelly}, {Morris}, {Rapetti}, {Schmidt}, {Adhikari}, {Allen},
  {Burchat}, {Burke}, {Cataneo}, {Donovan}, {Ebeling}, {Shandera}, \&
  {Wright}}]{Mantz2015cosmology}
{Mantz}, A.~B., {von der Linden}, A., {Allen}, S.~W., {et~al.} 2015, \mnras,
  446, 2205

\bibitem[{{Martini} {et~al.}(2013){Martini}, {Miller}, {Brodwin}, {Stanford},
  {Gonzalez}, {Bautz}, {Hickox}, {Stern}, {Eisenhardt}, {Galametz}, {Norman},
  {Jannuzi}, {Dey}, {Murray}, {Jones}, \& {Brown}}]{Martini2013ApJagnCLU}
{Martini}, P., {Miller}, E.~D., {Brodwin}, M., {et~al.} 2013, \apj, 768, 1

\bibitem[{{Marulli} {et~al.}(2021){Marulli}, {Veropalumbo},
  {Garc{\'\i}a-Farieta}, {Moresco}, {Moscardini}, \&
  {Cimatti}}]{Marulli2021ApJ...920...13M}
{Marulli}, F., {Veropalumbo}, A., {Garc{\'\i}a-Farieta}, J.~E., {et~al.} 2021,
  \apj, 920, 13

\bibitem[{{Marulli} {et~al.}(2018){Marulli}, {Veropalumbo}, {Sereno},
  {Moscardini}, {Pacaud}, {Pierre}, {Plionis}, {Cappi}, {Adami}, {Alis},
  {Altieri}, {Birkinshaw}, {Ettori}, {Faccioli}, {Gastaldello}, {Koulouridis},
  {Lidman}, {Le F{\`e}vre}, {Maurogordato}, {Poggianti}, {Pompei},
  {Sadibekova}, \& {Valtchanov}}]{Marulli2018A&A...620A...1M}
{Marulli}, F., {Veropalumbo}, A., {Sereno}, M., {et~al.} 2018, \aap, 620, A1

\bibitem[{{Mayer} \& {Bonoli}(2019)}]{Mayer2019RPPh...82a6901M}
{Mayer}, L. \& {Bonoli}, S. 2019, Reports on Progress in Physics, 82, 016901

\bibitem[{{McDonald} {et~al.}(2017){McDonald}, {Allen}, {Bayliss}, {Benson},
  {Bleem}, {Brodwin}, {Bulbul}, {Carlstrom}, {Forman}, {Hlavacek-Larrondo},
  {Garmire}, {Gaspari}, {Gladders}, {Mantz}, \&
  {Murray}}]{McDonald2017ApJ...843...28M}
{McDonald}, M., {Allen}, S.~W., {Bayliss}, M., {et~al.} 2017, \apj, 843, 28

\bibitem[{{McDonald} {et~al.}(2012){McDonald}, {Bayliss}, {Benson}, {Foley},
  {Ruel}, {Sullivan}, {Veilleux}, {Aird}, {Ashby}, {Bautz}, {Bazin}, {Bleem},
  {Brodwin}, {Carlstrom}, {Chang}, {Cho}, {Clocchiatti}, {Crawford}, {Crites},
  {de Haan}, {Desai}, {Dobbs}, {Dudley}, {Egami}, {Forman}, {Garmire},
  {George}, {Gladders}, {Gonzalez}, {Halverson}, {Harrington}, {High},
  {Holder}, {Holzapfel}, {Hoover}, {Hrubes}, {Jones}, {Joy}, {Keisler}, {Knox},
  {Lee}, {Leitch}, {Liu}, {Lueker}, {Luong-van}, {Mantz}, {Marrone}, {McMahon},
  {Mehl}, {Meyer}, {Miller}, {Mocanu}, {Mohr}, {Montroy}, {Murray}, {Natoli},
  {Padin}, {Plagge}, {Pryke}, {Rawle}, {Reichardt}, {Rest}, {Rex}, {Ruhl},
  {Saliwanchik}, {Saro}, {Sayre}, {Schaffer}, {Shaw}, {Shirokoff}, {Simcoe},
  {Song}, {Spieler}, {Stalder}, {Staniszewski}, {Stark}, {Story}, {Stubbs},
  {{\v{S}}uhada}, {van Engelen}, {Vanderlinde}, {Vieira}, {Vikhlinin},
  {Williamson}, {Zahn}, \& {Zenteno}}]{McDonald2012Natur.488..349M_Phoenix}
{McDonald}, M., {Bayliss}, M., {Benson}, B.~A., {et~al.} 2012, \nat, 488, 349

\bibitem[{{Merloni} {et~al.}(2012){Merloni}, {Predehl}, {Becker},
  {B{\"o}hringer}, {Boller}, {Brunner}, {Brusa}, {Dennerl}, {Freyberg},
  {Friedrich}, {Georgakakis}, {Haberl}, {Hasinger}, {Meidinger}, {Mohr},
  {Nandra}, {Rau}, {Reiprich}, {Robrade}, {Salvato}, {Santangelo}, {Sasaki},
  {Schwope}, {Wilms}, \& {German eROSITA Consortium}}]{Merloni2012}
{Merloni}, A., {Predehl}, P., {Becker}, W., {et~al.} 2012, arXiv e-prints,
  arXiv:1209.3114

\bibitem[{{Miyazaki} {et~al.}(2018){Miyazaki}, {Oguri}, {Hamana}, {Shirasaki},
  {Koike}, {Komiyama}, {Umetsu}, {Utsumi}, {Okabe}, {More}, {Medezinski},
  {Lin}, {Miyatake}, {Murayama}, {Ota}, \&
  {Mitsuishi}}]{Miyazaki2018PASJ_WLclu}
{Miyazaki}, S., {Oguri}, M., {Hamana}, T., {et~al.} 2018, \pasj, 70, S27

\bibitem[{{Moster} {et~al.}(2013){Moster}, {Naab}, \&
  {White}}]{Moster2013MNRAS.428.3121M}
{Moster}, B.~P., {Naab}, T., \& {White}, S. D.~M. 2013, \mnras, 428, 3121

\bibitem[{{Mullis} {et~al.}(2004){Mullis}, {Vikhlinin}, {Henry}, {Forman},
  {Gioia}, {Hornstrup}, {Jones}, {McNamara}, \&
  {Quintana}}]{Mullis2004ApJ...607..175M}
{Mullis}, C.~R., {Vikhlinin}, A., {Henry}, J.~P., {et~al.} 2004, \apj, 607, 175

\bibitem[{{Noordeh} {et~al.}(2020){Noordeh}, {Canning}, {King}, {Allen},
  {Mantz}, {Morris}, {Ehlert}, {von der Linden}, {Brandt}, {Luo}, {Xue}, \&
  {Kelly}}]{Noordeh2020MNRAS.498.4095N}
{Noordeh}, E., {Canning}, R.~E.~A., {King}, A., {et~al.} 2020, \mnras, 498,
  4095

\bibitem[{{Oguri}(2014)}]{Oguri2014MNRASCAMIRA}
{Oguri}, M. 2014, \mnras, 444, 147

\bibitem[{{Oguri} {et~al.}(2018){Oguri}, {Lin}, {Lin}, {Nishizawa}, {More},
  {More}, {Hsieh}, {Medezinski}, {Miyatake}, {Jian}, {Lin}, {Takada}, {Okabe},
  {Speagle}, {Coupon}, {Leauthaud}, {Lupton}, {Miyazaki}, {Price}, {Tanaka},
  {Chiu}, {Komiyama}, {Okura}, {Tanaka}, \& {Usuda}}]{Oguri2018PASJ...70S..20O}
{Oguri}, M., {Lin}, Y.-T., {Lin}, S.-C., {et~al.} 2018, \pasj, 70, S20

\bibitem[{{Pacaud} {et~al.}(2018){Pacaud}, {Pierre}, {Melin}, {Adami},
  {Evrard}, {Galli}, {Gastaldello}, {Maughan}, {Sereno}, {Alis}, {Altieri},
  {Birkinshaw}, {Chiappetti}, {Faccioli}, {Giles}, {Horellou}, {Iovino},
  {Koulouridis}, {Le F{\`e}vre}, {Lidman}, {Lieu}, {Maurogordato},
  {Moscardini}, {Plionis}, {Poggianti}, {Pompei}, {Sadibekova}, {Valtchanov},
  \& {Willis}}]{Pacaud2018XLL_cosmology}
{Pacaud}, F., {Pierre}, M., {Melin}, J.~B., {et~al.} 2018, \aap, 620, A10

\bibitem[{{Pacaud} {et~al.}(2006){Pacaud}, {Pierre}, {Refregier}, {Gueguen},
  {Starck}, {Valtchanov}, {Read}, {Altieri}, {Chiappetti}, {Gandhi}, {Garcet},
  {Gosset}, {Ponman}, \& {Surdej}}]{Pacaud2006XMM_det}
{Pacaud}, F., {Pierre}, M., {Refregier}, A., {et~al.} 2006, \mnras, 372, 578

\bibitem[{{Padovani} {et~al.}(2017){Padovani}, {Alexander}, {Assef}, {De
  Marco}, {Giommi}, {Hickox}, {Richards}, {Smol{\v{c}}i{\'c}},
  {Hatziminaoglou}, {Mainieri}, \& {Salvato}}]{Padovani2017A&ARv..25....2P}
{Padovani}, P., {Alexander}, D.~M., {Assef}, R.~J., {et~al.} 2017, \aapr, 25, 2

\bibitem[{{Pasini} {et~al.}(2021){Pasini}, {Br{\"u}ggen}, {Hoang},
  {Ghirardini}, {Bulbul}, {Klein}, {Liu}, {Shimwell}, {Hardcastle}, {Williams},
  {Botteon}, {Gastaldello}, {van Weeren}, {Merloni}, {de Gasperin}, {Bahar},
  {Pacaud}, \& {Ramos-Ceja}}]{Pasini2021arXiv210614524P}
{Pasini}, T., {Br{\"u}ggen}, M., {Hoang}, D.~N., {et~al.} 2021, arXiv e-prints,
  arXiv:2106.14524

\bibitem[{{Pierre} {et~al.}(2016){Pierre}, {Pacaud}, {Adami}, {Alis},
  {Altieri}, {Baran}, {Benoist}, {Birkinshaw}, {Bongiorno}, {Bremer}, {Brusa},
  {Butler}, {Ciliegi}, {Chiappetti}, {Clerc}, {Corasaniti}, {Coupon}, {De
  Breuck}, {Democles}, {Desai}, {Delhaize}, {Devriendt}, {Dubois}, {Eckert},
  {Elyiv}, {Ettori}, {Evrard}, {Faccioli}, {Farahi}, {Ferrari}, {Finet},
  {Fotopoulou}, {Fourmanoit}, {Gandhi}, {Gastaldello}, {Gastaud},
  {Georgantopoulos}, {Giles}, {Guennou}, {Guglielmo}, {Horellou}, {Husband},
  {Huynh}, {Iovino}, {Kilbinger}, {Koulouridis}, {Lavoie}, {Le Brun}, {Le
  Fevre}, {Lidman}, {Lieu}, {Lin}, {Mantz}, {Maughan}, {Maurogordato},
  {McCarthy}, {McGee}, {Melin}, {Melnyk}, {Menanteau}, {Novak}, {Paltani},
  {Plionis}, {Poggianti}, {Pomarede}, {Pompei}, {Ponman}, {Ramos-Ceja},
  {Ranalli}, {Rapetti}, {Raychaudury}, {Reiprich}, {Rottgering}, {Rozo},
  {Rykoff}, {Sadibekova}, {Santos}, {Sauvageot}, {Schimd}, {Sereno}, {Smith},
  {Smol{\v{c}}i{\'c}}, {Snowden}, {Spergel}, {Stanford}, {Surdej}, {Valageas},
  {Valotti}, {Valtchanov}, {Vignali}, {Willis}, \& {Ziparo}}]{Pierre2016XLL}
{Pierre}, M., {Pacaud}, F., {Adami}, C., {et~al.} 2016, \aap, 592, A1

\bibitem[{{Pillepich} {et~al.}(2012){Pillepich}, {Porciani}, \&
  {Reiprich}}]{Pillepich2012MNRAS.422...44P}
{Pillepich}, A., {Porciani}, C., \& {Reiprich}, T.~H. 2012, \mnras, 422, 44

\bibitem[{{Pillepich} {et~al.}(2018){Pillepich}, {Reiprich}, {Porciani},
  {Borm}, \& {Merloni}}]{Pillepich2018}
{Pillepich}, A., {Reiprich}, T.~H., {Porciani}, C., {Borm}, K., \& {Merloni},
  A. 2018, \mnras, 481, 613

\bibitem[{{Planck Collaboration} {et~al.}(2014{\natexlab{a}}){Planck
  Collaboration}, {Abergel}, {Ade}, {Aghanim}, {Alves}, {Aniano},
  {Armitage-Caplan}, {Arnaud}, {Ashdown}, {Atrio-Barandela}, \&
  et~al.}]{Planck_2014}
{Planck Collaboration}, {Abergel}, A., {Ade}, P.~A.~R., {et~al.}
  2014{\natexlab{a}}, \aap, 571, A11

\bibitem[{{Planck Collaboration} {et~al.}(2014{\natexlab{b}}){Planck
  Collaboration}, {Ade}, {Aghanim}, {Armitage-Caplan}, {Arnaud}, {Ashdown},
  {Atrio-Barandela}, {Aumont}, {Baccigalupi}, {Banday}, \&
  et~al.}]{Planck2014A&A_LSS_lensing}
{Planck Collaboration}, {Ade}, P.~A.~R., {Aghanim}, N., {et~al.}
  2014{\natexlab{b}}, \aap, 571, A17

\bibitem[{{Planck Collaboration} {et~al.}(2016{\natexlab{a}}){Planck
  Collaboration}, {Ade}, {Aghanim}, {Arnaud}, {Ashdown}, {Aumont},
  {Baccigalupi}, {Banday}, {Barreiro}, {Barrena}, \&
  et~al.}]{Planck2016A&A_SZ_sources}
{Planck Collaboration}, {Ade}, P.~A.~R., {Aghanim}, N., {et~al.}
  2016{\natexlab{a}}, \aap, 594, A27

\bibitem[{{Planck Collaboration} {et~al.}(2016{\natexlab{b}}){Planck
  Collaboration}, {Ade}, {Aghanim}, {Arnaud}, {Ashdown}, {Aumont},
  {Baccigalupi}, {Banday}, {Barreiro}, {Bartlett}, \&
  et~al.}]{Planck2016cluster_cosmology}
{Planck Collaboration}, {Ade}, P.~A.~R., {Aghanim}, N., {et~al.}
  2016{\natexlab{b}}, \aap, 594, A24

\bibitem[{{Pratt} {et~al.}(2019){Pratt}, {Arnaud}, {Biviano}, {Eckert},
  {Ettori}, {Nagai}, {Okabe}, \& {Reiprich}}]{Pratt2019SSRv..215...25P}
{Pratt}, G.~W., {Arnaud}, M., {Biviano}, A., {et~al.} 2019, \ssr, 215, 25

\bibitem[{{Predehl} {et~al.}(2021){Predehl}, {Andritschke}, {Arefiev},
  {Babyshkin}, {Batanov}, {Becker}, {B{\"o}hringer}, {Bogomolov}, {Boller},
  {Borm}, {Bornemann}, {Br{\"a}uninger}, {Br{\"u}ggen}, {Brunner}, {Brusa},
  {Bulbul}, {Buntov}, {Burwitz}, {Burkert}, {Clerc}, {Churazov}, {Coutinho},
  {Dauser}, {Dennerl}, {Doroshenko}, {Eder}, {Emberger}, {Eraerds},
  {Finoguenov}, {Freyberg}, {Friedrich}, {Friedrich}, {F{\"u}rmetz},
  {Georgakakis}, {Gilfanov}, {Granato}, {Grossberger}, {Gueguen}, {Gureev},
  {Haberl}, {H{\"a}lker}, {Hartner}, {Hasinger}, {Huber}, {Ji}, {Kienlin},
  {Kink}, {Korotkov}, {Kreykenbohm}, {Lamer}, {Lomakin}, {Lapshov}, {Liu},
  {Maitra}, {Meidinger}, {Menz}, {Merloni}, {Mernik}, {Mican}, {Mohr},
  {M{\"u}ller}, {Nandra}, {Nazarov}, {Pacaud}, {Pavlinsky}, {Perinati},
  {Pfeffermann}, {Pietschner}, {Ramos-Ceja}, {Rau}, {Reiffers}, {Reiprich},
  {Robrade}, {Salvato}, {Sanders}, {Santangelo}, {Sasaki}, {Scheuerle},
  {Schmid}, {Schmitt}, {Schwope}, {Shirshakov}, {Steinmetz}, {Stewart},
  {Str{\"u}der}, {Sunyaev}, {Tenzer}, {Tiedemann}, {Tr{\"u}mper}, {Voron},
  {Weber}, {Wilms}, \& {Yaroshenko}}]{Predehl2021A&A...647A...1P}
{Predehl}, P., {Andritschke}, R., {Arefiev}, V., {et~al.} 2021, \aap, 647, A1

\bibitem[{{Reiprich} \& {B{\"o}hringer}(2002)}]{Reiprich2002HIFLUGCS}
{Reiprich}, T.~H. \& {B{\"o}hringer}, H. 2002, \apj, 567, 716

\bibitem[{{Rosati} {et~al.}(2002){Rosati}, {Borgani}, \&
  {Norman}}]{Rosati2002xray_cluster_evo}
{Rosati}, P., {Borgani}, S., \& {Norman}, C. 2002, \araa, 40, 539

\bibitem[{{Rossetti} {et~al.}(2016){Rossetti}, {Gastaldello}, {Ferioli},
  {Bersanelli}, {De Grandi}, {Eckert}, {Ghizzardi}, {Maino}, \&
  {Molendi}}]{Rossetti2016dyn_state}
{Rossetti}, M., {Gastaldello}, F., {Ferioli}, G., {et~al.} 2016, \mnras, 457,
  4515

\bibitem[{{Rykoff} {et~al.}(2014){Rykoff}, {Rozo}, {Busha}, {Cunha},
  {Finoguenov}, {Evrard}, {Hao}, {Koester}, {Leauthaud}, {Nord}, {Pierre},
  {Reddick}, {Sadibekova}, {Sheldon}, \& {Wechsler}}]{Rykoff2014redmapper}
{Rykoff}, E.~S., {Rozo}, E., {Busha}, M.~T., {et~al.} 2014, \apj, 785, 104

\bibitem[{{Salvato} {et~al.}(2021){Salvato}, {Wolf}, {Dwelly}, {Georgakakis},
  {Brusa}, {Merloni}, {Liu}, {Toba}, {Nandra}, {Lamer}, {Buchner}, {Schneider},
  {Freund}, {Rau}, {Schwope}, {Nishizawa}, {Klein}, {Arcodia}, {Comparat},
  {Musiimenta}, {Nagao}, {Brunner}, {Malyali}, {Finoguenov}, {Anderson},
  {Shen}, {Ibarra-Mendel}, {Trump}, {Brandt}, {Urry}, {Rivera}, {Krumpe},
  {Urrutia}, {Miyaji}, {Ichikawa}, {Schneider}, {Fresco}, {Wilms}, {Boller},
  {Haase}, {Brownstein}, {Lane}, {Bizyaev}, \&
  {Nitschelm}}]{Salvato2021arXiv210614520S_efedsfollowup}
{Salvato}, M., {Wolf}, J., {Dwelly}, T., {et~al.} 2021, arXiv e-prints,
  arXiv:2106.14520

\bibitem[{{Sanders} {et~al.}(2018){Sanders}, {Fabian}, {Russell}, \&
  {Walker}}]{Sanders2018sptchandra}
{Sanders}, J.~S., {Fabian}, A.~C., {Russell}, H.~R., \& {Walker}, S.~A. 2018,
  \mnras, 474, 1065

\bibitem[{{Schellenberger} \&
  {Reiprich}(2017{\natexlab{a}})}]{Schellenberger2017MNRAS.469.3738S_scalingrel}
{Schellenberger}, G. \& {Reiprich}, T.~H. 2017{\natexlab{a}}, \mnras, 469, 3738

\bibitem[{{Schellenberger} \&
  {Reiprich}(2017{\natexlab{b}})}]{Schellenberger2017reiprich}
{Schellenberger}, G. \& {Reiprich}, T.~H. 2017{\natexlab{b}}, \mnras, 471, 1370

\bibitem[{{Schneider} {et~al.}(2021){Schneider}, {Freund}, {Czesla}, {Robrade},
  {Salvato}, \& {Schmitt}}]{Schneider2021arXivefedsstars}
{Schneider}, P.~C., {Freund}, S., {Czesla}, S., {et~al.} 2021, arXiv e-prints,
  arXiv:2106.14521

\bibitem[{{Sehgal} {et~al.}(2010){Sehgal}, {Bode}, {Das},
  {Hernandez-Monteagudo}, {Huffenberger}, {Lin}, {Ostriker}, \&
  {Trac}}]{Sehgal2010ApJ...709..920S_MWskysim}
{Sehgal}, N., {Bode}, P., {Das}, S., {et~al.} 2010, \apj, 709, 920

\bibitem[{{Seppi} {et~al.}(2021){Seppi}, {Comparat}, {Nandra}, {Bulbul},
  {Prada}, {Klypin}, {Merloni}, {Predehl}, \& {Ider
  Chitham}}]{Seppi2021A&A...652A.155S}
{Seppi}, R., {Comparat}, J., {Nandra}, K., {et~al.} 2021, \aap, 652, A155

\bibitem[{{Sherwin} {et~al.}(2012){Sherwin}, {Das}, {Hajian}, {Addison},
  {Bond}, {Crichton}, {Devlin}, {Dunkley}, {Gralla}, {Halpern}, {Hill},
  {Hincks}, {Hughes}, {Huffenberger}, {Hlozek}, {Kosowsky}, {Louis},
  {Marriage}, {Marsden}, {Menanteau}, {Moodley}, {Niemack}, {Page}, {Reese},
  {Sehgal}, {Sievers}, {Sif{\'o}n}, {Spergel}, {Staggs}, {Switzer}, \&
  {Wollack}}]{Sherwin2012PhRvDlensingcmb}
{Sherwin}, B.~D., {Das}, S., {Hajian}, A., {et~al.} 2012, \prd, 86, 083006

\bibitem[{{Smith} {et~al.}(2001){Smith}, {Brickhouse}, {Liedahl}, \&
  {Raymond}}]{Smith2001apec}
{Smith}, R.~K., {Brickhouse}, N.~S., {Liedahl}, D.~A., \& {Raymond}, J.~C.
  2001, \apjl, 556, L91

\bibitem[{{Springel} {et~al.}(2005){Springel}, {White}, {Jenkins}, {Frenk},
  {Yoshida}, {Gao}, {Navarro}, {Thacker}, {Croton}, {Helly}, {Peacock}, {Cole},
  {Thomas}, {Couchman}, {Evrard}, {Colberg}, \& {Pearce}}]{Springel2005b}
{Springel}, V., {White}, S.~D.~M., {Jenkins}, A., {et~al.} 2005, \nat, 435, 629

\bibitem[{{Staniszewski} {et~al.}(2009){Staniszewski}, {Ade}, {Aird}, {Benson},
  {Bleem}, {Carlstrom}, {Chang}, {Cho}, {Crawford}, {Crites}, {de Haan},
  {Dobbs}, {Halverson}, {Holder}, {Holzapfel}, {Hrubes}, {Joy}, {Keisler},
  {Lanting}, {Lee}, {Leitch}, {Loehr}, {Lueker}, {McMahon}, {Mehl}, {Meyer},
  {Mohr}, {Montroy}, {Ngeow}, {Padin}, {Plagge}, {Pryke}, {Reichardt}, {Ruhl},
  {Schaffer}, {Shaw}, {Shirokoff}, {Spieler}, {Stalder}, {Stark},
  {Vanderlinde}, {Vieira}, {Zahn}, \&
  {Zenteno}}]{Staniszewski2009ApJspt_clusters}
{Staniszewski}, Z., {Ade}, P.~A.~R., {Aird}, K.~A., {et~al.} 2009, \apj, 701,
  32

\bibitem[{{Suchyta} {et~al.}(2016){Suchyta}, {Huff}, {Aleksi{\'c}}, {Melchior},
  {Jouvel}, {MacCrann}, {Ross}, {Crocce}, {Gaztanaga}, {Honscheid}, {Leistedt},
  {Peiris}, {Rykoff}, {Sheldon}, {Abbott}, {Abdalla}, {Allam}, {Banerji},
  {Benoit-L{\'e}vy}, {Bertin}, {Brooks}, {Burke}, {Carnero Rosell}, {Carrasco
  Kind}, {Carretero}, {Cunha}, {D'Andrea}, {da Costa}, {DePoy}, {Desai},
  {Diehl}, {Dietrich}, {Doel}, {Eifler}, {Estrada}, {Evrard}, {Flaugher},
  {Fosalba}, {Frieman}, {Gerdes}, {Gruen}, {Gruendl}, {James}, {Jarvis},
  {Kuehn}, {Kuropatkin}, {Lahav}, {Lima}, {Maia}, {March}, {Marshall},
  {Miller}, {Miquel}, {Neilsen}, {Nichol}, {Nord}, {Ogando}, {Percival},
  {Reil}, {Roodman}, {Sako}, {Sanchez}, {Scarpine}, {Sevilla-Noarbe}, {Smith},
  {Soares-Santos}, {Sobreira}, {Swanson}, {Tarle}, {Thaler}, {Thomas},
  {Vikram}, {Walker}, {Wechsler}, {Zhang}, \& {DES
  Collaboration}}]{Suchyta2016MNRAS_Balrog_DES}
{Suchyta}, E., {Huff}, E.~M., {Aleksi{\'c}}, J., {et~al.} 2016, \mnras, 457,
  786

\bibitem[{{Tinker} {et~al.}(2008){Tinker}, {Kravtsov}, {Klypin}, {Abazajian},
  {Warren}, {Yepes}, {Gottl{\"o}ber}, \& {Holz}}]{Tinker2008}
{Tinker}, J., {Kravtsov}, A.~V., {Klypin}, A., {et~al.} 2008, \apj, 688, 709

\bibitem[{{Trudeau} {et~al.}(2020){Trudeau}, {Garrel}, {Willis}, {Pierre},
  {Gastaldello}, {Chiappetti}, {Ettori}, {Umetsu}, {Adami}, {Adams}, {Bowler},
  {Faccioli}, {H{\"a}u{\ss}ler}, {Jarvis}, {Koulouridis}, {Le Fevre}, {Pacaud},
  {Poggianti}, \& {Sadibekova}}]{Trudeau2020A&A...642A.124T}
{Trudeau}, A., {Garrel}, C., {Willis}, J., {et~al.} 2020, \aap, 642, A124

\bibitem[{{Veropalumbo} {et~al.}(2014){Veropalumbo}, {Marulli}, {Moscardini},
  {Moresco}, \& {Cimatti}}]{Veropalumbo2014MNRAS.442.3275V}
{Veropalumbo}, A., {Marulli}, F., {Moscardini}, L., {Moresco}, M., \&
  {Cimatti}, A. 2014, \mnras, 442, 3275

\bibitem[{{Viitanen} {et~al.}(2019){Viitanen}, {Allevato}, {Finoguenov},
  {Bongiorno}, {Cappelluti}, {Gilli}, {Miyaji}, \&
  {Salvato}}]{Viitanen2019A&A...629A..14V}
{Viitanen}, A., {Allevato}, V., {Finoguenov}, A., {et~al.} 2019, \aap, 629, A14

\bibitem[{{Vikhlinin} {et~al.}(2006){Vikhlinin}, {Kravtsov}, {Forman}, {Jones},
  {Markevitch}, {Murray}, \& {Van Speybroeck}}]{Vikhlinin2006ApJ...640..691V}
{Vikhlinin}, A., {Kravtsov}, A., {Forman}, W., {et~al.} 2006, \apj, 640, 691

\bibitem[{{Vikhlinin} {et~al.}(2009){Vikhlinin}, {Kravtsov}, {Burenin},
  {Ebeling}, {Forman}, {Hornstrup}, {Jones}, {Murray}, {Nagai}, {Quintana}, \&
  {Voevodkin}}]{Vikhlinin2009ApJ...692.1060V}
{Vikhlinin}, A., {Kravtsov}, A.~V., {Burenin}, R.~A., {et~al.} 2009, \apj, 692,
  1060

\bibitem[{{Vikhlinin} {et~al.}(1998){Vikhlinin}, {McNamara}, {Forman}, {Jones},
  {Quintana}, \& {Hornstrup}}]{Vikhlinin1998ApJ...502..558V}
{Vikhlinin}, A., {McNamara}, B.~R., {Forman}, W., {et~al.} 1998, \apj, 502, 558

\bibitem[{Virtanen {et~al.}(2020)Virtanen, Gommers, Oliphant, Haberland, Reddy,
  Cournapeau, Burovski, Peterson, Weckesser, Bright, {van der Walt}, Brett,
  Wilson, Millman, Mayorov, Nelson, Jones, Kern, Larson, Carey, Polat, Feng,
  Moore, {VanderPlas}, Laxalde, Perktold, Cimrman, Henriksen, Quintero, Harris,
  Archibald, Ribeiro, Pedregosa, {van Mulbregt}, \& {SciPy 1.0
  Contributors}}]{Virtanen2020SciPy-NMeth}
Virtanen, P., Gommers, R., Oliphant, T.~E., {et~al.} 2020, Nature Methods, 17,
  261

\bibitem[{{Voges} {et~al.}(1999){Voges}, {Aschenbach}, {Boller},
  {Br{\"a}uninger}, {Briel}, {Burkert}, {Dennerl}, {Englhauser}, {Gruber},
  {Haberl}, {Hartner}, {Hasinger}, {K{\"u}rster}, {Pfeffermann}, {Pietsch},
  {Predehl}, {Rosso}, {Schmitt}, {Tr{\"u}mper}, \&
  {Zimmermann}}]{Voges1999RASS}
{Voges}, W., {Aschenbach}, B., {Boller}, T., {et~al.} 1999, \aap, 349, 389

\bibitem[{{White} \& {Frenk}(1991)}]{White1991ApJ_hiercarchical}
{White}, S. D.~M. \& {Frenk}, C.~S. 1991, \apj, 379, 52

\bibitem[{{Wilms} {et~al.}(2000){Wilms}, {Allen}, \& {McCray}}]{Wilms2000tbabs}
{Wilms}, J., {Allen}, A., \& {McCray}, R. 2000, \apj, 542, 914

\end{thebibliography}

\appendix

\section{Simulated products}
\label{appendix:DR:sim}

\begin{table*}[]
    \centering
    \caption{Description of the columns for the input catalogs of the eRASS1 simulation.}
    \begin{tabular}{c|P{13.2cm}}
        \hline
        \hline
        \rule{0pt}{2.2ex} & \textbf{DESCRIPTION}  \\
         \hline
        \rule{0pt}{2.2ex} \textbf{NAME} & \textbf{Input Catalogs: Clusters, AGN, Stars} \\
        \hline
         \rule{0pt}{2.2ex} SRC\_ID & Source ID, STAR: >= 10$^7$ < 4$\times$10$^8$, CLUSTER: >= 4$\times$10$^8$ and < 10$^9$, AGN: >= 10$^9$  \\ 
         RA & Right Ascension [deg] \\
         DEC & Declination [deg] \\
         FLUX & Input Flux in the 0.5 - 2.0 keV band \\
         tile & Number of the eROSITA tile \\
         healpix & Number of the HEALPix field \\
         ID\_contam & ID of the contaminating source \\
         RA\_eSASS & Right Ascension of the corresponding eSASS detection [deg] \\
         DEC\_eSASS & Declination of the corresponding eSASS detection [deg] \\ Separation & Separation between the source and the corresponding eSASS detection [arcsec] \\
         DET\_LIKE\_0 & Detection likelihood of the corresponding eSASS detection \\
         EXT\_LIKE & extension likelihood of the corresponding eSASS detection \\
         EXT & Extent of the corresponding eSASS detection \\
         ML\_CTS\_0 & Maximum likelihood number of counts from eSASS \\
         ML\_RATE\_0 & Maximum likelihood count rate from eSASS \\
         ML\_FLUX\_0 & Maximum likelihood flux estimate from eSASS \\         
         RADEC\_ERR & Positional error from eSASS \\
         srcRAD & Source radius \\
         detected & Flag to identify simulated source that are detected by eSASS \\
        \hline
        \rule{0pt}{2.2ex} & \textbf{Input Catalogs: in common between Clusters and AGN} \\
        \hline
        \rule{0pt}{2.2ex} dL & Luminosity distance [cm] in the cosmology adopted by \citet{Comparat2020Xray_simulation} \\
        nH & Column density [cm$^{-2}$] \\
        redshift\_R & Redshift in real space \\
        redshift\_S & Redshift in redshift space \\
        FX\_soft\_attenuated & Observed flux in the 0.5--2.0 keV band, corrected by galactic absorption \\
        LX\_soft & Rest-frame X-ray luminosity in the 0.5--2.0 keV band \\
        Bg3Model & Average Value of the eSASS background maps in the nearest 20 pixels [cts/pixel] \\
        TexpModel & Average Value of the eSASS exposure maps in the nearest 20 pixels [sec] \\  
        \hline
        \rule{0pt}{2.2ex} & \textbf{Input Catalogs: only Clusters} \\
        \hline
        \rule{0pt}{2.2ex} g\_lat & Galactic latitude [deg] \\
        g\_lon & Galactic longitude [deg] \\
        HALO\_pid & Halo flag in the dark matter light cone (-1 for distinct halos, >0 for subhalos) \\
        HALO\_Mvir & Halo mass within the virial radius [M$_\odot$] \\
        HALO\_Rvir & Halo virial radius [kpc] \\
        HALO\_rs & Halo scale radius [kpc] \\        
        HALO\_M200c & Halo mass within the R$_{\rm 200c}$ [M$_\odot$] \\
        HALO\_500c & Halo mass within the R$_{\rm 500c}$ [M$_\odot$] \\    
        HALO\_Xoff & Halo offset parameter [kpc] \\
        HALO\_b\_to\_a\_500c & Halo ellipticity \\
        kT & Temperature of the cluster [keV] \\
        R500c\_kpc & Halo R$_{\rm 500c}$ [kpc] \\
        R500c\_arcmin & Halo R$_{\rm 500c}$ [arcmin] \\
        EM0 & Central Emissivity \\
        LX\_soft\_obs & Observer-frame X-ray luminosity in the 0.5--2.0 keV band \\
        COUNTS\_02\_23\_CLU\_CLU & Counts by clusters in [0.1, 0.2, 0.3, 0.4, 0.5, 0.6, 0.7, 0.8, 0.9, 1.0, 1.5, 2.0]$\times$R500c\_arcmin \\
        COUNTS\_02\_23\_CLU\_AGN & Counts by AGN in [0.1, 0.2, 0.3, 0.4, 0.5, 0.6, 0.7, 0.8, 0.9, 1.0, 1.5, 2.0]$\times$R500c\_arcmin \\
        COUNTS\_02\_23\_CLU\_STA & Counts by stars in [0.1, 0.2, 0.3, 0.4, 0.5, 0.6, 0.7, 0.8, 0.9, 1.0, 1.5, 2.0]$\times$R500c\_arcmin \\
        COUNTS\_02\_23\_CLU\_BKG & Counts by the background in [0.1, 0.2, 0.3, 0.4, 0.5, 0.6, 0.7, 0.8, 0.9, 1.0, 1.5, 2.0]$\times$R500c\_arcmin \\  
        ErsenModel & Average Value of the eSASS sensitivity maps for extended sources in the nearest 20 pixels [erg/s/cm$^2$] \\
        \hline
        \rule{0pt}{2.2ex} & \textbf{Input Catalogs: only AGN} \\
        \hline
        \rule{0pt}{2.2ex} LX\_hard & Rest-frame X-ray luminosity in the 2.0-10.0 keV band \\
        galaxy\_SMHMR\_mass & Host galaxy stellar mass \\
        COUNTS\_02\_23\_AGN\_AGN & Counts by AGN in [10, 20, 30, 40, 50, 60] arcsec \\
        COUNTS\_02\_23\_AGN\_CLU & Counts by clusters in [10, 20, 30, 40, 50, 60] arcsec \\
        COUNTS\_02\_23\_AGN\_STA & Counts by stars in [10, 20, 30, 40, 50, 60] arcsec \\
        COUNTS\_02\_23\_AGN\_BKG & Counts by the background in [10, 20, 30, 40, 50, 60] arcsec \\
        \hline
        \hline
    \end{tabular}
    \label{tab:column_descr}
\end{table*}

\begin{table*}[]
    \centering
    \caption{Description of the columns for the output catalogs of the eRASS1 simulation.}
    \begin{tabular}{c|c}
        \hline
        \hline
        \rule{0pt}{2.2ex} & \textbf{DESCRIPTION}  \\
         \hline
        \rule{0pt}{2.2ex} \textbf{NAME} & \textbf{Output Catalogs: Single band and Three band} \\
        \hline
        \rule{0pt}{2.2ex} ID\_cat & ID of the eSASS detection \\
        RA & Right Ascension [deg] \\
        DEC & Declination [deg] \\
        DET\_LIKE\_n & Detection likelihood of the corresponding eSASS detection \\
        EXT\_LIKE & extension likelihood of the corresponding eSASS detection \\
        EXT & Extent of the corresponding eSASS detection \\
        ML\_CTS\_n & Maximum likelihood number of counts from eSASS \\
        ML\_RATE\_n & Maximum likelihood count rate from eSASS \\
        ML\_FLUX\_0 & Maximum likelihood flux estimate from eSASS \\ 
        ID\_Uniq & ID of the unique input simulated counterpart \\
        ID\_Any & ID of the brightest input simulated counterpart, allowing for duplicates \\
        ID\_Any2 & ID of the secondary input simulated counterpart \\
        ID\_contam & ID of the simulated source that contaminates the unique simulated counterpart \\
        \hline
        \hline
    \end{tabular}
    \vskip.1cm
    \footnotesize{\textbf{Notes.} The eSASS properties are measured with photons in the 0.2--2.3 keV band for the single band catalog, and with photons in the 0.2--0.6, 0.6--2.3, 2.3--5.0 keV bands for the three band catalog.}

    \label{tab:column_descr_esass}
\end{table*}

\begin{table*}[]
    \centering
    \caption{Description of the columns for the catalogs of events in the eRASS1 simulation.}
    \begin{tabular}{c|c}
        \hline
        \hline
        \rule{0pt}{2.2ex} & \textbf{DESCRIPTION}  \\
         \hline
        \rule{0pt}{2.2ex} \textbf{NAME} & \textbf{Output Catalogs: Single band and Three band} \\
        \hline
        \rule{0pt}{2.2ex}
        RA & Right Ascension [deg] \\
        DEC & Declination [deg] \\
        SIGNAL & Photon energy [keV] \\
        \hline
        \hline
    \end{tabular}
    \vskip.1cm
    \footnotesize{\textbf{Notes.} We consider four catalogs for events generated by clusters, AGN, stars, and the background. The columns are the same for all four.}

    \label{tab:column_descr_events}
\end{table*}

We provide the input catalogs of simulated clusters, AGN, and stars, the events generated by the simulated sources, and the output source catalog from eSASS. They are available publicly at the CDS and at \footnote{\url{https://firefly.mpe.mpg.de/eROSITA_digitalTwin}}. We also provide the catalog for the three band detection, using the 0.2--0.6, 0.6--2.3, 2.3--5.0 keV bands \citep[see][for more details]{Liu2021teng_simulation, Brunner2022_efedscat}. The association between input and output can be built using the ID\_Uniq and the ID\_Any (see Sect. \ref{subsec:catalog_description}). The IDs of stars are >= 10$^7$ and < 4$\times$10$^8$, the ones for clusters are >= 4$\times$10$^8$ and < 10$^9$, and the ones for AGN are >= 10$^9$. The source IDs are assigned on the HEALPix fields. The IDs of the contaminating sources are saved as ID\_Contam.
See Sect. \ref{subsec:catalog_description} for the definition of the classes and additional details. The description of the columns for each file is given in Tables \ref{tab:column_descr} and \ref{tab:column_descr_esass}.

\section{Extension of the model to galaxy groups}
\label{app:model_extension}

\begin{figure}
    \centering
    \includegraphics[width=0.9\columnwidth]{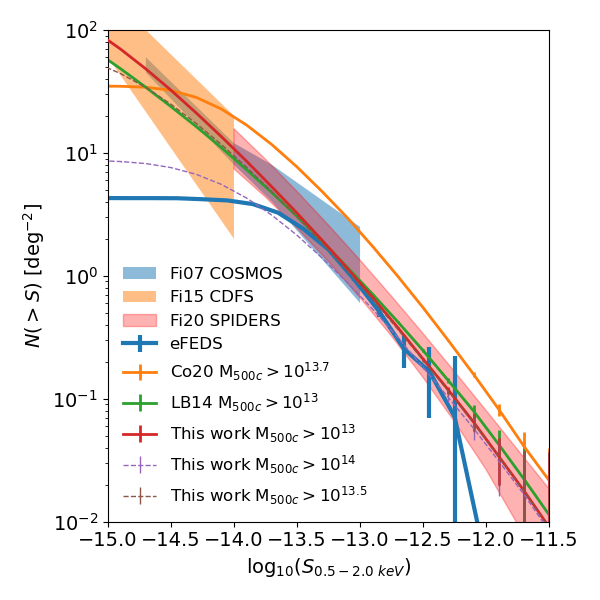}
    \includegraphics[width=0.9\columnwidth]{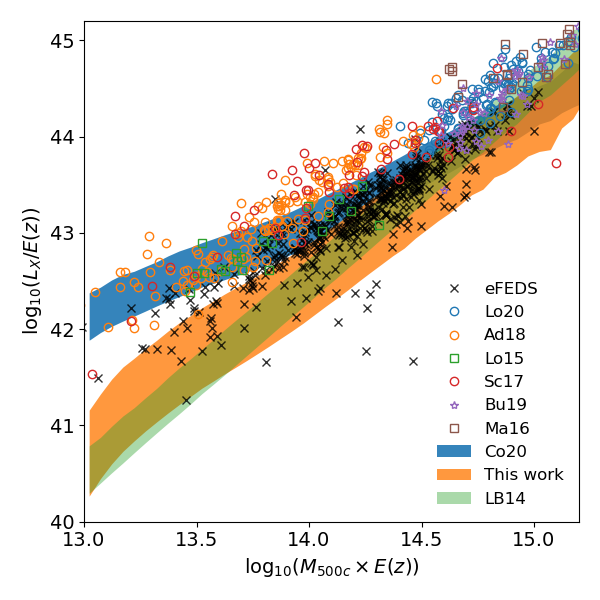}
	\caption{Improved cluster model. \textbf{Top panel}: Number density of sources as function of flux. The solid orange (red) line shows the prediction of the model before (after) applying the correction. The shaded areas in blue, orange, and red denote the logN--logS from \citet{Finoguenov2007ApJS..172..182F_cosmos, Finoguenov2015A&A_CDFS, Finoguenov2020A&A...638A.114F}. The green and blue lines show a comparison to \citet{Lebrun_2014MNRAS.441.1270L_agnfeedback} and \citet{2022A&A_LiuAng_eFEDS_clu}. The dashed pink and brown lines denote the model corrected for higher mass thresholds. \textbf{Bottom panel}: Relation between X-ray luminosity and mass. The blue (orange) shaded area shows the prediction of the model before (after) applying the correction. The green shaded area denotes the relation from \citet{Lebrun_2014MNRAS.441.1270L_agnfeedback}. Additional samples are shown by blue circles \citep[][]{Lovisari2020ApJ...892..102L_scalingrel}, orange circles \citep[][]{Adami2018A&A...620A...5A}, green squares \citep[][]{Lovisari2015A&A...573A.118L_scalingrel}, red circles \citep[][]{Schellenberger2017MNRAS.469.3738S_scalingrel}, pink stars \citep[][]{Bulbul2019ApJ...871...50B_scalingrel}, and brown squares \citep[][]{Mantz2016_scaling_relation}. 
	}
    \label{fig:logNlogS_model}
\end{figure}

\begin{figure}
    \centering
    \includegraphics[width=1.0\columnwidth]{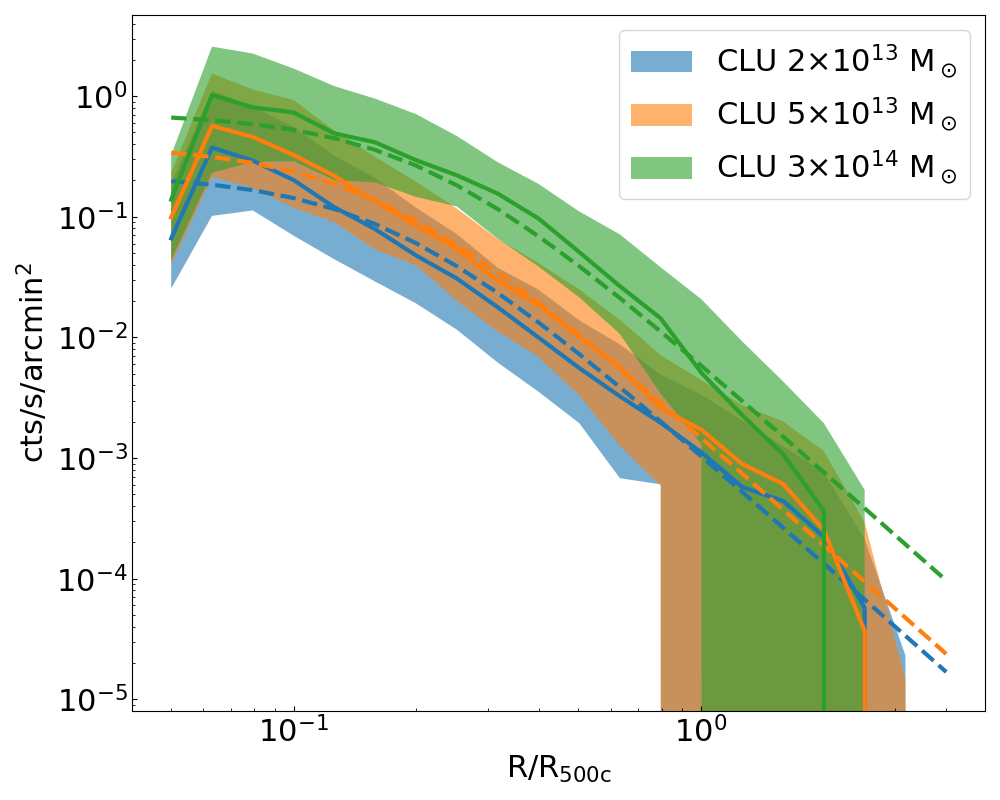}
    \caption{Surface brightness profiles of the simulated clusters. The radius is normalized to R$_{\rm 500c}$. The solid lines show the average profile, the shaded areas denote the 1$\sigma$ scatter around the mean. The dashed lines show the best-fitting beta model for each average profile.}
    \label{fig:SBprofiles}
\end{figure}

In this section, we provide further details about the extension of our improved cluster model to lower masses (see Sect. \ref{sec:simulation}), comparing it to the eFEDS cluster sample. Along with the \citep[][AN15]{Anderson2015Lx_stellarmass} correction using stellar mass, we also tested an improvement exploiting the X-ray luminosity - halo mass relation, following \citet{Lebrun_2014MNRAS.441.1270L_agnfeedback} (LB14). Such correction reads:
\begin{equation}
    \log_{\rm 10}L_{\rm x,(0.5-2.0 \text{keV})} = 2\log_{\rm 10}M_{\rm 500c} + 14.5.
    \label{eq:LB14}
\end{equation}
This correction gives a shallower slope in the cluster logN--logS (see Fig. \ref{fig:CLU_logNlogS}) than AN15. 
AN15 provides a better agreement to observations than LB14, especially at low luminosities < 1$\times$10$^{43}$ erg/s. LB14 underestimates observed values by a factor of $\sim$ 2 at 1$\times$10$^{42}$ erg/s.
AN15 provides a great correction for the X-ray luminosity to stellar mass relation by construction, while LB14 does not align well with observations.
The same holds for the X-ray luminosity to temperature relation. The AN15 version gives excellent agreement to eFEDS data for low luminosity clusters. The L$_{\rm x}$--T$_{\rm x}$ relation obtained from LB14 is too steep.
We ultimately choose the AN15 correction over LB14, as it produces a logN--logS and scaling relations that align better with observations (see Fig. \ref{fig:logNlogS_model}). 
The prediction of the L$_{\rm X}$--M$_{\rm 500c}$ relation is slightly underestimated at the high mass end compared to data, see Fig. \ref{fig:logNlogS_model}. This makes our approach conservative, since the most massive and luminous objects are detected more easily, see also Fig. \ref{fig:completeness}. On the other hand, the fact that observations suffer from the Malmquist bias at the low mass end possibly affects our correction using AN15. Nonetheless, the addition of the eFEDS cluster sample shows the ability of the model to reproduce observations also in the regime of galaxy groups. \\
We verify that the shape of the cluster profiles generated with the new model is on average compatible with a beta model. We measure the radial profile of events generated by three samples of 100 simulated clusters with masses of 2$\times$10$^{13}$, 5$\times$10$^{13}$, and 3$\times$10$^{14}$ M$_\odot$ as a function of R$_{\rm 500c}$. We fit each one of them with a beta model (see Sect. \ref{sec:data}). The result is shown in Fig. \ref{fig:SBprofiles}. The solid lines show the average surface brightness profile for each one of the three samples, the shaded areas denote the 1$\sigma$ scatter around the mean value. The dashed lines denote the best-fitting beta model to each average profile. We fix $\beta$=2/3, leaving the core radius as a free parameter. This is the same assumption taken by the \texttt{ermldet} task (see Sect. \ref{sec:data}). The agreement between the average profile and the beta models is good. Even if the profile of a single object can significantly deviate from a beta model, our model generates profiles that are on average compatible with the assumptions taken by eSASS in the source detection chain.

\section{Comparison to data}
\label{appendix:comp2data}
\begin{figure}
    \centering
    \includegraphics[width=\columnwidth]{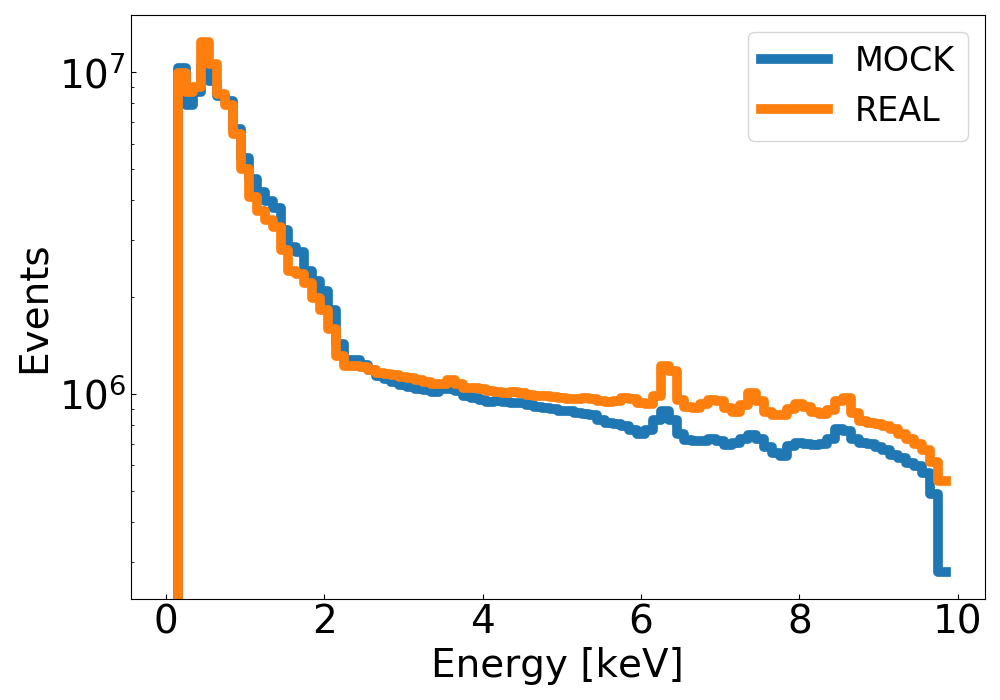}    
    \includegraphics[width=\columnwidth]{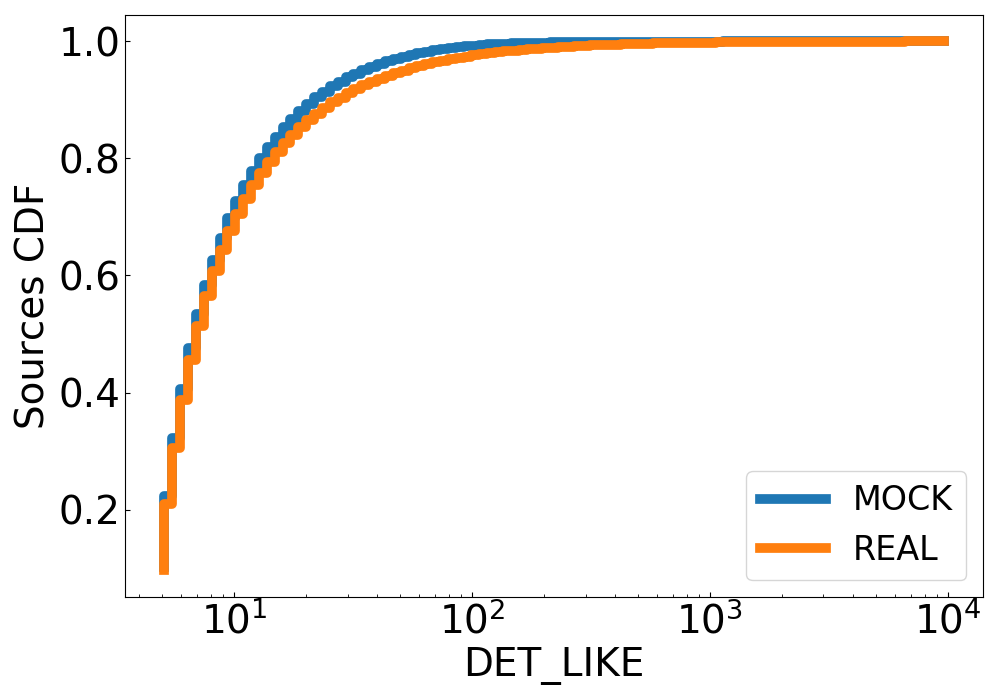}
    \includegraphics[width=\columnwidth]{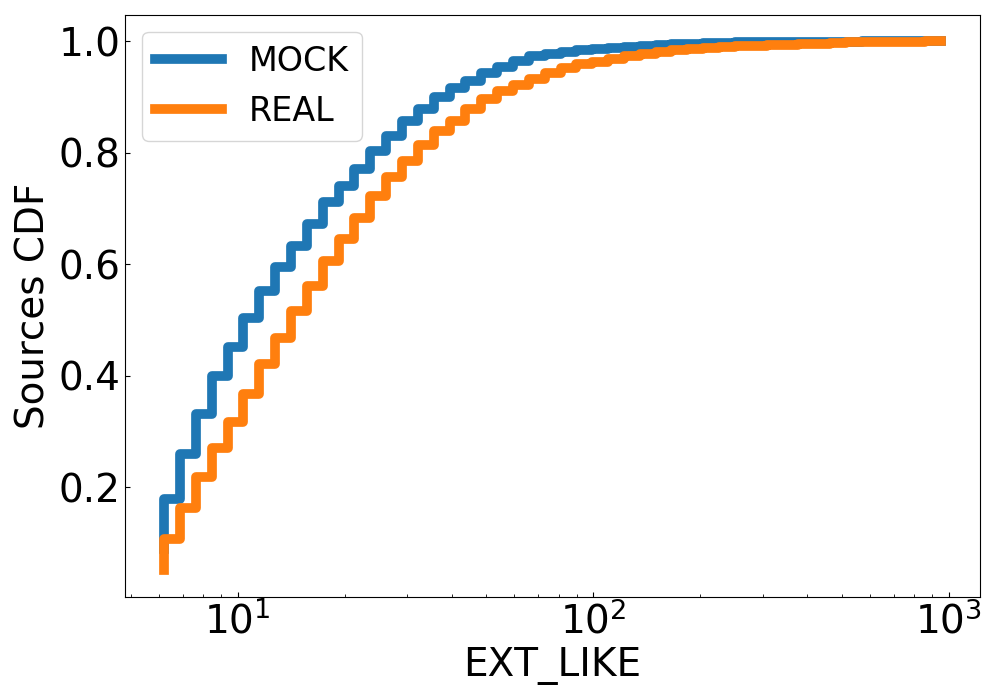}
	\caption{Comparison between the eRASS1 simulation and the real data. These are respectively denoted by the blue and the orange solid lines. \textbf{Top panel}: distribution of the photon energy.  \textbf{Central panel}: cumulative distribution of the sources as a function of detection likelihood. 
	\textbf{Bottom panel}: cumulative distribution of the sources as a function of extension likelihood.}
    \label{fig:comparison_to_eRASS1_data}
\end{figure}

\begin{figure}
    \centering
    \includegraphics[width=\columnwidth]{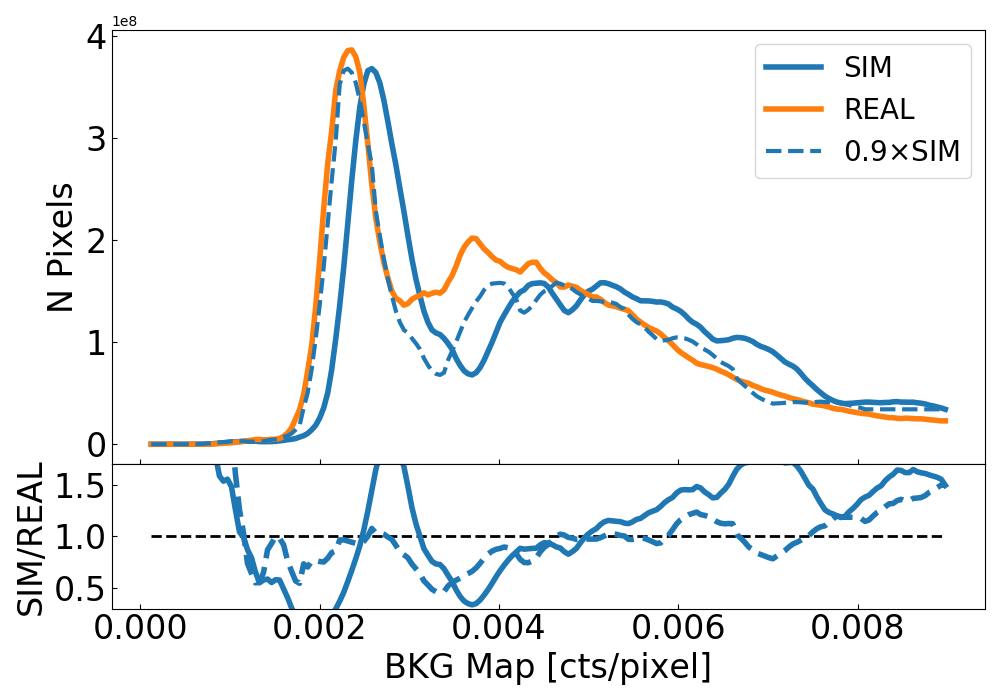}    \includegraphics[width=0.49\columnwidth]{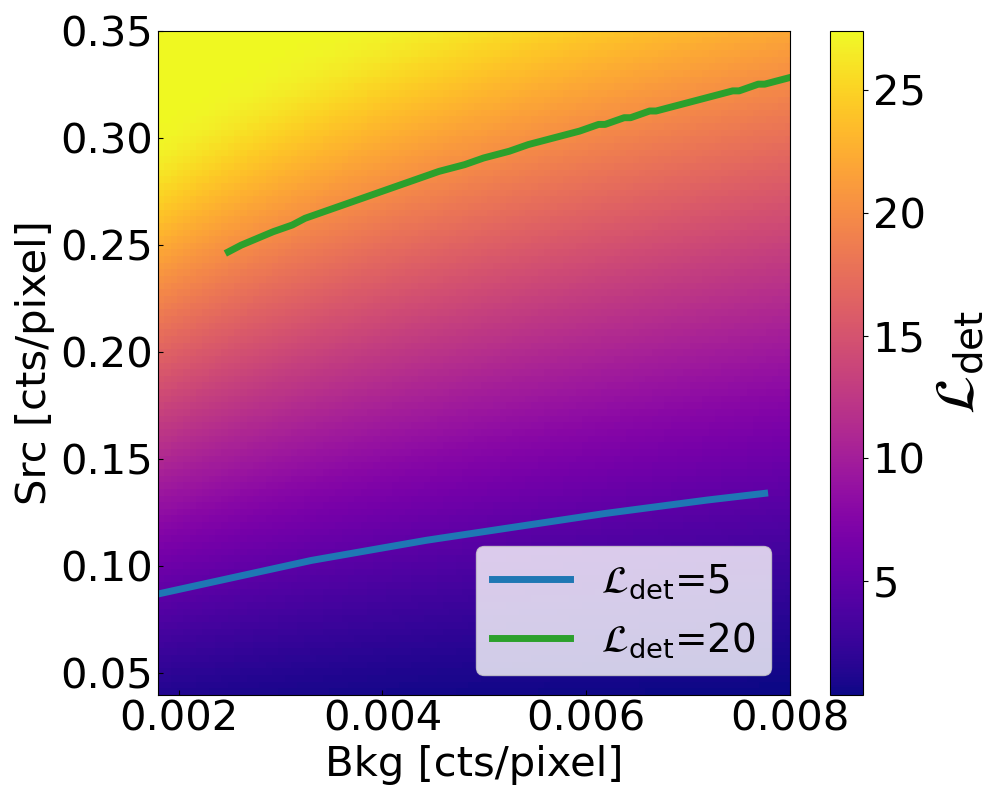}    \includegraphics[width=0.49\columnwidth]{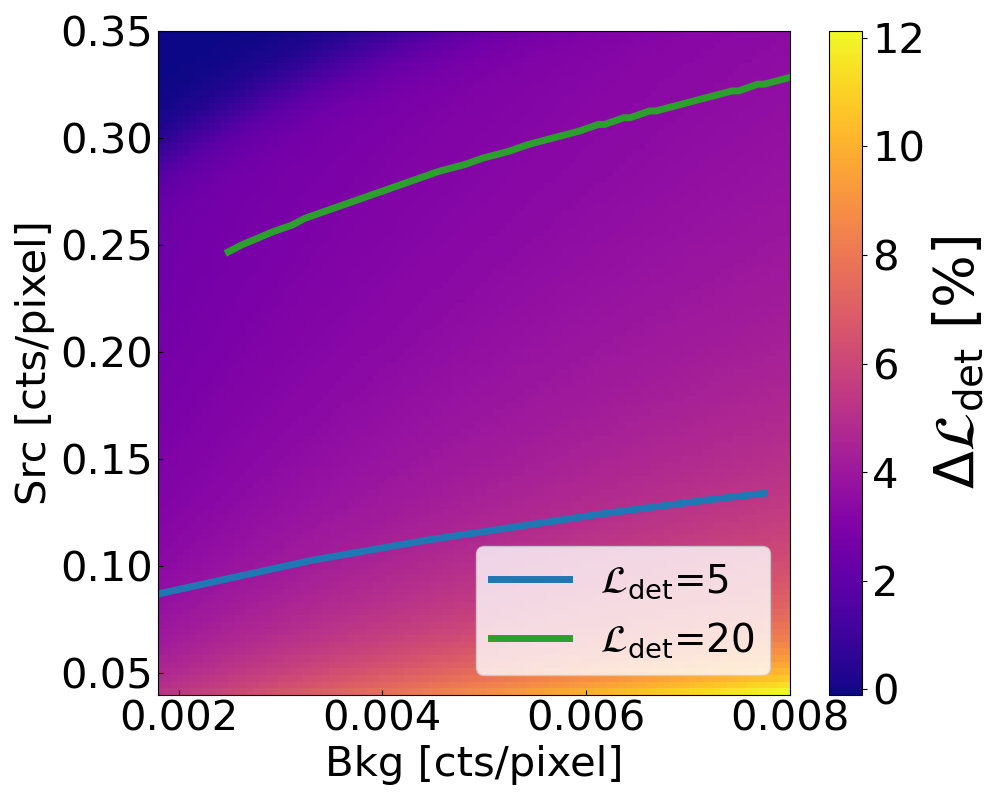}
    \caption{Background evaluation in the eRASS1 simulation. \textbf{Top panel}: Comparison between the mock and real background maps. The lines identify the number of pixels showing a given value of the background map. The mock data is denoted in blue, the real eRASS1 in orange. The dashed blue line shows the simulated background re-scaled by 0.9. The lower panel shows the ratio between the mock and real data. \textbf{Bottom panels}: Impact of a 10$\%$ overestimation of the background on the analytically computed value of detection likelihood. The left-hand panel shows DET\_LIKE as a function of counts in each pixel given by a source and by the background. The panel on the right shows the corresponding percentage error on detection likelihood caused by a 10$\%$ larger background. The blue and the green solid lines respectively denote DET\_LIKE = 5 and 20. }
    \label{fig:background}
\end{figure}


\begin{figure}
    \centering
    \includegraphics[width=\columnwidth]{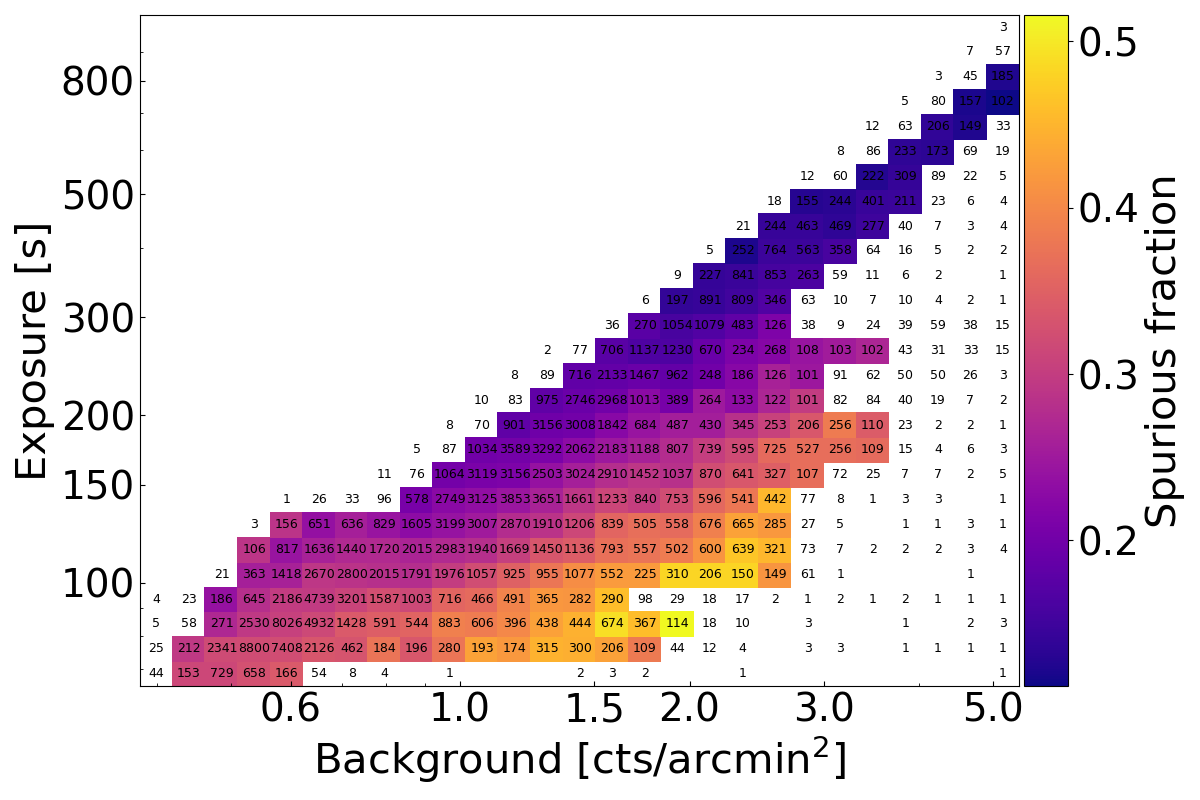}
    \includegraphics[width=\columnwidth]{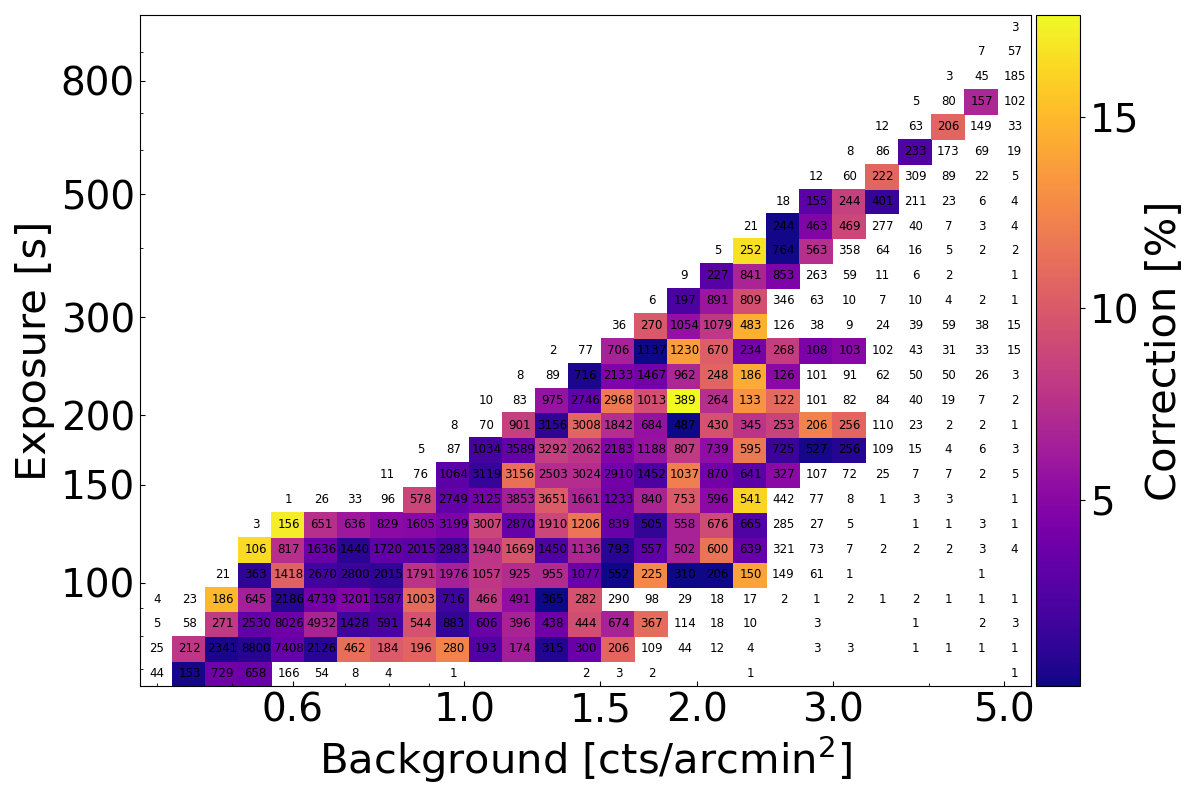}    
    \caption{Evolution of the false detection rate in the eRASS1 simulation. \textbf{Top panel}: Spurious fraction as a function exposure time and background level in the eRASS1 simulation. Each bin containing more than 100 sources is color-coded by the false detection rate. \textbf{Lower panel}: Correction of the prediction of the spurious fraction for the eRASS1 data using the simulation due to the 10$\%$ overestimate of the background.
    The x-axis is binned with a progressive 10$\%$ increment. The total number of spurious sources in each bin is written as text.}
    \label{fig:spurfrac_bkg_exp}
\end{figure}

We compare the source catalog of the eRASS1 simulation to the one obtained by processing the real data with the same eSASS set-up, described in Sect. \ref{sec:data}. There is good agreement between the mock and the real data. This is shown in Fig. \ref{fig:comparison_to_eRASS1_data}. The mock is denoted by the blue solid line and the real data by the orange one. The distributions of the photon energy shown in the top panel are in excellent agreement, especially for the soft energy range in our interest. The central and bottom panels show the cumulative distributions of detection and extension likelihood. A small difference between the two is expected, caused by the fewer number of bright simulated AGN due to the steep logN--logS at high flux (see Fig. \ref{fig:AGN_logNlogS}). This also contributes to the difference between the photon energy distributions at the hard end, above 5 keV. Nonetheless, the cumulative distributions show that the mock catalog and the real one have similar properties.

In addition, we compare the background maps measured on the eRASS1 simulation to the ones obtained from real data (see Sect. \ref{sec:data}). Figure \ref{fig:background} shows the total number of pixels with a given value of the background map, expressed in counts per pixel. The real data is identified by the orange line and the simulation by the solid blue one. On the one hand, the peaks of these two curves differ by about 10$\%$. In fact, a re-scaling of the simulated background by a factor of 0.9 (denoted by the dashed blue line) aligns well with the real eRASS1 maps. This is expected because the cosmic X-ray background component is slightly over-estimated in the simulation. The mock data contains the population of faint simulated AGN. However, this contribution is partially present also in the real eRASS1 maps that are used to create the background model. On the other hand, in some areas, the real background is higher than the mock data. This is because the model has been generated using a mean spectrum but in the eRASS1 data some local instabilities cause such higher background level. \\
In Sect. \ref{sec:data} we verified that such overestimation of the background has a negligible impact on the measured values of detection likelihood. This is reported in the bottom panels of Fig. \ref{fig:background}. We show the value of detection likelihood as a function of source and background counts per pixel (on the left), and the corresponding relative error due to the background over-estimate (on the right). The relative error is computed as 
\begin{equation}
    \Delta \mathcal{L}_{\rm det} = \frac{\mathcal{L}_{\rm det,UN} - \mathcal{L}_{\rm det,B} }{\mathcal{L}_{\rm det,UN}}, 
    \label{eq:delta_detlike}
\end{equation}
where $\mathcal{L}_{\rm det,UN}$ is the unbiased value of detection likelihood, and $\mathcal{L}_{\rm det,B}$ is the value of detection likelihood biased by a 10$\%$ overestimation of the background. The solid lines in blue and green denote values of $\mathcal{L}_{\rm det}$ = 5 and 20, respectively.
There is a $\sim$4$\%$ impact on the value of detection likelihood for faint sources with DET\_LIKE$\sim$5. \\
Finally, we quantify whether the 10$\%$ overestimate of the background significantly impacts our prediction of the false detection rate for the eRASS1 data using the digital twin. For this goal, we measure the spurious fraction on a two-dimensional grid of exposure time and background level. At fixed exposure, we build a binning scheme for the background level such that successive bins are 10$\%$ greater than the previous one, according to
X$_{\rm i+1}$ = 1.1 $\times$ X$_{\rm i}$, where X represents the background level bins. The upper panel of Fig. \ref{fig:spurfrac_bkg_exp} shows the spurious fraction in the exposure-background level plane. The grid contains 95$\%$ of the real eRASS1 catalog. At fixed background level, the spurious fraction decreases as a function of exposure time. Indeed the deeper data allows suppressing fluctuations of the background. At fixed exposure time, the false detection rate increases as a function of the background level, because the probability of picking up a random fluctuation is larger. This makes our prediction of the false detection rate conservative, because at fixed exposure, the real eRASS1 has a lower background compared to the simulation. \\
Given our choice of the binning scheme, we can compare successive background level bins at fixed exposure time to estimate a correction for the prediction of the spurious fraction in the eRASS1 data using the simulation. We compute the relative difference between the bins (f$_{\rm spur,i+1}$-f$_{\rm spur,i}$)/f$_{spur, i+1}$, where the index i runs on the background level bins for each exposure. The correction for each bin is shown the lower panel of Fig. \ref{fig:spurfrac_bkg_exp}. We average over the bins containing more than 100 spurious sources, in order not to be affected by noise. We find a mean correction of 5.7$\%$. We conclude that our measure of the spurious fraction in the digital twin is a conservative prediction of the false detection rate in the real data, and it is not significantly affected by the 10$\%$ overestimate of the background.

\section{Population histograms}
\label{appendix:histograms}
\begin{figure*}[h!]
    \centering
    \includegraphics[width=\columnwidth]{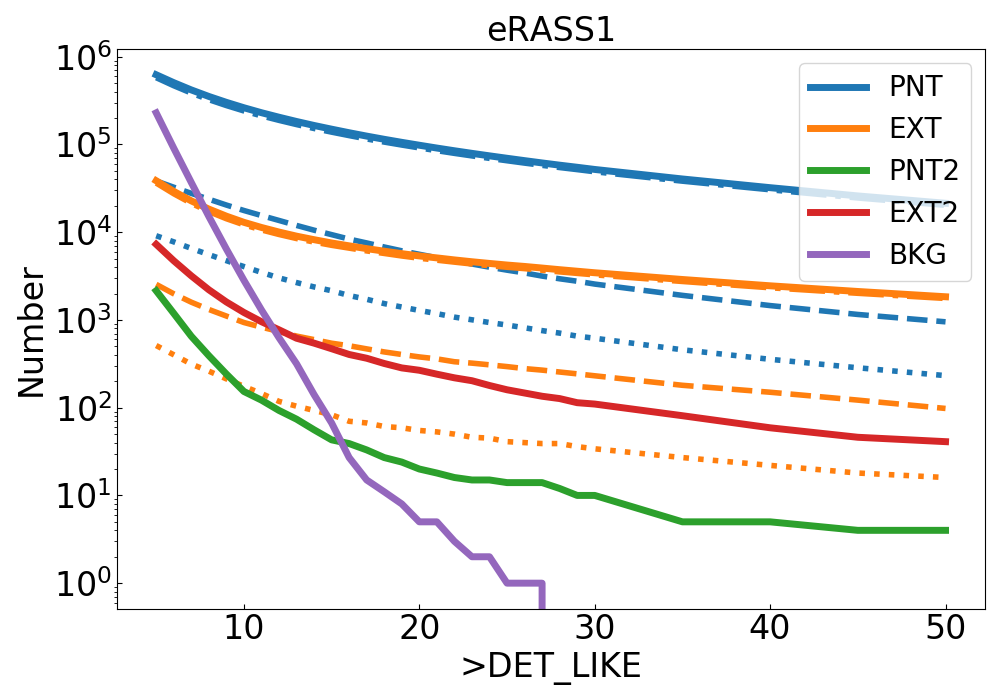}
    \includegraphics[width=\columnwidth]{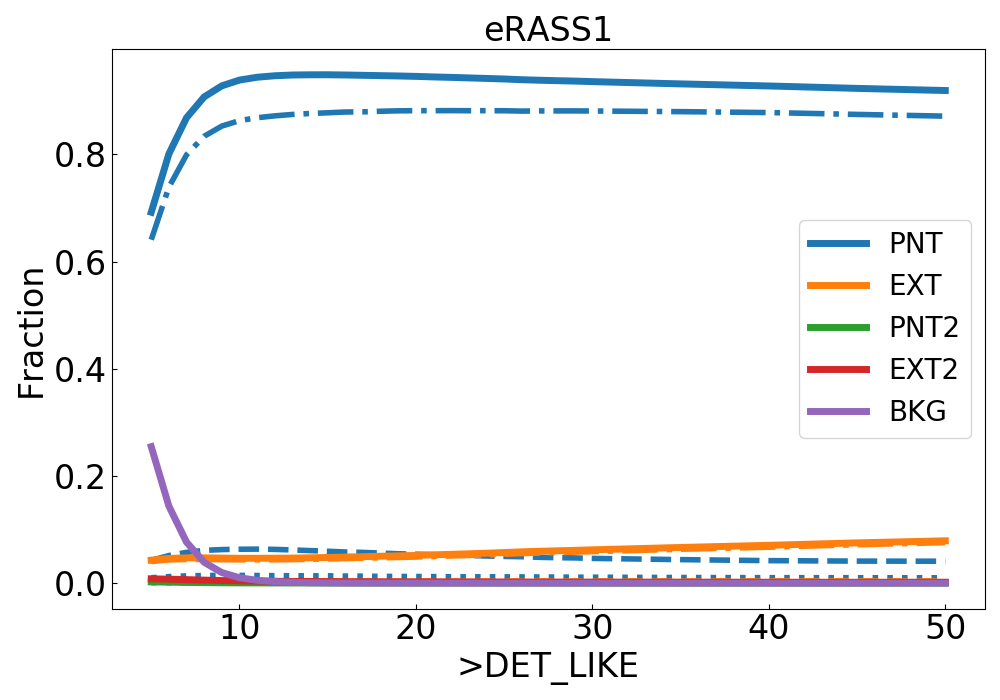}

    \includegraphics[width=\columnwidth]{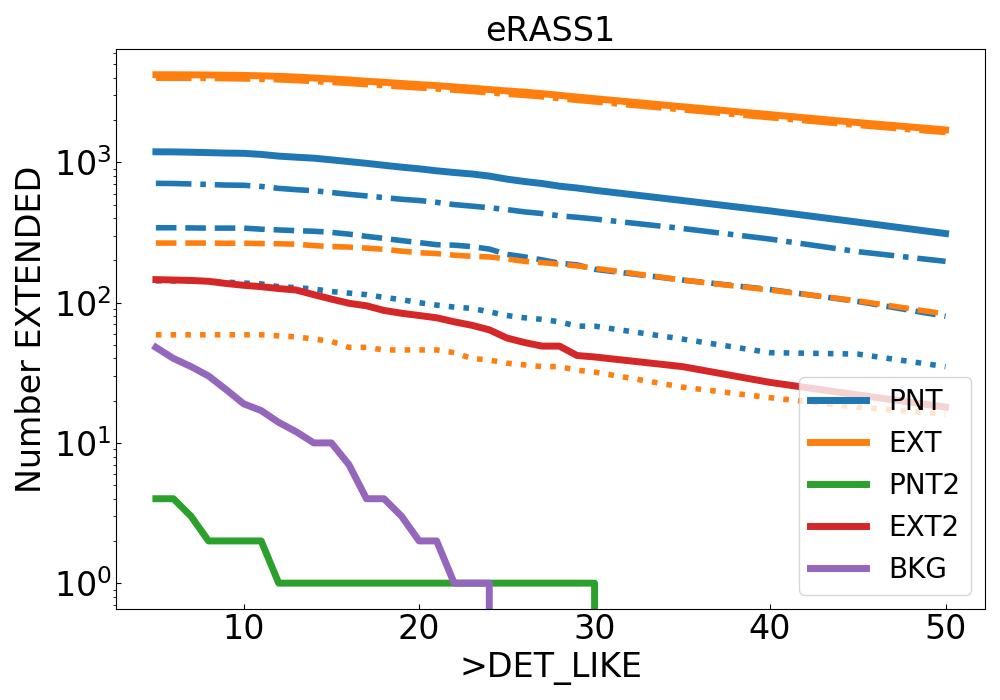}
    \includegraphics[width=\columnwidth]{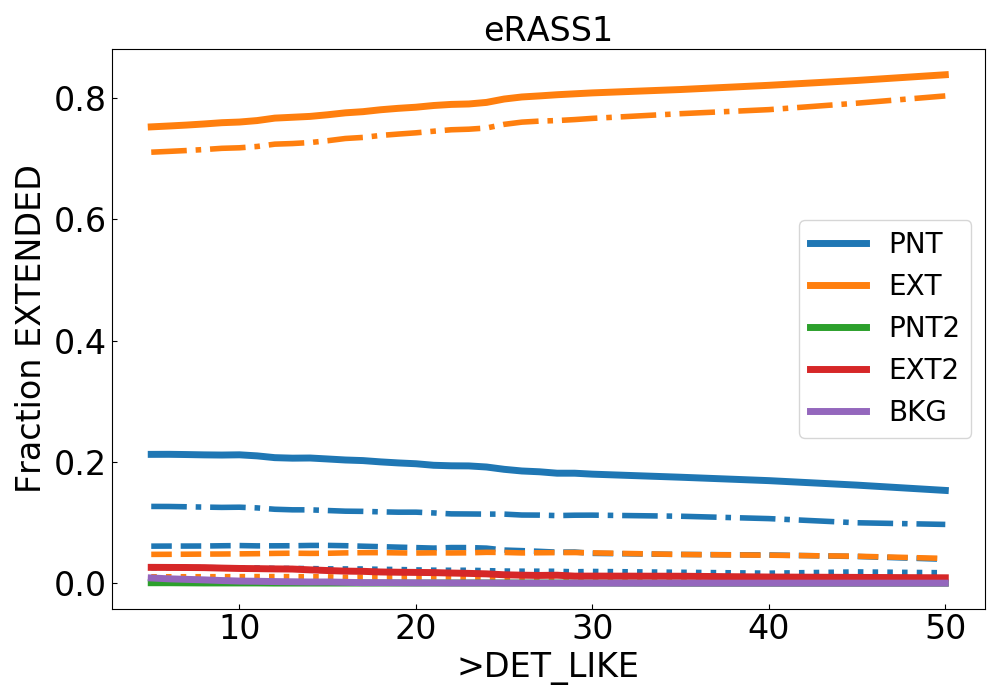}

    \includegraphics[width=\columnwidth]{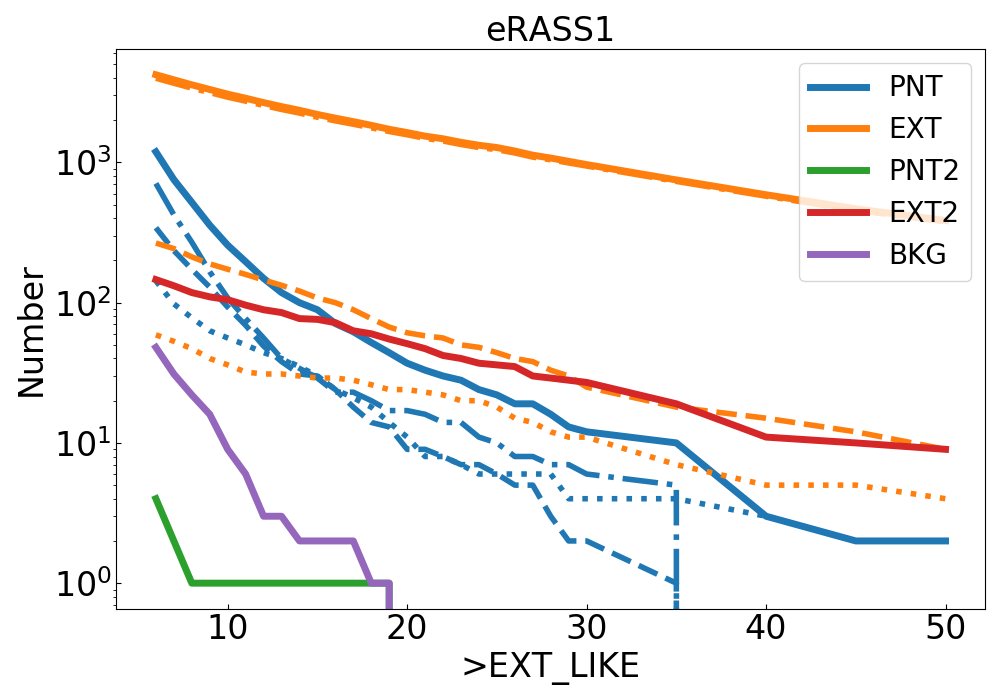}
    \includegraphics[width=\columnwidth]{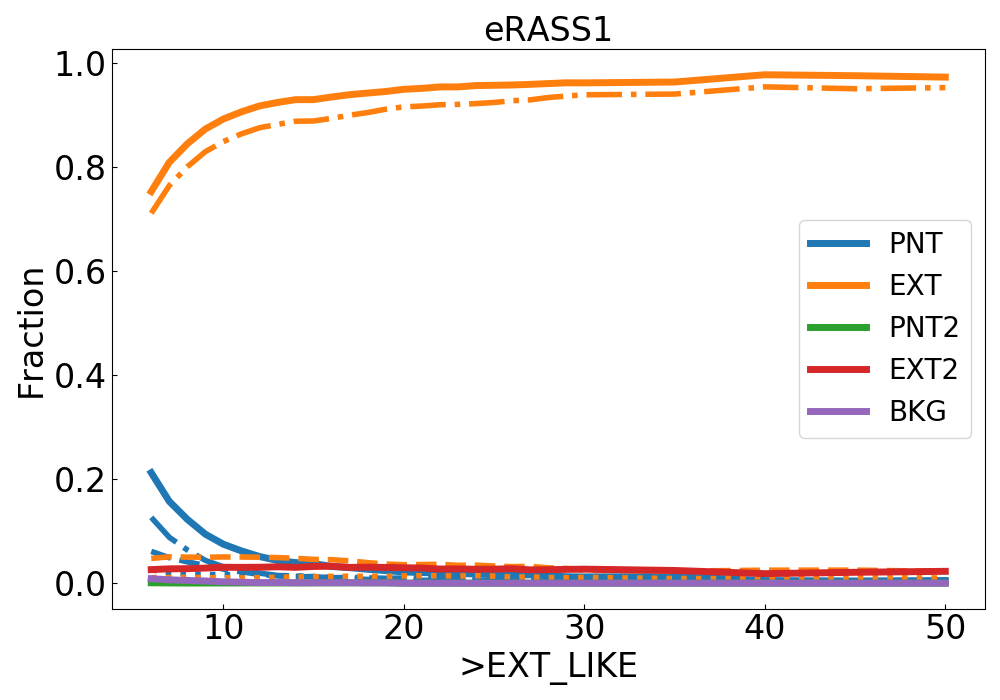}

	\caption{Population in the detected source catalog. The total number of sources in the cleaned catalog of the eRASS1 simulation is 901\,812. The number of extended sources is 5615.
	\textbf{Top panels}: Number of sources in the full catalog and fractions of the population classes in linear scale as a function of minimum detection likelihood.
	\textbf{Central panels}: Number of sources and fractions of the population classes in linear scale in the extent-selected sample (EXT\_LIKE >= 6) as a function of minimum detection likelihood. 
	\textbf{Bottom panels}: population in the source catalog and fractions of the population classes in linear scale as a function of minimum extension likelihood. Lines of different colors show the classes defined in Sect. \ref{sec:data}. The dashed-dotted lines denote sources that are not contaminated by photons of a secondary source (no blending), the dashed ones identify sources contaminated by a point source, and the dotted ones show sources blended with a cluster.
    }
    \label{fig:population_detlike_total}
\end{figure*}

In this appendix, we collect panels showing the histograms and linear fractions relative to the population in the source catalog, described in Sect. \ref{sect:population_source_cat} and Fig. \ref{fig:population_detlike}. These are shown in Fig. \ref{fig:population_detlike_total}. The panels on the left show the total number of sources for different cuts in detection or extension likelihood. The panels on the right show the relative fraction for each source class.

\section{AGN}
\label{appendix:agn}

\begin{figure}
    \centering
    \includegraphics[width=\columnwidth]{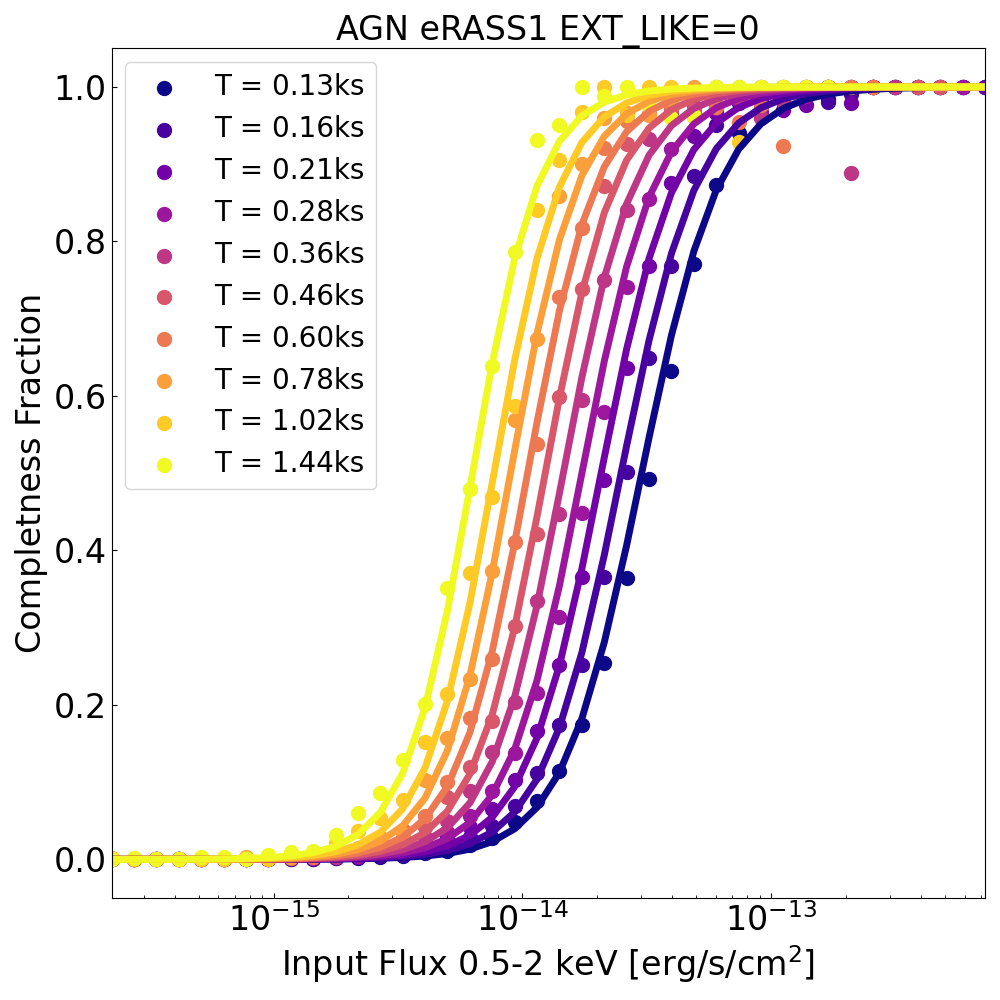}
    \includegraphics[width=\columnwidth]{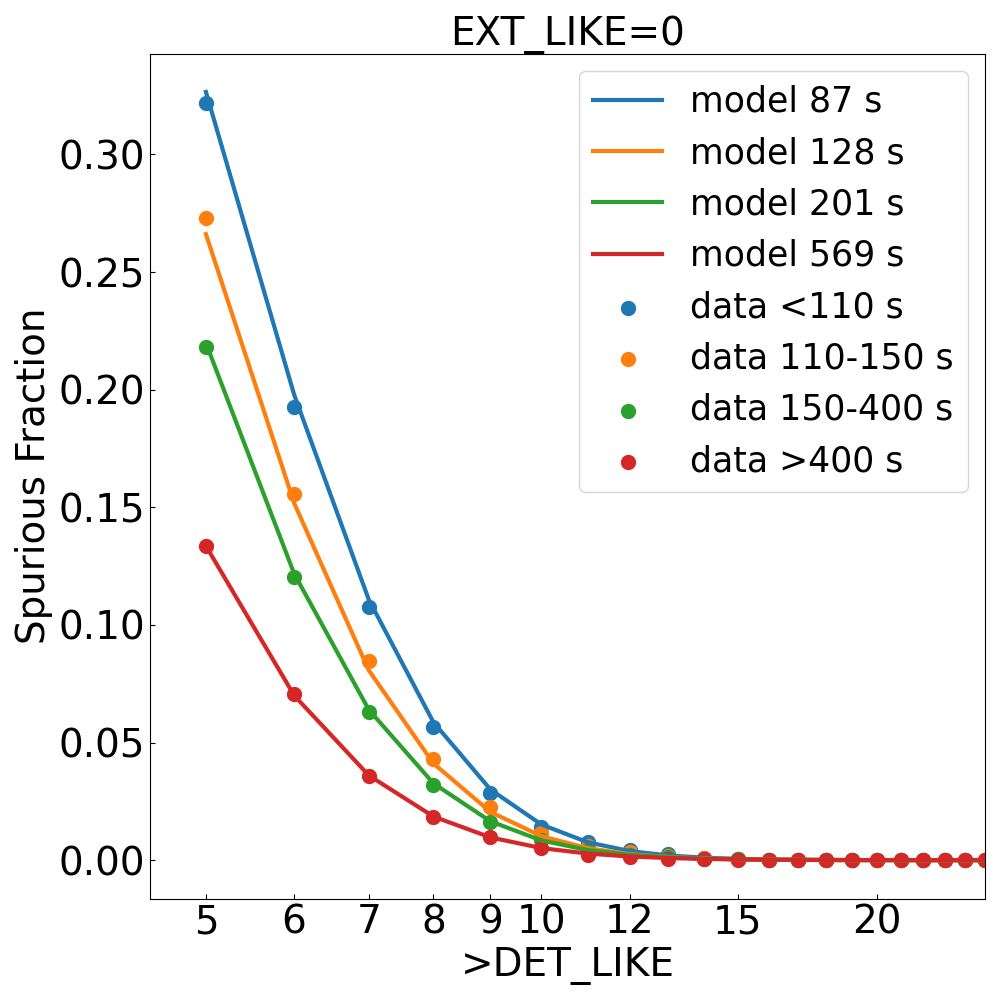}
	\caption{The point source sample. \textbf{Top panel}: fraction of AGN detected as point-like objects as a function of the input flux in the soft X-ray band for different exposure times. The circles show the values measured comparing input and source catalogs, the solid lines our best fit model in Eq.\ref{eq:modified_sigmoid_agn}. \textbf{Bottom panel}: fraction of spurious sources in the point source sample as a function of detection likelihood cuts for different exposure times. The full circles denote the false detection rate measured in the simulation, the solid lines identify the model described by Equation \ref{eq:spur_detlike_texp} computed at the average exposure time corresponding to each bin.}
    \label{fig:completeness_agn}
\end{figure}

\begin{figure}
    \centering
    \includegraphics[width=\columnwidth]{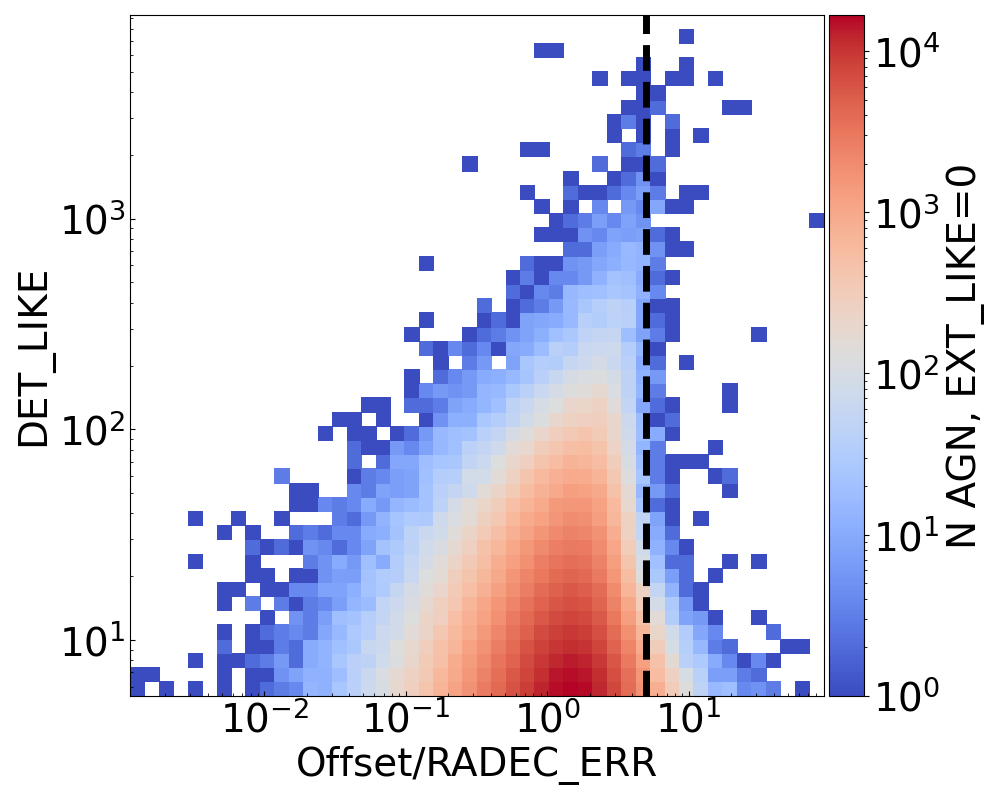}
	\caption{Positional accuracy of the AGN detected as point-like (EXT\_LIKE = 0). This figure shows a 2D histogram in the Offset/RADEC\_ERR -- DET\_LIKE parameter space, and the black dashed line denotes a cut at Offset/RADEC\_ERR = 5. The bins are color-coded according to the number of detected AGN in each bin.}
    \label{fig:radecerr_agn}
\end{figure}

We provide analytical fits to the completeness fraction of the sample of simulated AGN that are detected as point sources. Similarly to Sect. \ref{sec:results}, we measure the detected fraction in terms of input flux and exposure time. We model these trends according to a modified sigmoid function. Our model reads:

\begin{align}
    &b = q (\log_{\rm10}T)^{\rm w},  \nonumber\\
    &c = q1 (\log_{\rm10}T)^{\rm w1},  \nonumber\\
    &C(F,T) = \frac{1}{1+10^{\rm -3}{\rm e}^{-b\log_{\rm10}F + c}},
    \label{eq:modified_sigmoid_agn}
\end{align}
where q = 4.59, w = 0.41, q1 = 11.01, w1 = 0.16 for eRASS1. 
We measure the exposure time T in seconds and the flux F in erg/s/cm$^2$. 
We show the result in Fig. \ref{eq:modified_sigmoid_agn}. The values extracted by matching the source catalog and the simulated AGN are identified by circles, the best-fit model is shown by the solid lines, color coded by exposure time. This model is not intended to provide a complete description of the AGN selection function, but it gives a useful benchmark. We notice that it works particularly well for exposure times between 160 s and 600 s for eRASS1, containing most of the eROSITA coverage in terms of observing time. \\
In addition, we provide a functional form to describe the fraction of false detections for different cuts of detection likelihood and exposure time in the point-source sample. This is described by the following equation:
\begin{align}
    A &= a_1\times T^2 + b_1\times T + c_1, \nonumber \\
    B &= a_2\times T^2 + b_2\times T + c_2, \nonumber \\
    \text{BKG(A,B)} &= \dfrac{0.85}{(A \times \text{DET\_LIKE}^B + 1)^{4.2}}, 
    \label{eq:spur_detlike_texp}
\end{align}
where T is the exposure time in seconds, DET\_LIKE is a cut in detection likelihood, and the values of the parameters are reported in Table \ref{tab:spur_point_pars}. Such a model grasps the details of this trend. It is shown in the bottom panel of Fig. \ref{fig:completeness_agn}. The dots denote the false detection rate measured in the simulation in each exposure time interval as a function DET\_LIKE threshold, while the solid lines denote the model computed at the average exposure time corresponding to each interval.
\begin{table*}[]
    \centering
    \caption{Parameters describing the spurious fraction in the point source sample as a function of detection likelihood thresholds and exposure time.}    
    \begin{tabular}{|c|c|c|c|c|c|}
    \hline
    \hline
    \rule{0pt}{2.3ex} $a_1$ & $b_1$ & $c_1$ & $a_2$ & $b_2$ & $c_2$  \rule[-1.1ex]{0pt}{0pt} \\
    \hline
    \rule{0pt}{2.3ex} $-$1.9059$\times$10$^{-8}$ & 4.4167$\times$10$^{-5}$ & $-$1.2476$\times$10$^{-4}$ & 2.9317$\times$10$^{-6}$ & $-$3.070$\times$10$^{-3}$ & 2.8982 \\
    \hline    
    \end{tabular}
    
    \footnotesize{\textbf{Notes.} The model is described by Equation \ref{eq:spur_detlike_texp}.}
    \label{tab:spur_point_pars}
\end{table*}

Finally, we study the accuracy of the position of AGN detected as point sources (EXT\_LIKE=0). We study the offset between the simulated and detected positions and how it relates to the positional error computed by eSASS. Such error is the sum in quadrature of the error on the pixel position multiplied by the pixel scale and is named RADEC\_ERR. We find that 99.48$\%$ (99.75$\%$) of these point sources are contained by a ratio between the offset and RADEC\_ERR lower than 5 (6). This is especially true for secure detections with DET\_LIKE > 10, whereas sources with smaller values of detection likelihood show larger positional errors and populate the bottom right corner of Fig. \ref{fig:radecerr_agn}, which displays how AGN detected as point-like occupy the DET\_LIKE -- Offset/RADEC\_ERR parameter space. The figure is color-coded according to the number of sources in each bin. 

\section{Cluster characterization}
\label{appendix:clu_srcRAD}
\begin{figure}
    \centering
    \includegraphics[width=\columnwidth]{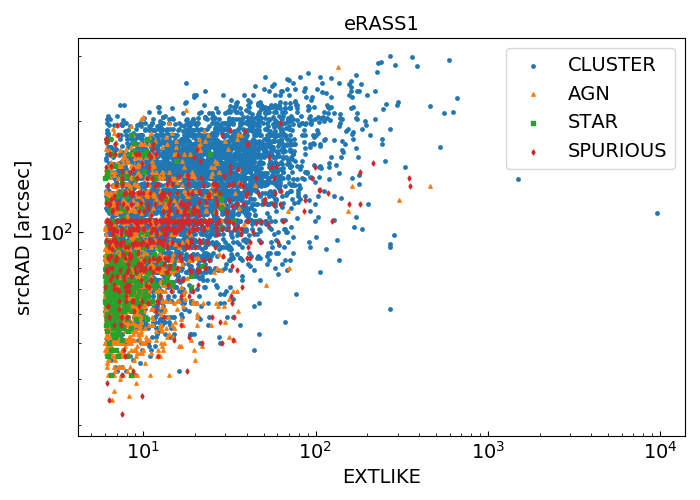}    \includegraphics[width=\columnwidth]{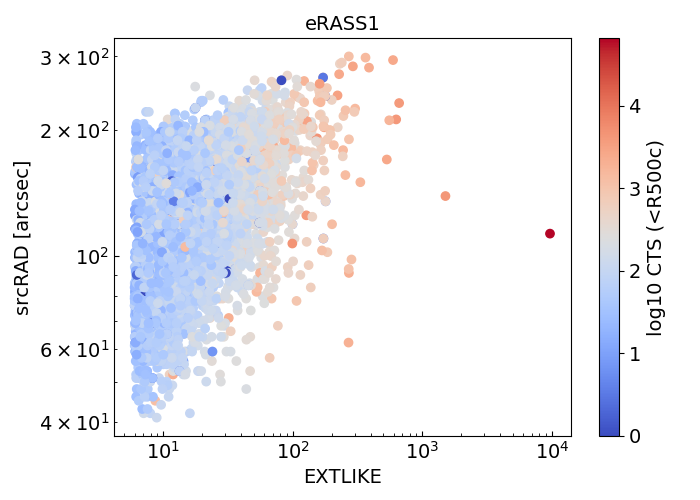}
	\caption{Distribution of the eSASS sources as a function of srcRAD and EXT\_LIKE. \textbf{Top panel}: entire source catalog for the eRASS1 simulation in the srcRAD-EXT\_LIKE parameter space. Clusters are identified by blue circles, AGN by yellow triangles, stars by green squares, and spurious sources plus secondary matches to simulated objects by red diamonds. \textbf{Bottom panel}: detected clusters color-coded by simulated counts in the 0.2--2.3 keV band inside R$_{\rm 500c}$.
	}
    \label{fig:extlike_srcrad}
\end{figure}

Our goal is to characterize a cluster sample that is as pure and complete as possible. On the one hand, we want to maximize the clusters detection rate, making sure that most of the simulated ones are recovered by eSASS. On the other hand, we want to keep the contamination low. This means not only rejecting spurious sources, that are entries in the source catalog that do not correspond to any physical object, but also reducing the contamination due to bright AGN and stars detected as extended objects. Simply applying a high threshold of detection likelihood is not enough to do this, as explained in Sect.\ref{sec:data} and Fig. \ref{fig:population_detlike}. Therefore, we now focus on the catalog of extended sources, with detection likelihood larger than 6. It contains 7731 entries, 75.2$\%$ are clusters, 21.2$\%$ are AGN, 3.6$\%$ are either spurious sources or secondary matches to simulated objects ($\sim$ 0.9$\%$ and 2.7$\%$ respectively, see Fig. \ref{fig:population_detlike}), and 5$\%$ are stars. Our goal is to single out a complete and pure cluster sample in terms of observables, such as properties measured by the eSASS detection algorithm. We focus on two parameters: the source radius and the extension likelihood. We show the entire source population in this parameter space in the left-hand panel of Fig. \ref{fig:extlike_srcrad}. Clusters are identified by blue circles, AGN by yellow triangles, stars by green squares, and spurious sources by red diamonds. Although most of the sources seem to span the entire srcRAD interval, only clusters reach very high values larger than 200 arcseconds. In addition, galaxy clusters populate the high EXT\_LIKE end of this panel. We conclude that a double selection in terms of source radius and extension likelihood is relevant for future cosmological experiments using eROSITA galaxy clusters.
We further study the population of detected clusters in terms of extension likelihood, srcRAD and counts in the right-hand panel of Fig. \ref{fig:extlike_srcrad}.
Clusters with a larger amount of counts are detected at higher values of EXT\_LIKE and show a larger srcRAD. This suggests once again how focusing on the top-right corner of this parameter space, selecting sources with large extension likelihood and source radius, allows one to identify secure clusters emitting a large number of photons. Such correlation also shows the impact on clusters selection of srcRAD. In particular, high count clusters are all located at the high srcRAD end: there are 255 detections with srcRAD > 200 arcseconds and 250 are uniquely matched to a cluster. 
However, objects with less than 100 counts are detected at different values of srcRAD, indicating that this parameter is less relevant in the selection of low count clusters.

\end{document}